\documentclass{aa}  
\usepackage{graphicx}
\usepackage{txfonts}

\newcommand{\orcid}[1]{\href{https://orcid.org/#1}{\includegraphics[width=10pt]{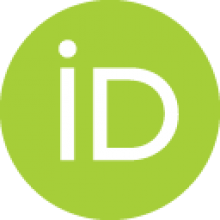}}}
\usepackage{natbib}
\usepackage{multirow}
\usepackage{hyperref}
\hypersetup{
    colorlinks=true,
    linkcolor=black,
    filecolor=magenta,      
    urlcolor=cyan,
    citecolor=blue,
    }
\defcitealias{queiroz21}{Q21}
\defcitealias{perezvillegas2020}{PV20}
\defcitealias{kharchenko16}{KC16}
\defcitealias{cohen21b}{C21}
\defcitealias{vandenberg13}{VB13}
\defcitealias{marinfranch09}{MF09}
\defcitealias{dotter10}{DO10}
\begin{document}

   \title{Chrono-chemodynamical analysis of the globular cluster NGC~6355: Looking for the fundamental bricks of the Bulge
\thanks {Based on observations from ESO Programs 083.D-0063 (A) (PI: S. Ortolani) and 099.D-0136 (A) (PI: M. Valentini), and HST Project GO-11628 (PI:Noyola).}}
  \author{S. O. Souza\inst{1,2}\orcid{0000-0001-8052-969X}
          \and
          H. Ernandes\inst{3,2}\orcid{0000-0001-6541-1933}
          \and
          M. Valentini\inst{1}\orcid{0000-0003-0974-4148}
          \and
          B. Barbuy\inst{2}\orcid{0000-0001-9264-4417}
          \and 
          C. Chiappini\inst{1}\orcid{0000-0003-1269-7282}
          \and
          A. Pérez-Villegas\inst{4}\orcid{0000-0002-5974-3998}
          \and
          S. Ortolani\inst{5,6,7}\orcid{0000-0001-7939-5348}
          \and
          A. C. S. Fria\c ca\inst{2}
          \and
          A. B. A. Queiroz\inst{1,8}\orcid{0000-0001-9209-7599}
          \and
          E. Bica\inst{9}
          }

\institute{
Leibniz-Institut für Astrophysik Potsdam (AIP), An der Sternwarte 16, Potsdam, D-14482, Germany \\ \email{ssouza@aip.de}
\and
Universidade de S\~ao Paulo, IAG, Rua do Mat\~ao 1226, Cidade Universit\'aria, S\~ao Paulo 05508-900, Brazil \\ \email{stefano.souza@usp.br}
\and
Lund Observatory, Department of Astronomy and Theoretical Physics, Lund University, Box 43, SE-221 00 Lund, Sweden
\and
Instituto de Astronom\'ia, Universidad Nacional Aut\'onoma de M\'exico, A. P. 106, C.P. 22800, Ensenada, B. C., M\'exico
\and
Universit\`a di Padova, Dipartimento di Astronomia, Vicolo dell'Osservatorio 2, I-35122 Padova, Italy
\and
INAF-Osservatorio Astronomico di Padova, Vicolo dell'Osservatorio 5, I-35122 Padova, Italy
\and
Centro di Ateneo di Studi e Attivit\`a Spaziali "Giuseppe Colombo" - CISAS. Via Venezia 15, 35131 Padova, Italy
\and
Institut f\"{u}r Physik und Astronomie, Universit\"{a}t Potsdam, Haus 28 Karl-Liebknecht-Str. 24/25, D-14476 Golm (Potsdam), Germany
\and
Universidade Federal do Rio Grande do Sul, Departamento de Astronomia,CP 15051, Porto Alegre 91501-970, Brazil
             }

   \date{Received September 15, 1996; accepted March 16, 1997}

 
  \abstract
   {The information on Galactic assembly time is imprinted on the chemodynamics of globular clusters. This makes them important probes that help us to understand the formation and evolution of the Milky Way. Discerning between in-situ and ex-situ origin of these objects is difficult when we study the Galactic bulge, which is the most complex and mixed component of the Milky Way. To investigate the early evolution of the Galactic bulge, we analysed the globular cluster NGC~6355. We derived chemical abundances and kinematic and dynamic properties by gathering information from high-resolution spectroscopy with FLAMES-UVES, photometry with the Hubble Space Telescope, and Galactic dynamic calculations applied to the globular cluster NGC~6355. We derive an age of $13.2\pm1.1$ Gyr and a metallicity of [Fe/H]$=-1.39\pm0.08$ for NGC~6355, with $\alpha$-enhancement of [$\alpha$/Fe]$=+0.37\pm0.11$. The abundance pattern of the globular cluster is compatible with bulge field RR Lyrae stars and in-situ well-studied globular clusters. The orbital parameters suggest that the cluster is currently confined within the bulge volume when we consider a heliocentric distance of $8.54\pm0.19$ kpc and an extinction coefficient of $R_V = 2.84\pm0.02$. NGC~6355 is highly likely to come from the main bulge progenitor. {Nevertheless, it still} has a low probability of being formed from an accreted event because its age is uncertain and because of the combined [Mg/Mn] [Al/Fe] abundance. Its relatively low metallicity with respect to old and moderately metal-poor inner Galaxy clusters may suggest a low-metallicity floor for globular clusters  that formed in-situ in the early Galactic bulge.}

\titlerunning{Globular cluster NGC~6355}
\authorrunning{Souza et al.}

   \maketitle
%

\section{Introduction}

{The $\Lambda$CDM hierarchical theory of galaxy formation predicts that a galaxy forms from successive mergers of low-mass objects that are absorbed by more massive objects \citep{peebles74,white78,kauffmann93,springel06}. The less massive objects are gradually absorbed while orbiting the massive objects. The Milky Way (MW) contains remnants of this early history that can be divided into two groups: those still orbiting the Galaxy, with their structures entirely or almost intact (e.g. the Magellanic Clouds); and another group of objects that were already dissolved by the MW after several encounters {and were completely} accreted. The latter objects could have retained the dynamic signatures of their progenitor if the merger event occurred during the recent evolution of the Galaxy. An example is Gaia-Sausage-Enceladus (GSE), known as the remnant of the last major merger of the MW with a dwarf galaxy \citep[][]{belokurov18,helmi18}. {Because} the estimated merger time is $\sim 8$ Gyr \citep{gallart19,montalban21}, the remnant stars did not have enough time to change their dynamical properties completely.} 

{In addition to the dynamic properties, mergers influence the chemical properties of the Galaxy \citep[e.g.][]{grand20}. It is expected that an old, moderately metal-poor stellar population will be formed upon the halt in the star formation history. Many authors have tried to identify the chemodynamical imprints of the early assembly steps that are {left on} the Galactic stellar populations with the aim to constrain these important events in the Galaxy history. This is now possible based on the joint information from large spectroscopic surveys and the Gaia proper motion data \citep[e.g.][ among many others]{anders14,hayden15,anders17,kordopatis20,queiroz20,queiroz21,buder22}.}

{The Galactic bulge is one of the most complex regions of the Galaxy because in addition to the high extinction, it contains stellar populations from several parts of the MW. {However}, a study of the bulge can provide information about its complex formation processes \citep[e.g.][]{barbuy2018a,queiroz20,queiroz21,rojas-arriagada20}. In order to distinguish the different stellar  populations, we have to study the object orbit \citep{perez-villegas18} together with ages and chemical composition. The orbits are a key ingredient that provides information whether the object always lived in the bulge. \citet[][hereafter \citetalias{queiroz21}]{queiroz21} mapped and analysed the stellar populations of the bulge  from a chemodynamical point of view, which allowed them to describe the stellar content of the bulge field.}

{Another way to characterize the Galactic bulge comes from the old stellar population, such as RR Lyrae. \cite{savino20} analysed the stellar population in the inner spheroid of the Galaxy and reported that this structure is very old, with an age of $13.41\pm0.54$ Gyr, and it is also metal poor, with a metallicity of [Fe/H]$\sim-1.02$ \citep{pietrukowicz15}, [Fe/H]$\sim-1.0$ \citep{minniti16}, [Fe/H]$\sim-1.55$ \citep{crestani21}, and [Fe/H]$\sim-1.35$ \citep{dekany22}.} The globular clusters (GCs) are also important tracers of the formation and evolution of the Galaxy because they are old and retain the chemodynamical signatures of the first stages of the MW formation. Some studies have demonstrated that the metallicity distribution of bulge GCs peaks at [Fe/H]$\sim -1.0$ \citep[][and references therein]{bica16,perezvillegas2020}, and that they are mostly older than $12.5$ Gyr \citep[][]{miglio16,barbuy16,barbuy2018a,kerber19,ortolani19,oliveira20,fernandez-trincado20,fernandez-trincado21,souza21}.

{The assignment to which Galactic component a GC belongs to is made depending on {its orbital integration} in order to verify the most probable regions of its trajectory. While part of the GCs could have formed in the main progenitor of the Galaxy \citep[e.g. main bulge or main disk;][]{massari19}, others could come from accreted progenitors. To study the origin of a GC, we therefore need to analyse its chemical, photometric, and dynamical properties \citep[e.g.][]{souza21}. }

{The age-metallicity relation (AMR) of the MW GCs shows a bifurcation that splits it into two main groups \citep{marinfranch09,forbes2010,leaman13}. A steeper branch in which  more older GCs are concentrated is associated with an in-situ population. In contrast, the other component is broader and includes very young to old ages. It is also associated with accretion events during the early evolution of the Galaxy \citep{kruijssen19,massari19,forbes20,limberg22,callingham22}. When an axisymmetric Galactic potential is employed, the so-called integrals of motion space (IOM) can be used together with the AMR. By studying the total energy ($E$) versus Z-component of the angular momentum ($L_Z$), it is possible to investigate the dynamic history of the Galaxy. For example, the region of lower $E$ with almost zero $L_Z$ can be associated with the inner part of the MW, the bounded objects. On the other hand, the Galactic halo accreted objects (e.g. GSE) are in the region of high $E$. Therefore, the combination of the AMR with the IOM space has improved the knowledge about the origin of the GCs system, helping us to understand the Galactic evolutionary history, particularly that of the Galactic bulge.}

 Observing GCs within the Galactic bulge is difficult because the extinction tends to {hide} the objects. One example is NGC~6355 (also called GCl-63 and ESO 519-SC15), projected towards the direction of the Galactic bulge ($l = 359.58^{\circ}$, $b = +5.43^{\circ}$) with a relatively high extinction \citep[$E(B-V)=0.79$;][]{harris96}. NGC~6355 is a well-known cluster that has been studied since the 1900s. It was classified as a probable open cluster \citep{shapley1919}. However, it did not take long before its globular nature was confirmed based on its relatively high mass, which according to \cite{baumgardt18} is $1.01\times10^{5}$ M$_{\odot}$. \cite{djorgovski86} classified NGC~6355 as a core-collapse cluster. This result was recently confirmed by \cite{cohen21a} using the Hubble Space Telescope (HST) filters F606W and F814W from the Advanced Camera for Survey (ACS).

{\cite{ortolani03} analysed the horizontal branch (HB) and the red giant branch (RGB) of NGC~6355 using a [$V$,$V-I$] colour-magnitude diagram (CMD). They obtained 
a reddening of $E(B-V)=0.78$, a distance of d$_{\odot} = 8.8$ kpc, and a metallicity of [Fe/H]$\sim -1.3$. This was deduced by comparing the cluster mean locus with the mean loci of the well-studied clusters NGC~6171 and M~5. 
Assuming their distance derivation, the authors concluded that the cluster is near the Galactic center \citep[see also][]{bica06}. \cite{valenti07} analysed the RGB slope and the K magnitude of the RGB tip using the [$K$,$J-K$] and [$H$, $J-H$] {CMDs}. They found $E(B-V)=0.82$, d$_{\odot}=8.7$kpc, and [Fe/H]$=-1.42$. Both results agree with the metallicity scales of \cite{carretta09b} and \cite{zinn84} of [Fe/H]$= -1.33\pm0.14$ and [Fe/H]$=-1.50\pm0.15$, respectively. Subsequent metallicity derivation by \cite{vasquez15} and \cite{dias16} of [Fe/H]$\sim-1.49$ and [Fe/H]$\sim-1.46$, respectively, are also within the range of both metallicity scales. }

{ \cite{barbuy09} identified NGC~6355 as a blue horizontal branch (BHB) metal-poor GC, located in the ring at $-6^{\circ}$ -- $-12^{\circ}$ around the Galactic centre. This suggested that NGC~6355 belonged to the BHB moderately metal-poor clusters of the Galactic bulge, such as NGC~6558 \citep{barbuy07,barbuy2018b}, HP~1 \citep{barbuy06,barbuy16}, AL~3 \citep{ortolani06,barbuy2021}, Terzan~9 \citep{ernandes19}, and UKS~1 \citep{fernandez-trincado20}. Nevertheless, when examined from the orbital viewpoint, it was suggested that NGC~6355 is more compatible with the Galactic thick disk with
a probability of $93\%$ \citep[][hereafter \citetalias{perezvillegas2020}]{rossi15,perez-villegas18,perezvillegas2020}, assuming a distance of $8.70\pm0.87$ kpc. It
also has a probability of $7\%$ to be part of the Galactic bulge. Here we stress the importance of having a precise distance derivation.}

{\citet[][hereafter \citetalias{kharchenko16}]{kharchenko16} analysed $147$ GCs including NGC~6355 using integrated $JHK_s$ magnitudes. They derived its age as $\log\,t = 10.10$ ($\sim 12.5$ Gyr). Assuming this age derivation and the distances derived by \cite{baumgardt18}, \cite{massari19} found that NGC~6355 may have been formed from the main-bulge progenitor and might therefore be an in-situ cluster. Their result for NGC~6355 was confirmed with a more realistic approach adopted in \cite{moreno22}, who employed the formalism for dynamical friction. More recently, \citet[][hereafter \citetalias{cohen21b}]{cohen21b} derived a relative age of $1.1$ Gyr by comparing the CMDs of NGC~6355 and NGC~6205.
The authors give an absolute age of $\sim 13.2$ Gyr for NGC~6355 and assume an age of 12.1 Gyr for NGC~6205 \citep[][hereafter \citetalias{vandenberg13}]{vandenberg13}. This relatively older age compared to the previous one by \citetalias{kharchenko16} was used by \cite{callingham22} to reclassify NGC~6355 as compatible with the main-bulge progenitor and {also with the} Kraken accreted structure as an alternative origin. It is worth noting that a possible accreted structure within the Galactic bulge was hypothesized also by \cite{massari19} (low-energy progenitor), \cite{kruijssen19} (Kraken), \cite{forbes20} (Koala), and \cite{horta21} (Heracles).}

{In the present work, we combine the chemical information with photometric and dynamical properties of the cluster to constrain its history. 
The chemical information is based on the UVES spectrograph \citep{dekker2000} in FLAMES-UVES mode at the ESO-VLT, the photometry on HST data, and the dynamical properties are provided by orbital integration employing the \cite{mcmillan17} Galactic potential. }

{This paper is organized as follows. The photometry processing, reduction of spectra, and membership analysis of the observed stars are described in Section \ref{sec_data}. Section \ref{sec:photo_param} gives the derivation of fundamental parameters. The analysis of individual line abundances and the comparison with the literature are described in Section \ref{sec:chemical_analysis}. The orbital analysis and dynamical properties of NGC~6355 are presented in Section \ref{sec:dynamics}. In Section \ref{sec:discussion} we discuss the origin of the cluster. Finally, the conclusions are drawn in Section \ref{sec:conclusions}.}

\section{Data}\label{sec_data}

\subsection{HST photometry processing}\label{sec:phot}

\begin{figure}
    \centering
    \includegraphics[width=\columnwidth]{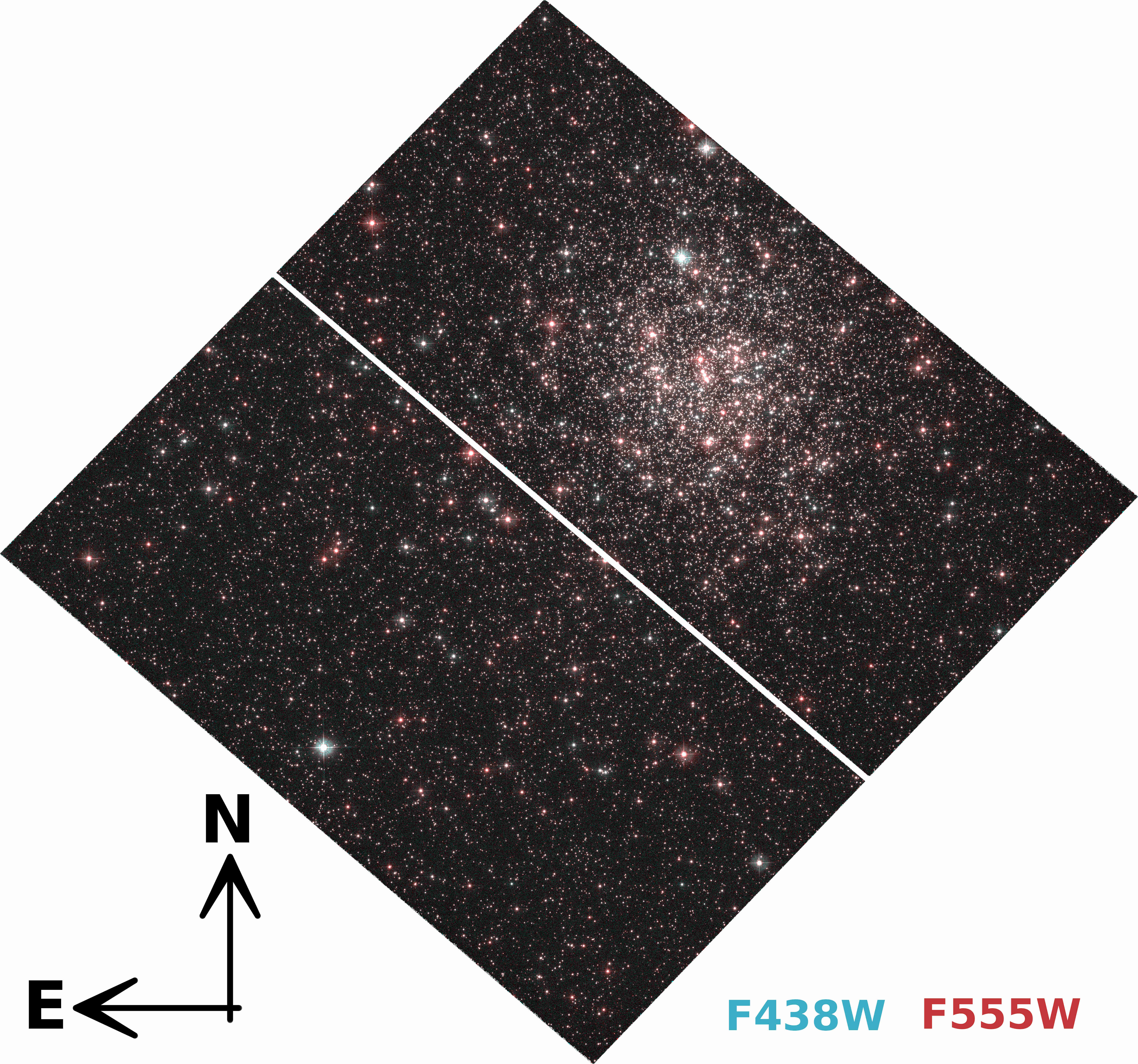}
    \caption{$F438W/F555W$ combined colour image from the HST WFC3 camera for NGC~6355.} 
    \label{fig:image_hst}
\end{figure}

\begin{figure*}
    \centering
    \includegraphics[width=1\textwidth]{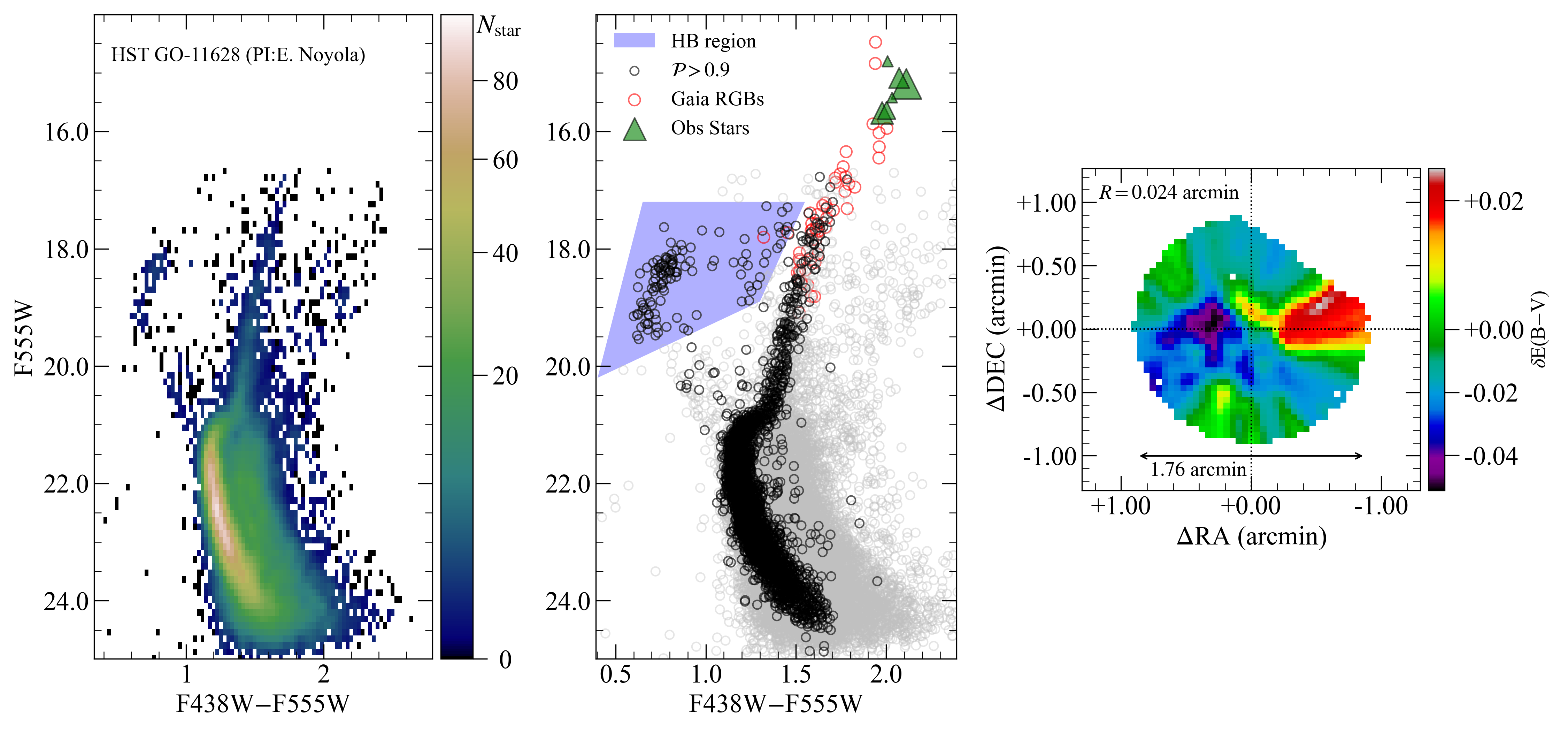}
    \caption{{Photometric data processing. Left panel: All stars within the FOV obtained from the HST Project (GO-11628, PI: Noyola). Middle panel: Final differential-reddening-corrected CMD with selected stars (black) and discarded stars (grey). The Gaia EDR3 member stars from the \cite{vasiliev21} catalogue matched with 2MASS to obtain the HST are shown in red. The green triangles represent the observed stars with HST magnitudes obtained from the isochrone calibration, and the sizes are from the S/N. The HB region is plotted in blue. Right panel: Differential reddening map for stars within a half-light radius. The resolution of the map is 0.024 arcminutes (1.44 arcseconds). }}
    \label{fig:phot}
\end{figure*}

{The photometric data for NGC~6355 were retrieved from the HST Project (GO-11628, PI:Noyola), which used the Wide Field Camera for Surveys 3 (WFC3) with the filters F438W and F555W. {The observation consists of three F438W images with an exposure time of $440$ s,
and three F555W images with an exposure time of $80$ s. } Figure \ref{fig:image_hst} shows the colour image composed of the combined HST images. We performed a further selection based on the pipeline described in \cite{nardiello18} using the quality-of-fit and photometric error parameters to select well-measured stars and reject poor measurements (top left panel of Figure \ref{fig:phot}). Additionally, we selected stars within a half-light radius of $0.88$ arcmin \citep{harris96} to avoid a substantial number of field stars. For the resulting sample, we computed a simple membership probability by combining the {stars offset from} the fiducial line on the CMD with the star distance to the cluster centre.}

{The extinction towards the cluster is relatively high, and it increases the CMD spread. To reduce the effect of differential reddening, we used the same method as was applied to Palomar~6 in \cite{souza21} \citep[adapted from ][]{milone12,bedin17}. The differential reddening map (bottom left panel of Figure \ref{fig:phot}) shows that $\delta E(B-V) \sim -0.04,$ which is approximately $5\%$ of the expected reddening \citep[E(B-V)$=0.79;$][]{harris96}, and which we can convert into a magnitude difference of $\delta {\rm m}_{\rm F438W}=+0.17$ and $\delta {\rm m}_{\rm F555W}=+0.13$, and into a difference in colour $\delta \left( {\rm m}_{\rm F438W} - {\rm m}_{\rm F555W} \right) = +0.04$. }

{Finally, to scale the photometry to the same zero-point as in the evolutionary models, we converted the AB magnitudes into the Vega system. The final sample corrected for differential reddening shows a smaller spread and a clear morphology from the RGB and HB to the lower MS (top right panel of Figure \ref{fig:phot}).}

{The ACS F438W/F555W photometry is saturated for magnitudes brighter than F555W$\sim 17$. Therefore, our {spectroscopic targets} were not observed for these filters. To estimate the position of our stars in the CMD, we derived an approximation of their F438W and F555W magnitudes. We fixed the reddening, metallicity, and distance modulus from \cite{harris96}. For each star, we fitted the magnitudes $J$, $K_S$, G, G$_{BP}$, and G$_{RP}$ (green triangles in Figure \ref{fig:phot}). We also used this method for a sample of RGB stars of the Gaia EDR3 from the \cite{vasiliev21} catalogue (open red circles). It is worth noting that the F438W filter is affected by variations in C, N, and O abundances. Hence, this filter can better be estimated via spectral convolution and integration with the filter response curve.}


\subsection{Spectral data reduction}\label{sec:specs}

The UVES spectra were obtained using the FLAMES-UVES setup centred at 580 nm, covering the wavelength range 480  - 680 nm, from the ESO Programs 083.D-0063 (A) (PI: S. Ortolani) and 099.D-0136 (A) (PI: M. Valentini). The latter ESO program was coordinated with the program GO11126 (PI: M. Valentini) for campaign 11 of the K2 satellite \citep[K2 is the repurposed \textit{Kepler} mission;][]{Howell2014}: the goal was to obtain asteroseismology for the giants in the sample GCs. However, obtaining reliable light curves for these stars was not possible. The log of observations is given in Table \ref{tab:logobs}. 

 \begin{table}
    \caption{Log of the spectroscopic FLAMES-UVES observations of programs 083.D-0063 (A) and 099.D-0136 (A), carried out in 2009 and 2017, respectively. The reported seeing and airmass are the mean values in the exposures. The last column contains the corresponding GIRAFFE setup, in which additional stars were observed.}
    \small
    \centering
    \begin{tabular}{lccccc}
    \noalign{\smallskip}
    \hline
    \noalign{\smallskip}
    \hline
    \noalign{\smallskip}
    Date & UT & exp    & Airmass  & Seeing & SETUP \\
         &    &   ( s )  &          & ($''$) & GIRAFFE  \\
    \noalign{\smallskip}
    \hline
    \noalign{\smallskip}
    \multicolumn{6}{c}{ Program 083.D-0063 (A) } \\
    \hline
    \noalign{\smallskip}
    2009-09-02 & 02:48:43 & 2700 & 1.455 & 1.88 & H13-1  \\
    2009-09-01 & 01:03:00 & 2700 & 1.184 & 0.87 & H13-2  \\
    2009-09-01 & 01:50:54 & 2700 & 1.191 & 0.72 & H13-3  \\
    2009-09-13 & 23:32:32 & 2700 & 1.091 & 0.91 & H13-4  \\
    2009-09-14 & 00:31:12 & 2700 & 1.182 & 0.82 & H14-1  \\
    2009-09-14 & 01:17:51 & 2700 & 1.467 & 0.78 & H14-2  \\
    2009-09-14 & 02:04:21 & 2700 & 1.848 & 0.75 & H14-3  \\
    \noalign{\smallskip}
    \hline
    \noalign{\smallskip}
    \multicolumn{6}{c}{ Program 099.D-0136 (A) } \\
    \hline
    \noalign{\smallskip}
    2017-07-14 & 06:21:39 & 2400 & 1.751 & 0.75 & H11-1  \\
    2017-07-14 & 04:34:44 & 2400 & 1.172 & 0.67 & H11-2  \\
    2017-09-02 & 01:50:12 & 2400 & 1.279 & 0.61 & H11-4  \\
    2017-09-07 & 02:53:12 & 2400 & 1.831 & 0.54 & H13-1  \\
    \noalign{\smallskip}
    \hline
    \noalign{\smallskip}
    \hline
    \end{tabular}
    \label{tab:logobs}
\end{table}

We performed the FLAMES-UVES data reduction procedure using the ESO-Reflex software with the UVES-Fibre pipeline \citep{ballester00,modigliani04}. The corresponding spectra of each star were corrected for the radial velocity computed using the Python library \texttt{PyAstronomy}. The radial velocities were obtained by cross-correlating the stellar spectra with the Arcturus spectrum \citep{hinkle00}. The values of the heliocentric radial velocity of each spectrum and their mean values are presented in Table \ref{tab:velocities} for the member stars, selected from the membership analysis (see section \ref{sec:memb}). 

{The spectra of stars 1546 and 1239 from ESO Program 083.D-0063, have a low signal-to-noise ratio (S/N; < 15), which is significantly lower than those obtained from ESO Program 099.D-0136. The spectra of these two stars are therefore strongly affected by noise, {which makes it very difficult} to distinguish strong lines and prevents a satisfactory radial velocity derivation from the cross-correlation method. For consistency, they can therefore not be confirmed as
members of NGC~6355 given the uncertainties in their radial velocity values even though these stars are considered members from the proper-motion membership check. Consequently, the final observed star sample is composed of the four stars of ESO Program 099.D-0136.}

{Based on our final sample, we found a mean heliocentric radial velocity for NGC~6355 of $-193.2\pm1.1$ km s$^{-1}$, which agrees well with the value of $-194.6\pm1.2$ km s$^{-1}$ obtained from the individual stars of Gaia DR2\footnote{\url{https://people.smp.uq.edu.au/HolgerBaumgardt/globular/appendix/ngc6355.txt}}. Finally, the normalized spectra were combined and were weighted by the median flux to obtain the final stellar spectra.}

\begin{table}
\caption{Heliocentric radial velocity obtained for each extracted spectrum and the average value for each star. }
\centering
\begin{tabular}{lcclcc}

\hline
\hline
Target     &  V$_{r}^{hel}$   & $\sigma_{V_{r}}$ & Target  & V$_{r}^{hel}$    & $\sigma_{V_{r}}$ \\
           &  km s$^{-1}$     &  km s$^{-1}$     &         &  km s$^{-1}$     &  km s$^{-1}$     \\

\hline
1546\_1    &   $-227.40$    &   $1.40$   &  1239\_1      &  $-66.95 $  &   $0.19$   \\
1546\_2    &   $-29.98 $    &   $0.70$   &  1239\_2      &  $-192.62$  &   $0.56$   \\
1546\_3    &   $-216.96$    &   $0.34$   &  1239\_3      &  $-68.47 $  &   $0.31$   \\
1546\_4    &   $-318.81$    &   $0.42$   &  1239\_4      &  $-192.17$  &   $0.90$   \\
1546\_5    &   $-166.55$    &   $0.35$   &  1239\_5      &  $-187.30$  &   $0.22$   \\
\hline
1546  & ${-216.96}$ & ${94.72}$   &  {1239}   & ${-187.30}$ & ${60.28}$ \\
\hline
133\_1    &   $-192.92$    &   $0.47$   &  1176\_1      &  $-196.35$  &   $0.56$   \\
133\_2    &   $-191.65$    &   $0.52$   &  1176\_2      &  $-196.85$  &   $0.65$   \\
133\_3    &   $-192.80$    &   $0.48$   &  1176\_3      &  $-193.25$  &   $0.69$   \\
133\_4    &   $-192.03$    &   $0.45$   &  1176\_4      &  $-193.79$  &   $0.68$   \\
\hline
{133}  & ${-192.41}$ & ${0.53}$  &  {1176}    & ${-195.07}$ & ${1.56}$ \\
\hline
1539\_1     &   $-192.25$   &   $0.41$   &  1363\_1      &  $-193.93$  &   $0.41$   \\
1539\_2     &   $-192.36$   &   $0.40$   &  1363\_2      &  $-194.01$  &   $0.40$   \\
1539\_3     &   $-192.30$   &   $0.41$   &  1363\_3      &  $-192.44$  &   $0.40$   \\
1539\_4     &   $-191.90$   &   $0.41$   &  1363\_4      &  $-192.34$  &   $0.39$   \\
\hline
{1539}  & ${-192.27}$ & ${0.18}$   &  {1363}    & ${-193.18}$ & ${0.79}$ \\
\hline
\hline
\end{tabular}
\label{tab:velocities}
\end{table}

\subsection{Membership selection}\label{sec:memb}

{The power of Gaia astrometry has been demonstrated in different ways, such as in the search for new open clusters and in the selection of the most probable members of a GC. In particular, regarding the latter, Gaia was not available until recent years, and now the membership probabilities should be verified in all samples preceding the Gaia era, and in particular for our sample stars.}

{To remove bias from our sample, we performed a membership analysis to determine which stars observed in both ESO programs are members of NGC~6355. Considering both programs, we have a total of nine stars. We selected the Gaia DR3 stars within $10'$ from the cluster center, and we applied the Gaussian mixture models \citep[GMM;][]{Pedregosa11} clustering method to separate the cluster members from the field stars. The derived mean proper-motion for NGC~6355 is $<\mu_{\alpha}^{*}> = -4.76 \pm 0.06$ mas~yr$^{-1}$ and $<\mu_{\delta}> = -0.58\pm0.05$ mas~yr$^{-1}$. This agrees very well with the new values computed by \cite{vasiliev21}. }

\begin{figure}
    \centering
    \includegraphics[width=\columnwidth]{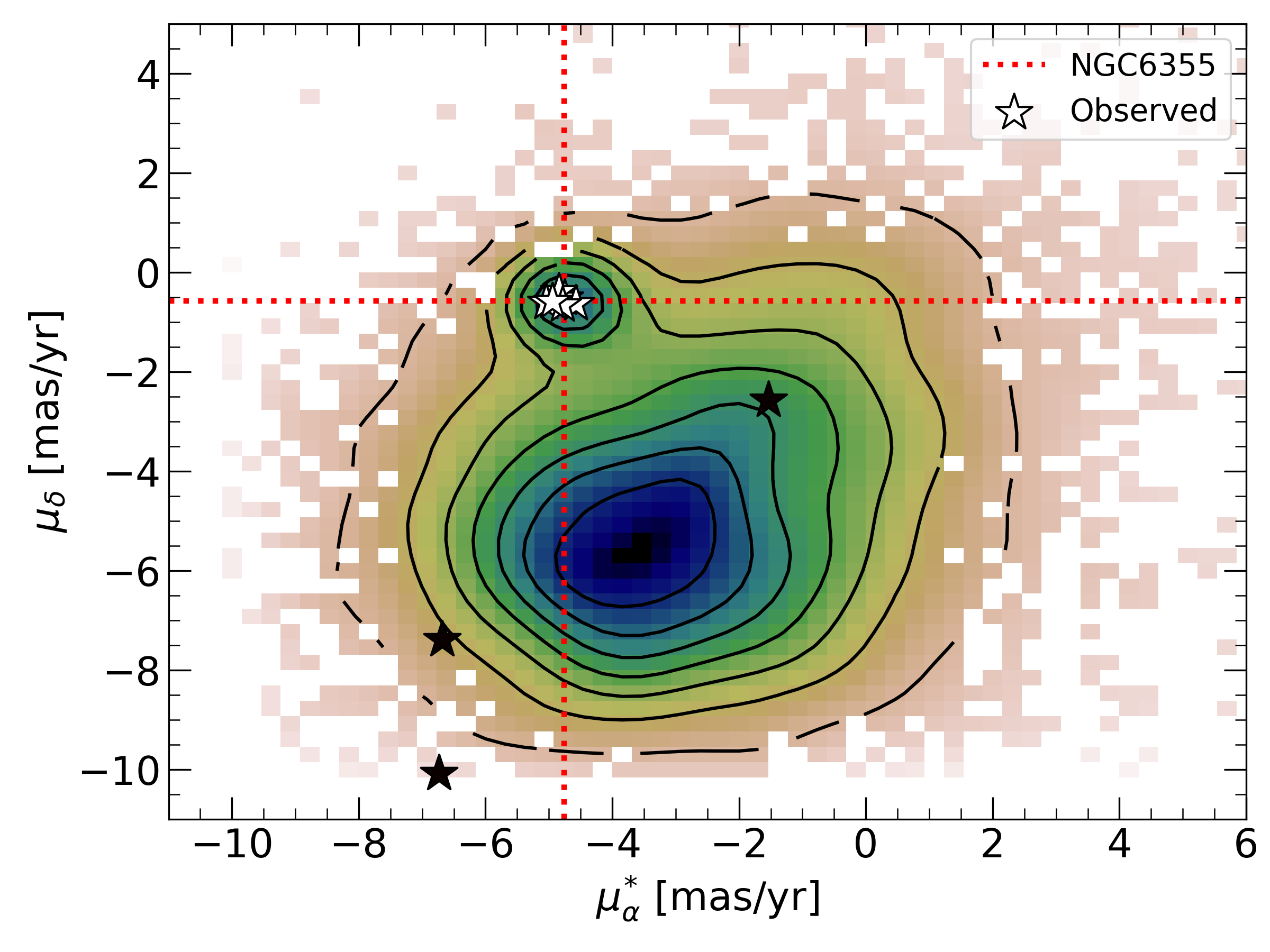}
    \caption{{Proper-motion density map from Gaia DR3. The stars show all the observed stars in both programs (members are plotted in white, and non-members are given in black). The red lines show the position of the mean proper motion of NGC 6355.}}
    \label{fig:vpd}
\end{figure}

{The membership probabilities were computed considering cluster and field distributions, following the method presented in \cite{bellini09}. When we had determined the membership probability, we cross-matched our sample stars with the Gaia data (Table \ref{starmag}), which are indicated with stars in Figure \ref{fig:vpd}. We found that six of nine stars from both programs have membership probabilities above $80\%$. Combining the information of radial velocity and the proper-motion membership probability, we therefore disregard the non-member stars in the following analysis.}

\section{Fundamental parameters} \label{sec:photo_param}

\subsection{Atmospheric stellar parameters}

\subsubsection{Stellar magnitudes}

{The photometric effective temperature ($T_{\rm eff}$) and surface gravity ($\log g$) were derived from  the $VIJHK_S$ magnitudes given in Table \ref{starmag}. For comparison purposes, we obtained the $T_{\rm eff}$ from the Transiting Exoplanet Survey Satellite (TESS) input catalogue \citep[TIC;][]{stassun18} for our sample. The 2MASS $J$, $H$, and $K_S$ magnitudes were taken from \citet{skrutskie06}. To obtain the $T_{\rm eff}$ from a wide wavelength range, we calculated the colour $V-I$ employing the photometric systems relations $G - V = f(G_{BP} - G_{RP})$ and $G - I = f(G_{BP} - G_{RP})$ from Gaia EDR3 \citep{riello2021}.}

\begin{table*}
\caption[1]{\label{starmag}Identifications, coordinates, magnitudes from $JHK_{s}$ 2MASS survey, $VI$, HST/ACS, and matched Gaia DR3 information. The first two stars are from program 083.D-0063 (A), and the four last stars are from 099.D-0136 (A).}
\resizebox{\linewidth}{!}{%
\small
\tabcolsep 0.15cm
\begin{tabular}{lccccccccccccccc}
\hline
\hline
ID          &     ID              &   RA          &       DEC     &   $V$   &   $V-I$ &  $J$     &  $H$   &  $K_S$    &   F438W       &   F555W          & $^\dag \mu_{\alpha}^{*}$   & $\mu_{\delta}$   &   $G$   &  BP$-$RP  & S/N   \\
            &    2MASS            &   (deg)       &      (deg)    &         &         &  \multicolumn{3}{c}{ 2MASS }  &  \multicolumn{2}{c}{ HST/WFC3 }   &         \multicolumn{2}{c}{(mas yr$^{-1}$)}   &         &           &         \\
\hline                                                                                                                                                                    
1546        &  $17235883-2620183$ & $260.996$     & $-26.338$     & $15.06$ & $2.24$  & $11.359$ & $10.45$ & $10.19$  &    $17.46$    &    $15.42$       & $-4.747$                   & $-0.523$         & $14.32$ & $2.39$    & $10.33$  \cr
1239        &  $17240227-2621267$ & $261.010$     & $-26.357$     & $14.40$ & $2.50$  & $10.284$ &  $9.25$ &  $8.92$  &    $16.81$    &    $14.81$       & $-4.839$                   & $-0.394$         & $13.51$ & $2.66$    & $12.05$ \cr
\hline                                                                                                                                                                                                    
1539        &  $17235356-2620223$ & $260.973$     & $-26.339$     & $14.82$ & $2.25$  & $10.942$ & $10.12$ &  $9.73$  &    $17.30$    &    $15.19$       & $-4.780$                   & $-0.659$         & $14.08$ & $2.40$    & $79.19$  \cr  
1363        &  $17240101-2620597$ & $261.004$     & $-26.349$     & $14.74$ & $2.34$  & $10.892$ & $9.944$ &  $9.63$  &    $17.16$    &    $15.09$       & $-4.942$                   & $-0.591$         & $13.95$ & $2.49$    & $44.93$  \cr
1176        &  $17235712-2621441$ & $260.988$     & $-26.362$     & $15.28$ & $2.11$  & $11.684$ & $10.90$ & $10.59$  &    $17.66$    &    $15.69$       & $-5.041$                   & $-0.609$         & $14.60$ & $2.26$    & $51.33$  \cr
133         &  $17235528-2621088$ & $260.980$     & $-26.352$     & $15.30$ & $2.24$  & $11.435$ & $10.62$ & $10.21$  &    $17.64$    &    $15.64$       & $-4.572$                   & $-0.635$         & $14.56$ & $2.39$    & $36.36$  \cr
\noalign{\hrule}
 \hline
\end{tabular}    }                                                     
  
$^\dag \mu_{\alpha}^{*} = \mu_{\alpha} \cos \delta. $             
\end{table*}

\subsubsection{Photometric effective temperatures $T_{\rm eff}$ and gravities $\log~g$}\label{sec:tefflogg}

{The $T_{\rm eff}$ values were derived from $V-I$, $V-K_S$, and $J-K_S$  colour-temperature calibrations of \citet{casagrande10}. To use the calibrations, we must perform the reddening corrections. For NGC~6355, we assumed the metallicity [Fe/H]$\, = -1.33$, $E(B-V)=0.77$, and $(m-M)_V=17.21$  from
\cite{harris96} . 
Table \ref{tabteff} lists the derived photometric effective temperatures. The $<T_{\rm eff}>$ value given in the fifth column is the mean effective temperature without the TESS values (which are too hot).}

\begin{table*}
\centering
\caption{ Photometric parameters derived using calibrations by \citet{casagrande10} for $V-I$, $V-K$, $J-K$ colours are given in columns 2-8. In columns 9-14 are given the spectroscopic stellar parameters.}
\label{tabteff}
\small
\resizebox{\linewidth}{!}{%
\tabcolsep 0.1cm
\begin{tabular}{lcccccccccccccc}
\hline
\hline
 & \multicolumn{7}{c}{Photometric parameters} & & \multicolumn{6}{c}{Spectroscopic parameters} \cr
 \cline{2-8}  \cline{10-15}   
\noalign{\vskip 0.1cm}

ID & T$_{(V-I)}$ & T$_{(V-K_S)}$ & T$_{(J-K_S)}$ & $<$T$_{\rm eff}>$ & ${\rm BC_V}$ & ${\rm M_{bol}}$ & $\log g$ & & T$_{\rm eff}$ & $\log g$ & [\ion{Fe}{I}/H] & [\ion{Fe}{II}/H] & [Fe/H] &  $v_t$ \cr
   &   (K)       &   (K)       &   (K)       &   (K)             &              &                 &          & &      (K)      &          &                 &                  &        & (km s$^{-1}$)    \cr
\hline
\noalign{\vskip 0.1cm}
1539 & $4359$ & $4330$ & $4297$ & $4330$ & $-0.615$ & $-3.02$ & $0.74$ & &    $4300\pm65$ & $0.87\pm0.23$ & $-1.35\pm0.11$ & $-1.33\pm0.18$ & $-1.34\pm0.15$ & $1.0\pm0.1$  \cr
1363 & $4246$ & $4315$ & $4152$ & $4246$ & $-0.702$ & $-3.19$ & $0.64$ & &    $4296\pm76$ & $0.84\pm0.24$ & $-1.36\pm0.09$ & $-1.35\pm0.02$ & $-1.36\pm0.07$ & $1.2\pm0.1$  \cr
1176 & $4573$ & $4642$ & $4660$ & $4642$ & $-0.481$ & $-2.43$ & $1.10$ & &    $4580\pm69$ & $1.20\pm0.26$ & $-1.48\pm0.08$ & $-1.48\pm0.23$ & $-1.48\pm0.17$ & $1.0\pm0.1$  \cr
133  & $4373$ & $4328$ & $4250$ & $4328$ & $-0.606$ & $-2.53$ & $0.94$ & &    $4378\pm76$ & $1.24\pm0.19$ & $-1.46\pm0.07$ & $-1.44\pm0.17$ & $-1.45\pm0.13$ & $0.9\pm0.1$  \cr
\noalign{\vskip 0.01cm}
\noalign{\hrule}
 \hline
\end{tabular}}
\end{table*}


To derive the photometric $\log g$ value, we used the classical ratio $\log(g_* / g_\odot),$ where $\log g_\odot = 4.44$ is
\begin{eqnarray}
\log g_* = 4.44+4\log \frac{T_{\rm eff}*}{T_{\odot}}+0.4(M_{\rm bol}-M_{\rm bol\odot})+\log \frac{M_*}{M_{\odot}}.
\end{eqnarray}

We adopted the values of <$T_{\rm eff}$> from Table \ref{tabteff}, $M_*=0.85 M_{\odot}$ and $M_{\rm bol \odot} = 4.75$. The derived values of the photometric $T_{\rm eff}$ and $\log g$ are given in the left columns of Table \ref{tabteff}.

\subsubsection{Spectroscopic stellar parameters}

{ The final spectroscopic stellar parameters T$_{\rm eff}$, $\log g$, and the microturbulence velocity v$_{\rm t}$ of NGC~6355 were derived together with [Fe/H] based on excitation and ionization equilibria. Equivalent widths (EW) for a list of lines of \ion{Fe}{I} and \ion{Fe}{II} lines were measured using DAOSPEC \citep{stetson08}. Using a visual inspection of the stellar spectrum, we remeasured some lines with the IRAF routine to evaluate the impact of blending lines, mainly for \ion{Fe}{II}, and some lines that were poorly fitted with DAOSPEC. The employed lines are listed in the appendix (Table \ref{tab:fe}) with the adopted oscillator strengths ($\log gf$) for \ion{Fe}{I} lines obtained from the VALD3 and NIST databases \citep{piskunov95, martin02} and for \ion{Fe}{II} lines from \citet{melendez09}.}

{ We extracted 1D photospheric models for our sample using the MARCS grid of atmospheric models \citep{gustafsson08}. The adopted CN-mild models consider [$\alpha$/Fe]$=+0.20$ for [Fe/H]$=-0.50$ and [$\alpha$/Fe]$=+0.40$ for [Fe/H]$\leq -1.00$. For the solar Fe abundance, we adopted $\epsilon(\rm Fe) = 7.50$ \citep{grevesse98}.}

\begin{figure}
    \centering
    \includegraphics[width=\columnwidth]{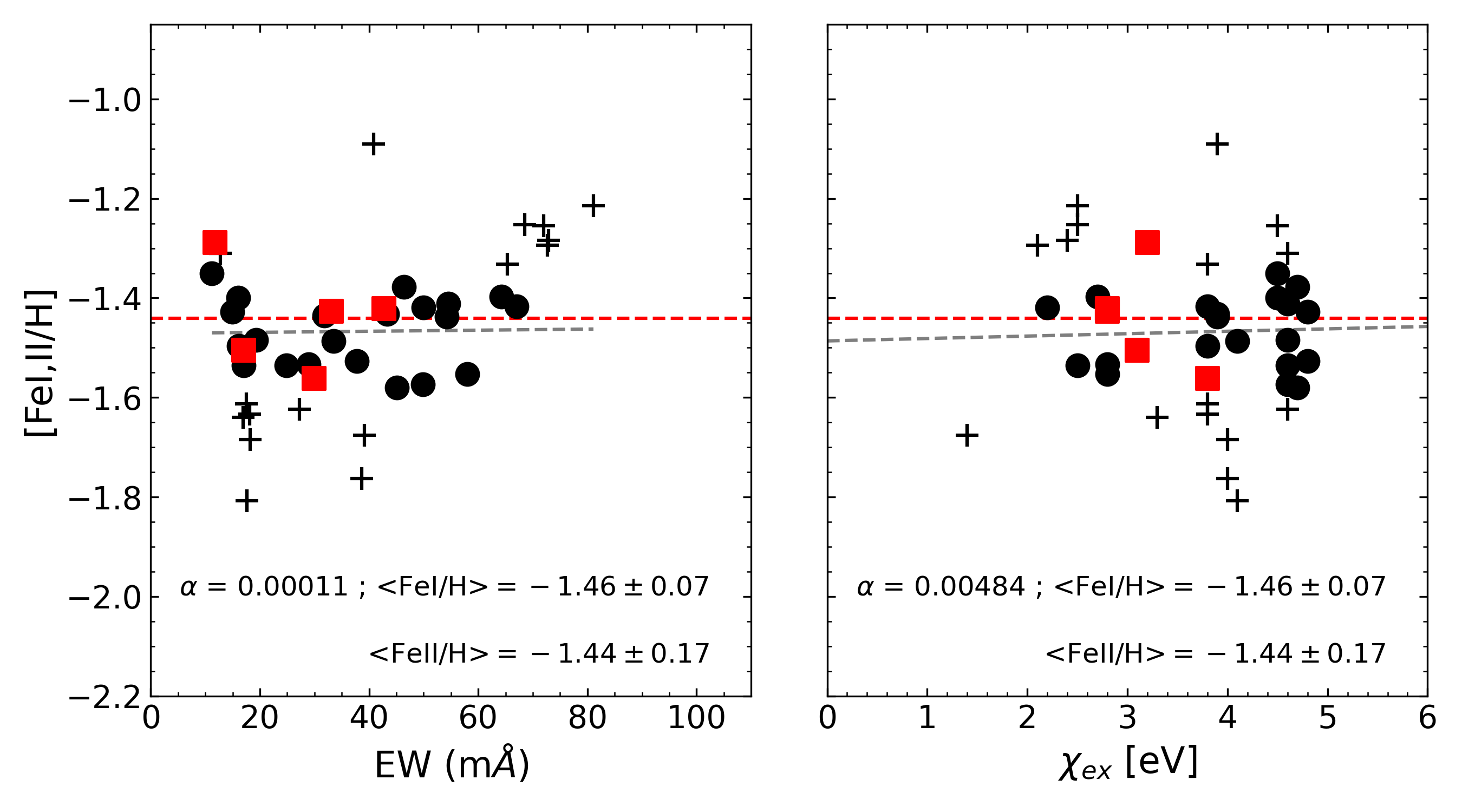}
    \caption{{Ionization and excitation equilibria for NGC~6355 star 133. The black dots and red squares correspond to the  [FeI/H]  and [FeII/H] lines, respectively. {The crosses are the FeI lines that were excluded through a $3\sigma$ clipping method.}  }}
    \label{fig:equi}
\end{figure}

{The mean photometric <T$_{\rm eff}$> and $\log g$ values calculated in Section \ref{sec:tefflogg} were assumed as initial guesses to derive the spectroscopic parameters. The method consists of obtaining the excitation and ionization equilibrium of \ion{Fe}{I} and \ion{Fe}{II} lines. Figure \ref{fig:equi} shows the excitation and ionization equilibrium for star 133. The derived spectroscopic parameters T$_{\rm eff}$, $\log g$, [\ion{Fe}{I}/H], [\ion{Fe}{II}/H], [Fe/H], and $v_t$ are presented in the right columns of Table \ref{tabteff}.}

\begin{figure}
    \centering
    \includegraphics[width=0.99\columnwidth]{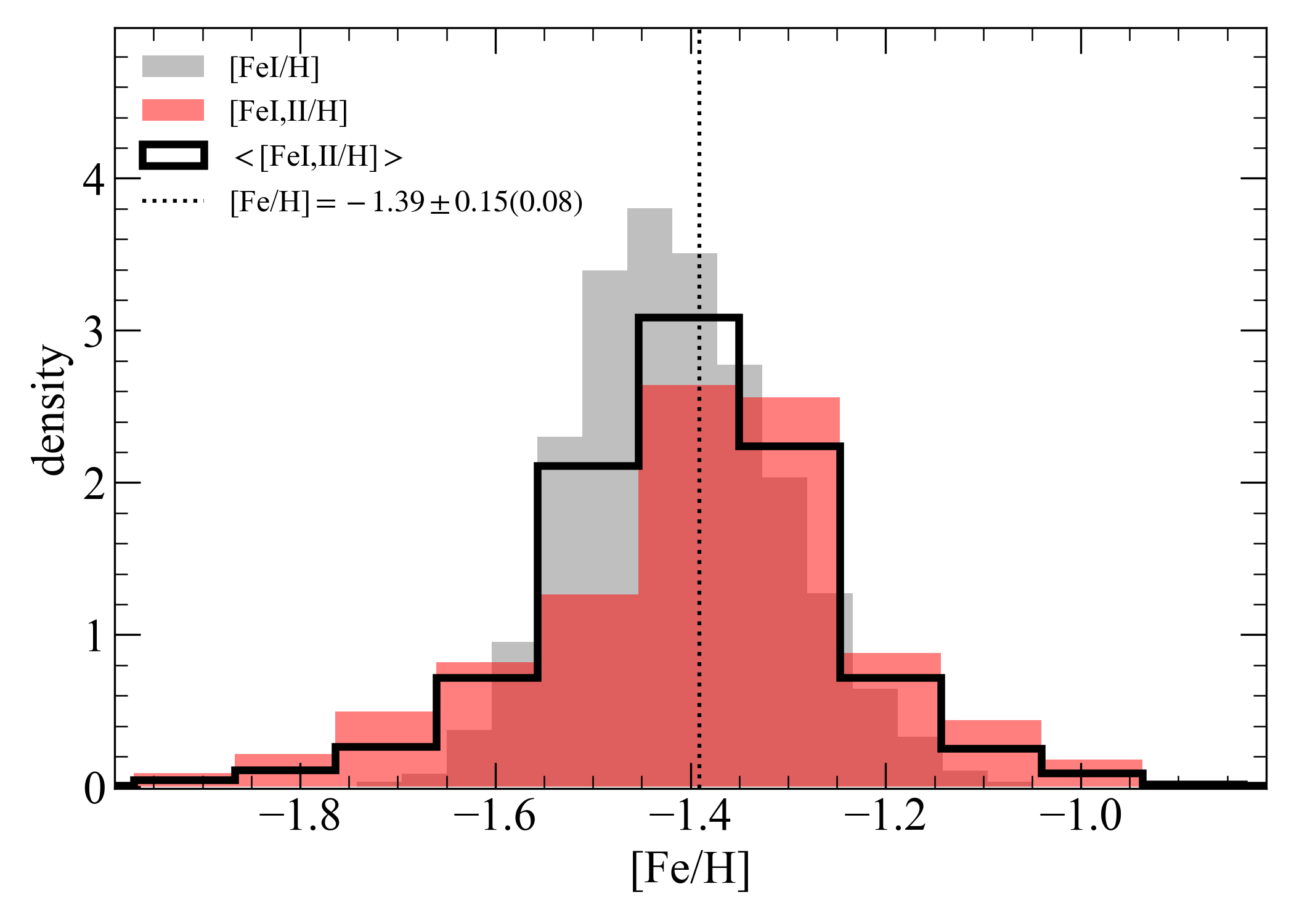}
    \caption{{Metallicity distribution from sample stars of NGC~6355. The final distribution (black step histogram) considers both [FeI/H] (grey) and [FeII/H] (red) for all lines of our sample member stars.}}
    \label{fig:feh_dist}
\end{figure}

{To derive the final metallicity, we generated a Monte Carlo (MC) sample for each star to construct their [FeI/H] and [FeII/H] distributions. The distributions composed of the individual MC sample of each star are shown in Figure \ref{fig:feh_dist} as grey and red for [FeI/H] and [FeII/H], respectively. Finally, the cluster metallicity distribution was obtained by combining the two distributions (grey and red). The best metallicity value, the corresponding standard deviation, and the error of the mean are [Fe/H]$= - 1.39\pm0.15 \left(0.08\right)$. This metallicity agrees well with the \citet{carretta09}  metallicity scale, which gives a value of [Fe/H] = $-1.33\pm0.02$ for NGC~6355.}

\subsection{Age and distance}

{We employed the \texttt{SIRIUS} code \citep{souza20} to perform the isochrone fitting to the CMD [F555W, F438W-F555W]
of NGC~6355. The code can provide a Bayesian view of the fundamental parameters age, reddening ($E(B-V)$), $d_{\odot}$, and metallicity ([Fe/H]). We adopted the isochrones from the Dartmouth Stellar Evolutionary Database \citep{dotter08} with a further linear interpolation in age and [Fe/H] with the random values given by the algorithm. As a Gaussian prior for the metallicity, we employed the value derived in this work, while for the other parameters, we adopted uniform priors: $10$ Gyr $\leq$ age $\leq 14$ Gyr, E(B$-$V) $\geq 0.0$, and  $d_{\odot} \leq 20$ kpc. We used the CMD structure constraints similar to the procedure described by \citetalias{vandenberg13} to improve the code. Nevertheless, we kept the Bayesian nature of the code and used the structure pattern of the CMD as priors.}

{The direct comparison between observational data and isochrones cannot give an accurate physical interpretation of the cluster \citep{dantona18} {because the likelihood in this case is purely geometrical}. Therefore, the prior distributions are of great importance to improve the method. In that sense, we adopted a more robust prior to the magnitude of the horizontal branch (HB). This prior is crucial to give a more precise distance derivation when it is very close to the magnitude level of RR Lyrae stars. To constrain the HB magnitude, we employed the relation by \cite{recio-blanco05},}

\begin{eqnarray}
 \displaystyle M^{ZAHB}_{F555W} = 0.981 + 0.410\times[M/H] + 0.061\times[M/H]^2
,\end{eqnarray}
{where $\displaystyle {\rm [M/H]}={\rm [Fe/H]} + \log\left(0.638\times10^{[\alpha/\rm Fe]} + 0.362\right)$. We assumed [$\alpha$/Fe]$=+0.4$ because this is the expected value for GCs with a similar metallicity \citep{barbuy2018a}. Then, we recalculated the magnitude level for each iteration of the Markov chain Monte Carlo (McMC) sampling. For the apparent magnitude of the HB, we assumed  m$_{F555W}^{ZAHB} = 17.9\pm0.1$ by a visual inspection, which is very close to the value derived by \cite{ortolani03} of V$_{\rm HB} = 17.8\pm0.2$.}

{Another morphological parameter is the magnitude difference between zero-age HB (ZAHB) and the turn-off point (TO), also known as \emph{\textup{vertical parameter}} \citep{vandenberg90,rosenberg99}. However, this parameter is strongly dependent on the ZAHB level. Because of this, we decided to use the \emph{\textup{horizontal parameter}} \citep{vandenberg90,rosenberg99}. The horizontal parameter is the colour difference between the TO and the point at the RGB that is $2.5$ magnitude brighter than the TO.}

{In order to implement the horizontal method in the observed CMD, we computed the fiducial or ridge line of NGC~6355 using the method described in \citet[][hereafter \citetalias{marinfranch09}]{marinfranch09}. The procedure is briefly described as follows. We first computed a simple fiducial line by binning the cluster magnitude and calculating the median colour for each bin. We applied a differential binning method to have more points around the TO. The second step was to derive the median colour perpendicular to each bin. This method is most important for the subgiant branch (SGB) because this sequence is almost horizontal for bluer filters. Finally, the algorithm computes the horizontal parameter for the cluster fiducial line and each McMC isochrone. }

\begin{figure*}
    \centering
    \includegraphics[width=0.9\textwidth]{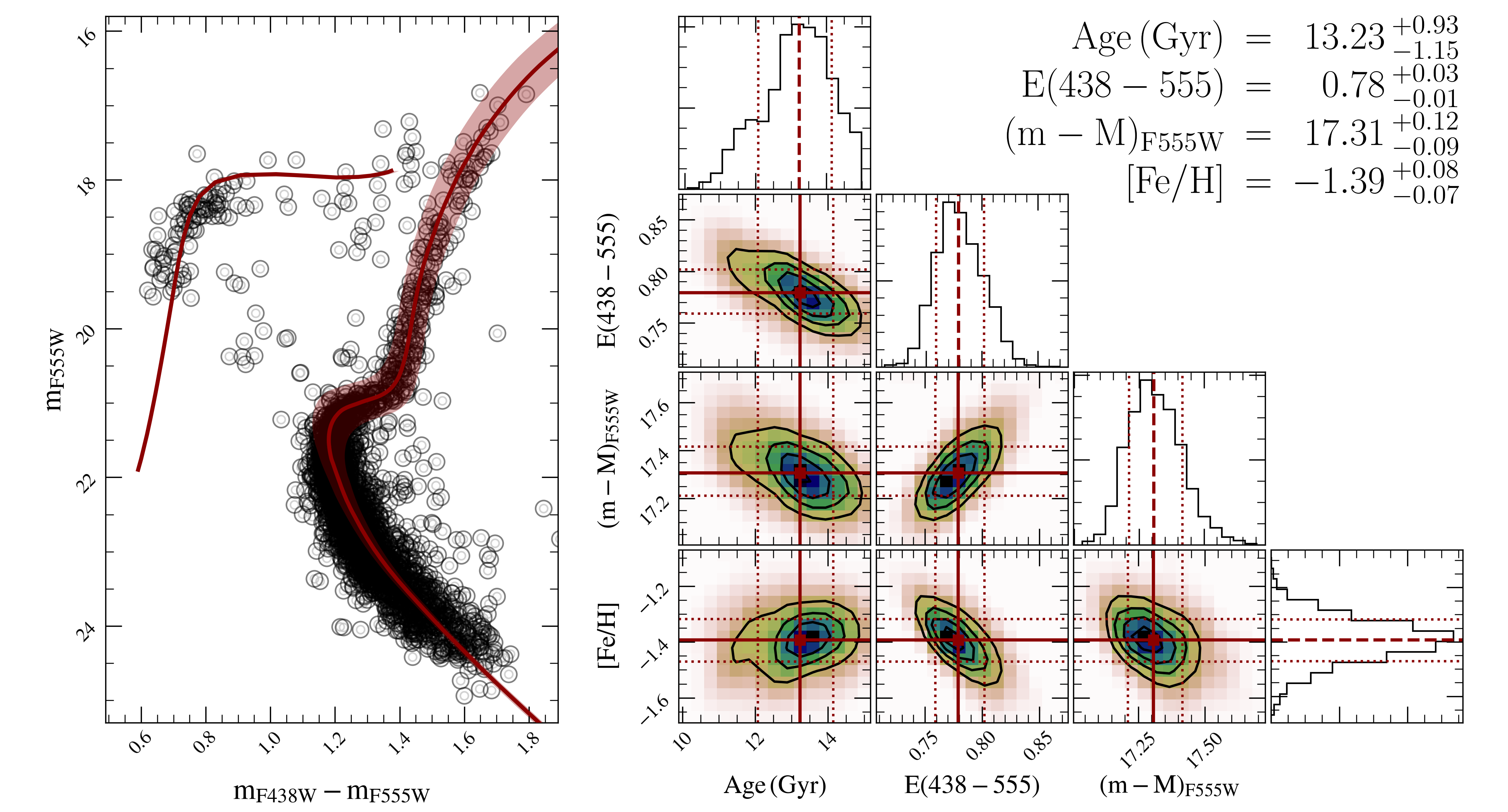}
    \caption{{Isochrone fitting for NGC~6355. The best solution is composed of the median values of the posterior distributions (solid dark red line), and the $1\sigma$ extrapolation is constructed from the $16th^{}$ and $84th^{}$ percentiles (shaded dark red region). The corner plot shows the correlations among the parameters.}}
    \label{fig:isoc}
\end{figure*}

{The posterior distributions of the parameters are given by the $50th^{}$ percentile as the best value, and the $16th^{}$ and $84th^{}$ percentiles to provide the uncertainties (right corner plots of Figure \ref{fig:isoc}). In Figure \ref{fig:isoc}, the NGC~6355 CMD (left panel) is over-plotted by the best solution of the isochrone fitting composed of the median value (solid line) and the $1\sigma$ region (shaded region). }

{Because the expected extinction is relatively high, it is necessary to consider the $T_{\rm eff}$ correction to the isochrones. It is worth noting that the $T_{\rm eff}$ correction effect increases with the temperature and changes the isochrone morphology. The method is well described in \cite{oliveira20} and \cite{souza20}. 
We found the following equations: }
\begin{eqnarray}
 \label{eq:teff1}
 & A_{\rm F438W}/A_{V}   =  7.688 -  86.606 x + 325.254 x^2 - 407.219 x^3 \\
 & A_{\rm F555W}/A_{V}   = 12.043 - 135.394 x + 507.496 x^2 - 634.233 x^3\\
 & A_{\rm J}/A_{V}       = -0.128 +   1.428 x -   5.309 x^2 +   6.573 x^3\\
 & A_{\rm K_{S}}/A_{V}   =  0.061 -   0.677 x +   2.522 x^2 -   3.134 x^3\\
 & A_{\rm G}/A_{V}  =  4.346 -  48.867 x + 183.277 x^2 - 229.296 x^3\\
 & A_{\rm G_{BP}}/A_{V} =  6.899 -  77.627 x + 291.243 x^2 - 364.345 x^3\\
 & A_{\rm G_{RP}}/A_{V} = -0.154 -   1.695 x -   6.175 x^2 +   7.449 x^3
 \label{eq:teff3}
,\end{eqnarray}
{where $x$ is  $\log {\rm T}_{\rm eff}$. The immediate effect on the isochrone is an offset in the direction of the CMD blue-brighter region. Therefore, the horizontal ($E(438-555)$) and vertical ($(m-M)_{\rm F555W}$) displacements should be different from those without a correction. In addition, the morphology is defined essentially by the age and metallicity when the helium mass fraction (Y) is fixed \citep[see][]{souza20}. In our case, the metallicity was constrained to the value derived here from high-resolution spectroscopy. Therefore, only age changes the isochrone morphology. Because of this, the age considering the $T_{\rm eff}$ correction tends to be older than the simple isochrone fitting. The result is shown in Figure \ref{fig:isoc}.}

{In this work, we derived the absolute age of $13.2\pm1.1$ Gyr for NGC~6355. The considerable uncertainty on the age derivation is due to the narrow colour baseline adopted in this work (F438W-F555W), which {spread} the TO region slightly more. Although we provide the first absolute age for NGC~6355 through isochrone fitting, \citetalias{kharchenko16} derived an age of $\sim 12.5$ Gyr using integrated magnitudes, and \citetalias{cohen21b} reported the age as $13.2$ Gyr for NGC~6355 by comparing its CMD with that of NGC~6205. The age derived in this work assuming the $T_{\rm eff}$ correction agrees very well with the age in \citetalias{cohen21b}$+$\citetalias{vandenberg13}. This illustrates the importance of this correction for highly reddened clusters in the central part of the Galaxy.}

\begin{figure}
    \centering
    \includegraphics[width=\columnwidth]{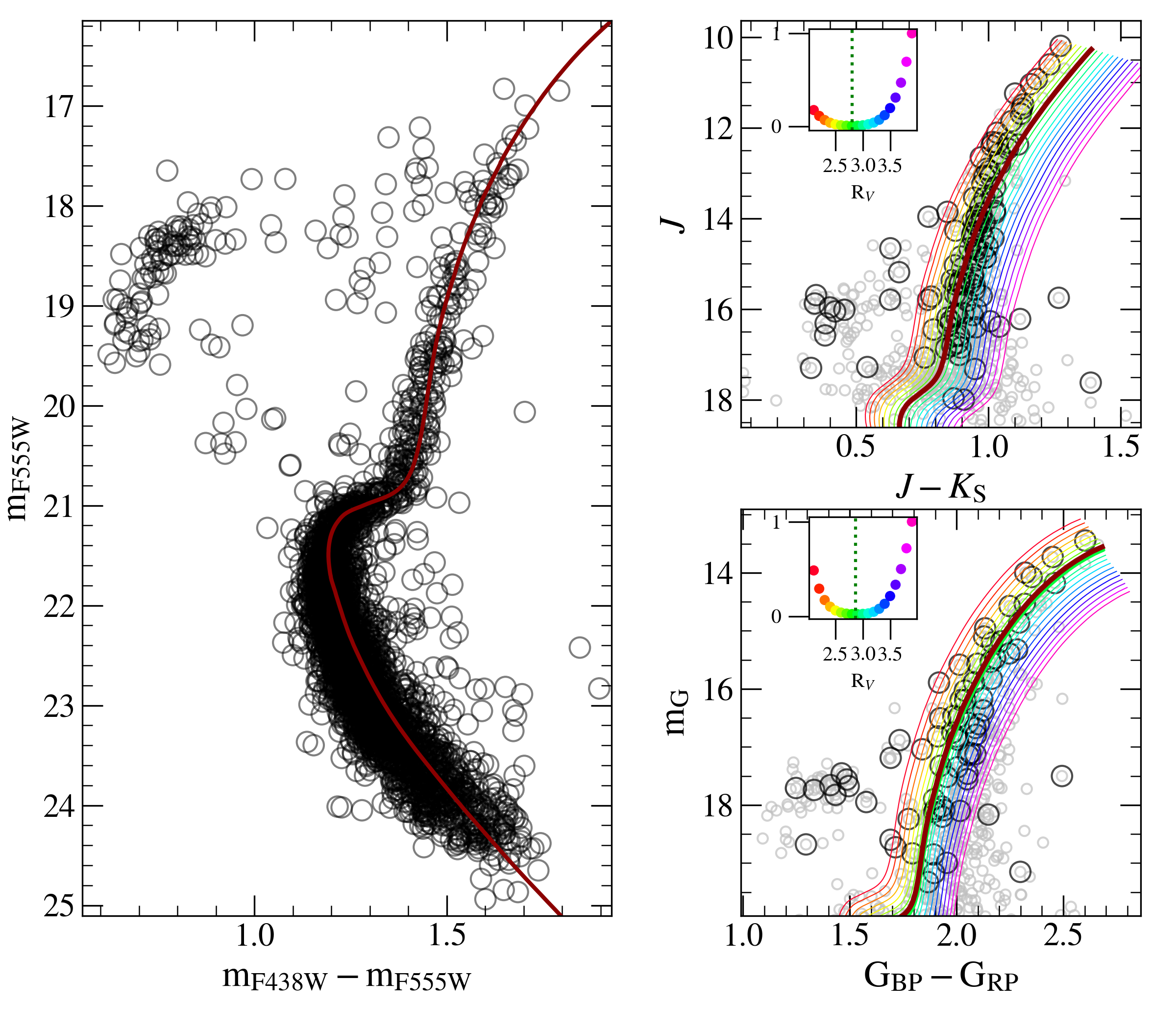}
    \caption{Simultaneous isochrone fitting to derive the cluster $R_V$ using three CMDs: HST (left panel), 2MASS $JK_S$ from \protect\cite{valenti07} (top right panel), and Gaia DR3 (bottom right panel). The isochrones are coloured according to their $R_V$ value. In each panel, the best solution is represented by the solid dark red isochrone. For the two right panels, the $\chi^2$ analysis is plotted in the inset plot, and the dots are coloured by the same colour as the corresponding isochrone.}
    \label{fig:get-rv}
\end{figure}

{\cite{nataf16} discussed the extinction towards GCs located in the Galactic bulge, where the $R_V$ value can be as low as $2.5$. \cite{pallanca21} reported a straightforward method for determining the best value of $R_V$ for highly reddened clusters. The method was also applied by \cite{souza20}, who derived a value of $2.6$ for Pal 6. The method for deriving the $R_V$ consists of simultaneously fitting CMDs with different colour baselines with the same set of reddening and distance. Here we fitted (in addition to the HST CMD) the CMDs [$J$, $J-K_S$] from \cite{valenti07} and [$G$, $G_{\rm BP}-G_{\rm RP}$] from Gaia DR3. From the HST CMD, we found $E(438-555)=0.78\pm0.03$ and $(m-M)_{\rm F555W}=17.31\pm0.12$. These values were converted into $E(B-V)$ and $(m-M)_0$ for different values of $R_V$, as shown in Figure \ref{fig:get-rv}. The best $R_V$ is the mean between the best values for \cite{valenti07} and Gaia DR3 CMDs. We find $R_V = 2.84\pm0.02$. Hence, for NGC~6355 with the derived $R_V$, we find $E(B-V)=0.89\pm0.03$ and d$_{\odot} = 8.54\pm0.19$ kpc.}

\begin{figure}
    \centering
    \includegraphics[width=\columnwidth]{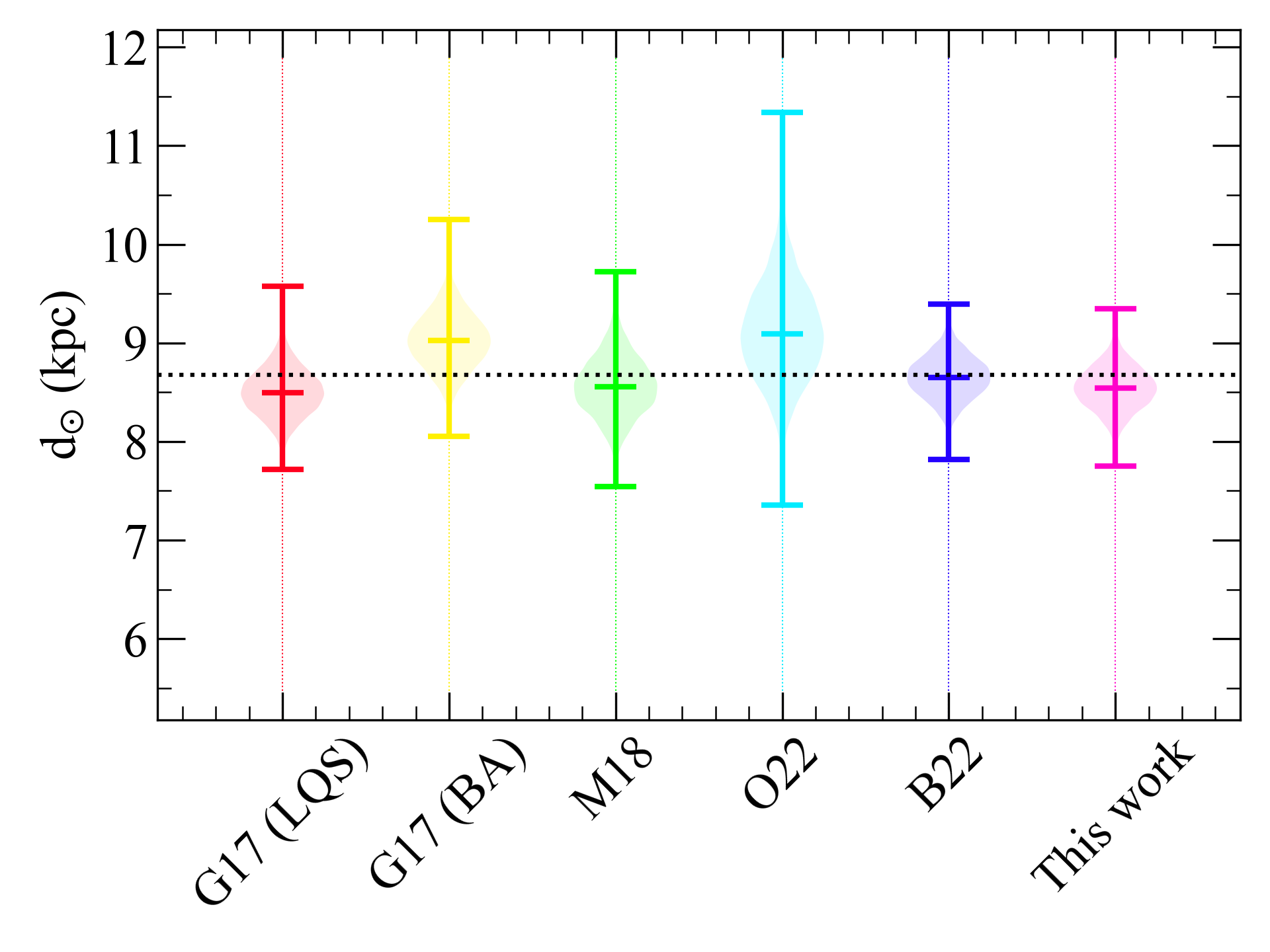}
    \caption{{Our distance derivation compared with the literature. The violins show the distance distribution using RR Lyrae stars, the recent distance derivation by \protect\cite{baumgardt21}, and the distance found in this work through isochrone fitting. For the derived RR Lyrae distances, four calibrations were adopted that are represented by the first four violins (see the text).}}
    \label{fig:distances}
\end{figure}

{The distance value is crucial for deriving the orbital parameters of the clusters, as demonstrated by \citetalias{perezvillegas2020} and illustrated by the case of
Palomar~6, as discussed in \cite{souza21}. To verify our distance derivation, we collected the RR Lyrae star members of NGC~6355 from the fourth data release of the Optical Gravitational Lensing Experiment \citep[OGLE-IV;][]{Soszynski2019}. We adopted the calibrations from \citet[][G17]{gaia17} using the least-squares (LQS) and Bayesian (BA) methods (\citet[][M18]{muraveva18}, and \citet[][O22]{oliveira22}). All distances are displayed in Figure \ref{fig:distances}, including the derivation by \citet[][B22]{baumgardt21} \footnote{\url{https://people.smp.uq.edu.au/HolgerBaumgardt/globular/fits/disfit/ngc6355_dist.pdf}}. The value of $8.54\pm0.19$ kpc derived in this work agrees well with the others, particularly the B22 value of $8.65\pm0.22$ kpc, which is the most recent value. }

\section{Abundance analysis}\label{sec:chemical_analysis}

{
We carried out a detailed abundance analysis employing line-by-line spectrum synthesis. We employed the spectrum synthesis code PFANT \citep{barbuy2018c} to derive the abundances of the elements C, N, O, Na, Mg, Al, Si, Ca, Ti, V, Mn, Co, Cu, Zn, Y, Zr, Ba, La, Nd, and Eu. The line list with the abundance ratios for each line are given in the appendix (Table \ref{tab:lines_ratios}). The code PFANT is an update of the Meudon code by M. Spite and adopts local thermodynamic equilibrium (LTE). The atomic line list is from VALD3 \citep{Ryabchikova15}. }

\begin{figure}
    \centering
    \includegraphics[width=\columnwidth]{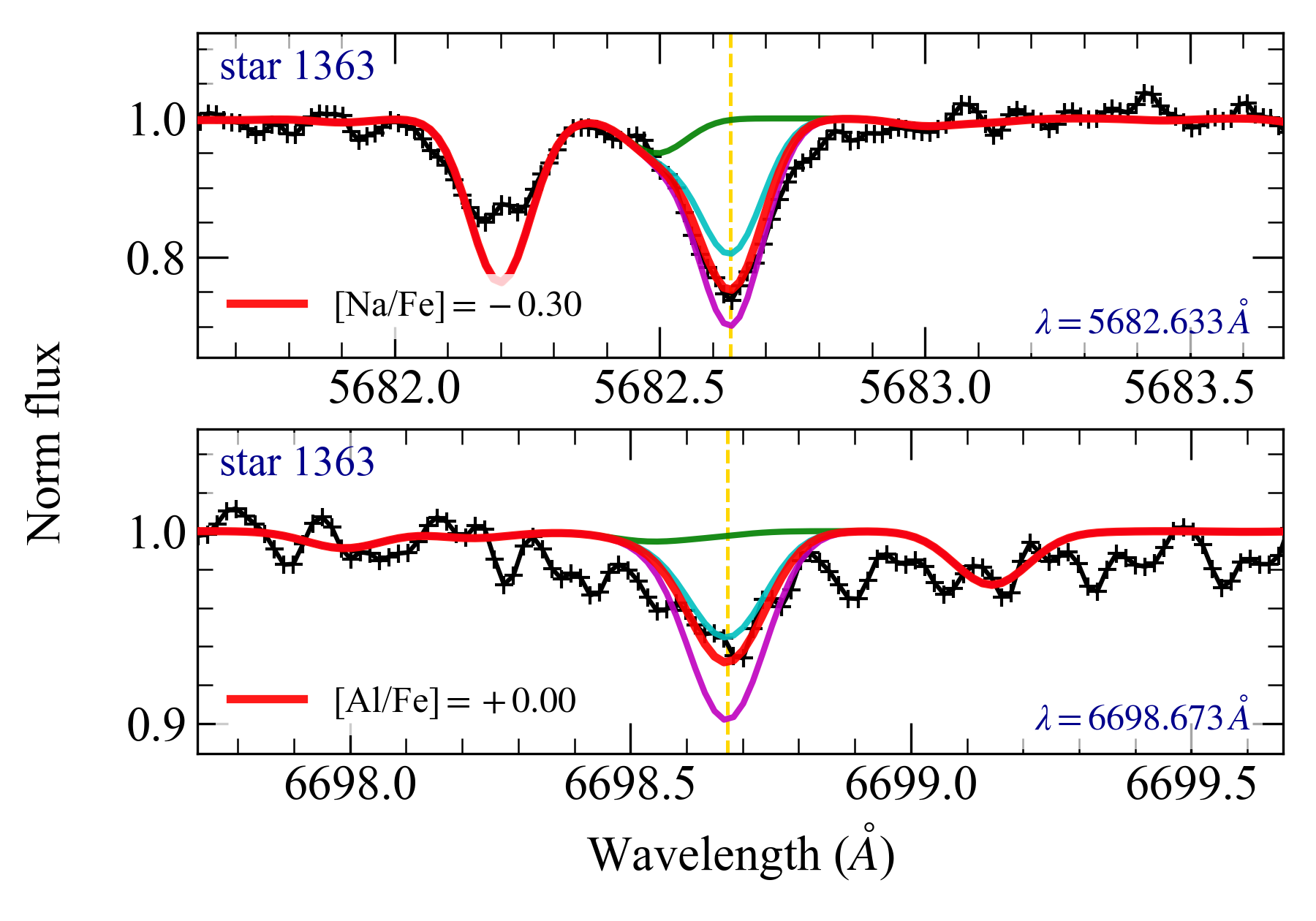}
    \caption{Example of line-profile fitting for star 1363. The upper panel shows the result for the \ion{Na}{I} $5682.633$ \AA, and the bottom panel shows the fit for the \ion{Al}{I} $6698.673$ \AA line. The black lines correspond to the observed spectra. The solid red line shows the best-fit solution as the median. For comparison purposes, we also plot the best-fit solution with a variation of $\pm 0.15$ (solid cyan and magenta lines) and the spectrum without the element abundance (green line). }
    \label{fig:na-al_fit}
\end{figure}

{The abundance values were derived through the $\chi^{2}$ minimization algorithm described in detail in \cite{souza21}. Figure \ref{fig:na-al_fit} gives a visual {illustration} of the method for star 1363, where the observed spectrum around the lines \ion{Na}{I} $5682.633$ {\AA} and \ion{Al}{I} $6698.673$ {\AA} is shown in black. {The best-fit solution is the solid red line.} For completeness, we also compare the spectrum without the abundance contribution of the current element (solid green line), the best fit plus 0.15 (solid magenta line), and the best fit minus 0.15 (solid cyan line). Finally, we adopted the solar abundances from \cite{grevesse15}.}

\subsection{C, N, and O abundances}

\begin{figure}
    \centering
    \includegraphics[width=\columnwidth]{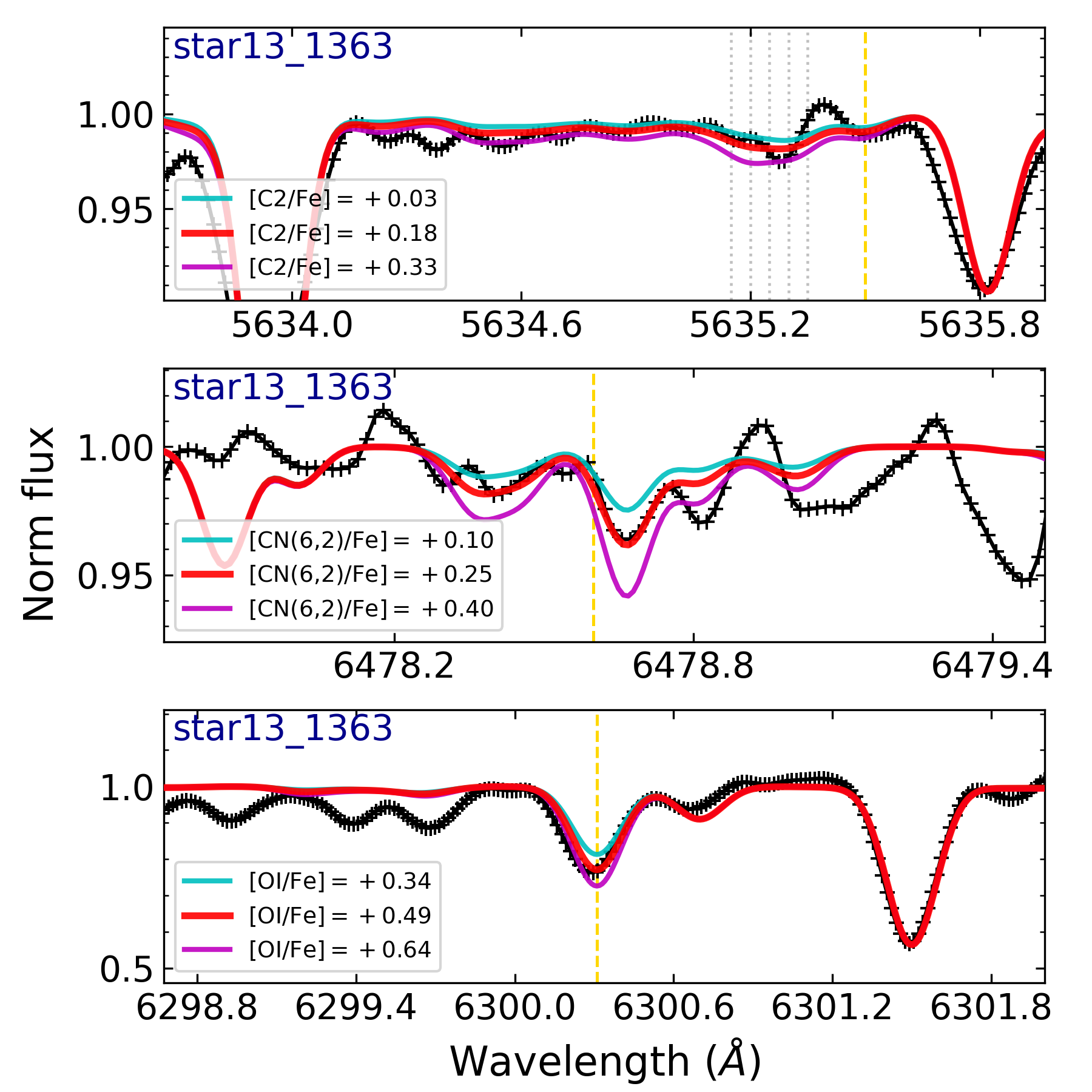}
    \caption{Spectral fitting of C, N, and O for star 1363. The observed spectrum is given in black. The solid red line is the best fit, and the cyan and magenta lines show the best fit $\pm 0.15$, respectively. The yellow line shows the line region. For C$_{2}$ (upper panel) we also show the bandhead lines in dotted silver lines.}
    \label{fig:cno-fits}
\end{figure}

{The CNO abundances were derived through an iterative fitting of the C$_2$(1,0) Swan bandhead at $5635.3$ \AA, and CN(6,2) at $6478.48$ \AA\, of the $A^2 \Pi X^2 \Sigma$ system  band heads and the forbidden oxygen line [OI] $6300.31$ \AA. The algorithm fits the three lines simultaneously and takes the interdependent continuum variation due to changes in C, O, and N values into account. Table \ref{tab:cno} lists the derived abundances. Because the region of the C$_2$(1,0) bandhead is strongly affected by the S/N and the line is weak, we assumed the C abundances as upper limits. Finally, before fitting the [OI] line, we verified the contamination by telluric lines in this region and concluded that for our sample, none of the stars has telluric line contamination on the [OI] line. The spectral fitting for C, N, and O for star 1363 are shown in Figure \ref{fig:cno-fits}.}

{As expected for most GCs \citep{piotto15,milone17}, NGC~6355 seems to host multiple stellar populations \citep[MPs; see the reviews][]{gratton04,gratton12,bastian18,milone22}. The relatively high nitrogen abundance of stars 1176 and 133 with relatively low values of carbon abundances indicates the presence of MPs in NGC~6355. {The other two stars have a relatively low N abundance and relatively normal (solar) C}. Because stellar evolution theory predicts an N-C anti-correlation, we must further investigate to confirm the presence of MPs in NGC~6355. This is further analysed below.}

\begin{table}
\caption{Carbon, nitrogen, and oxygen abundances [X/Fe] from C$_2$, CN bandhead, and [OI], respectively.}
\centering
\tabcolsep 0.15cm
\begin{tabular}{lccc}
\hline
\hline
     &     [C/Fe]   & [N/Fe]  & [O/Fe] \\
Star &     C$_2$    & CN(6,2) &   [OI]   \\
     & $5635.50$ {\AA} & $6478.60$ \AA & $6300.31$ {\AA} \\  
\hline
1539  &    $\leq+0.10$  &  $+0.21$     &  $+0.43$     \\
1363  &    $\leq+0.18$  &  $+0.25$     &  $+0.49$     \\
1176  &    $\leq+0.00$  &  $+0.87$     &  $+0.37$     \\
133   &    $\leq-0.09$  &  $+0.70$     &  $+0.24$     \\
 \hline
\end{tabular}  
\label{tab:cno}
\end{table}

\subsection{alpha-elements}

\begin{figure}
    \centering
    \includegraphics[width=\columnwidth]{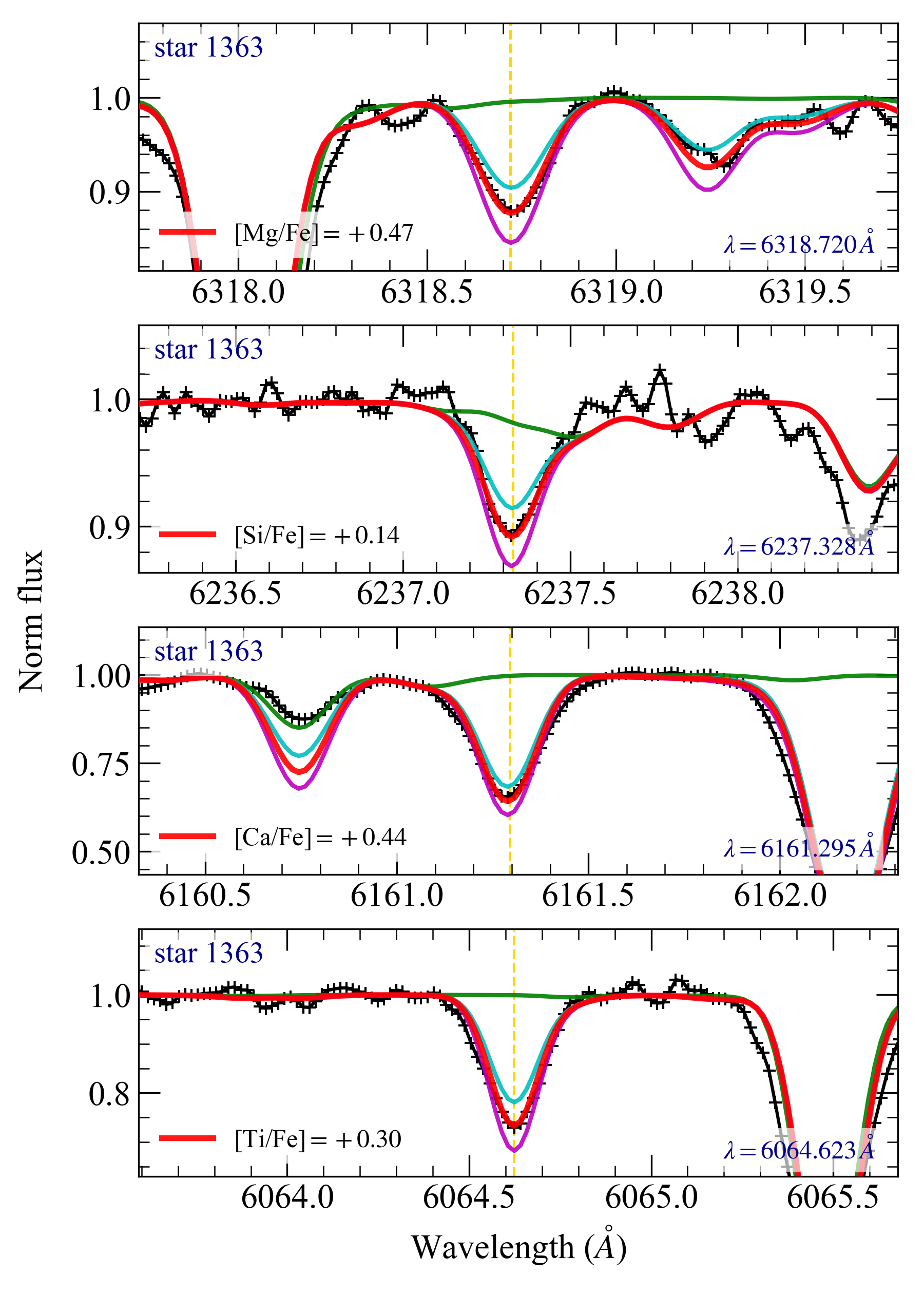}
    \caption{Same as figure \ref{fig:cno-fits} for Mg, Si, Ca, and Ti. The solid red line is the best fit, and the cyan and magenta lines show the best fit $\pm 0.15$, respectively.}
    \label{fig:alpha-fits}
\end{figure}

{The $\alpha$-elements O and Mg are the most reliable indicators of enrichment in $\alpha$-elements from hydrostatic phases of massive stars \citep{Woosley95}. Together with the explosive $\alpha$-elements  Si and Ca, they are good indicators of  a fast early enrichment of the proto-cluster gas by supernovae type II (SNII). 
Ti is classified as an iron-peak element
\citep{Woosley95}, but shows a similar $\alpha$-element behaviour and is often included as
another $\alpha$-element. The spectral fitting results for Mg, Si, Ca, and Ti of star 1363 are shown in Figure \ref{fig:alpha-fits}, {and the results are presented in Table \ref{tab:mean_abunds}}.}

\subsection{Odd-Z elements}

{The sodium abundances were derived from \ion{Na}{I} $5682.633$  {\rm \AA}, $5688.194$  {\rm \AA}, $6154.23$ {\rm \AA}, and $6160.753$ {\rm \AA} lines. The Al abundances were derived from lines \ion{Al}{I} $6696.185$ {\rm \AA}, $6698.673$ {\rm \AA}.} 

\begin{figure}
    \centering  
        \includegraphics[width=\columnwidth]{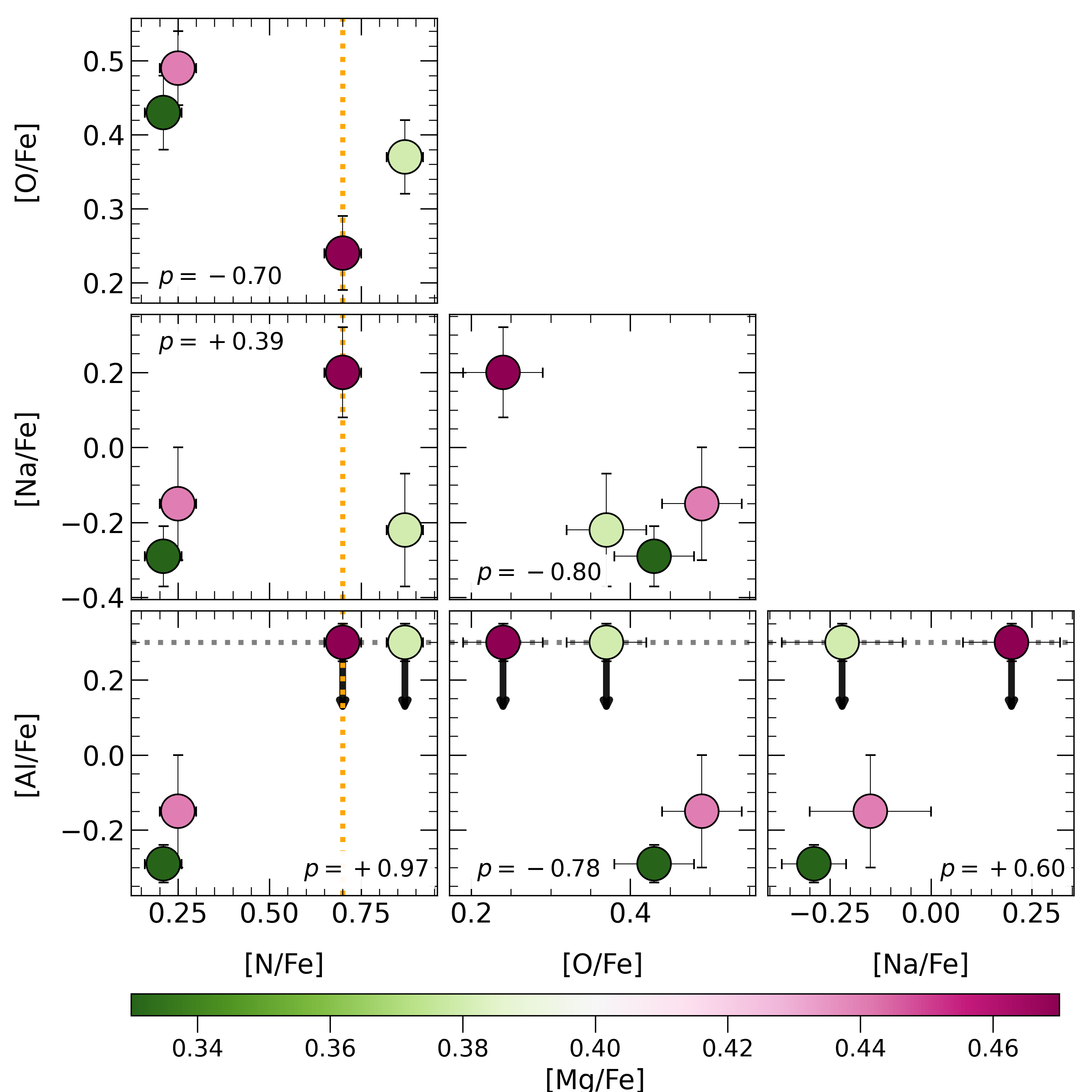}
    \caption{{(Anti-)Correlations indicating effects of multiple stellar populations. The dotted orange line in both left panels represents the transition to the N-rich regime at [N/Fe]$\sim 0.7$ for [Fe/H] around the NGC~6355 value \citep{fernandez-trincado22}. Additionally, the grey line in the two bottom panels shows the upper limit for first-generation stars \citep{meszaros20}. The colour bar shows the Mg abundances. }}
    \label{fig:anticorrs}
\end{figure}

{The (anti-)correlations indicating the effect of MPs are shown in Figure \ref{fig:anticorrs}. We also calculated the Spearman correlation parameter for each combination. For N-Al, we found a strong correlation, and the anti-correlation for N-O, Na-O, and {Al-O is high}. Moreover, the main correlations come from the nitrogen abundances \citep{fernandez-trincado22}. However, [Al/Fe]$=+0.30$ is also a threshold for second-generation (2G) stars \citep{meszaros20}. The figure shows a visible separation of our sample into two groups: two stars are moderately rich in N and Al, and two stars have {low} values of [Al/Fe]. This affects their mean abundances (Table \ref{tab:mean_abunds}). This is further discussed below.}

\subsection{Iron-peak elements}

\begin{figure}
    \centering
    \includegraphics[width=\columnwidth]{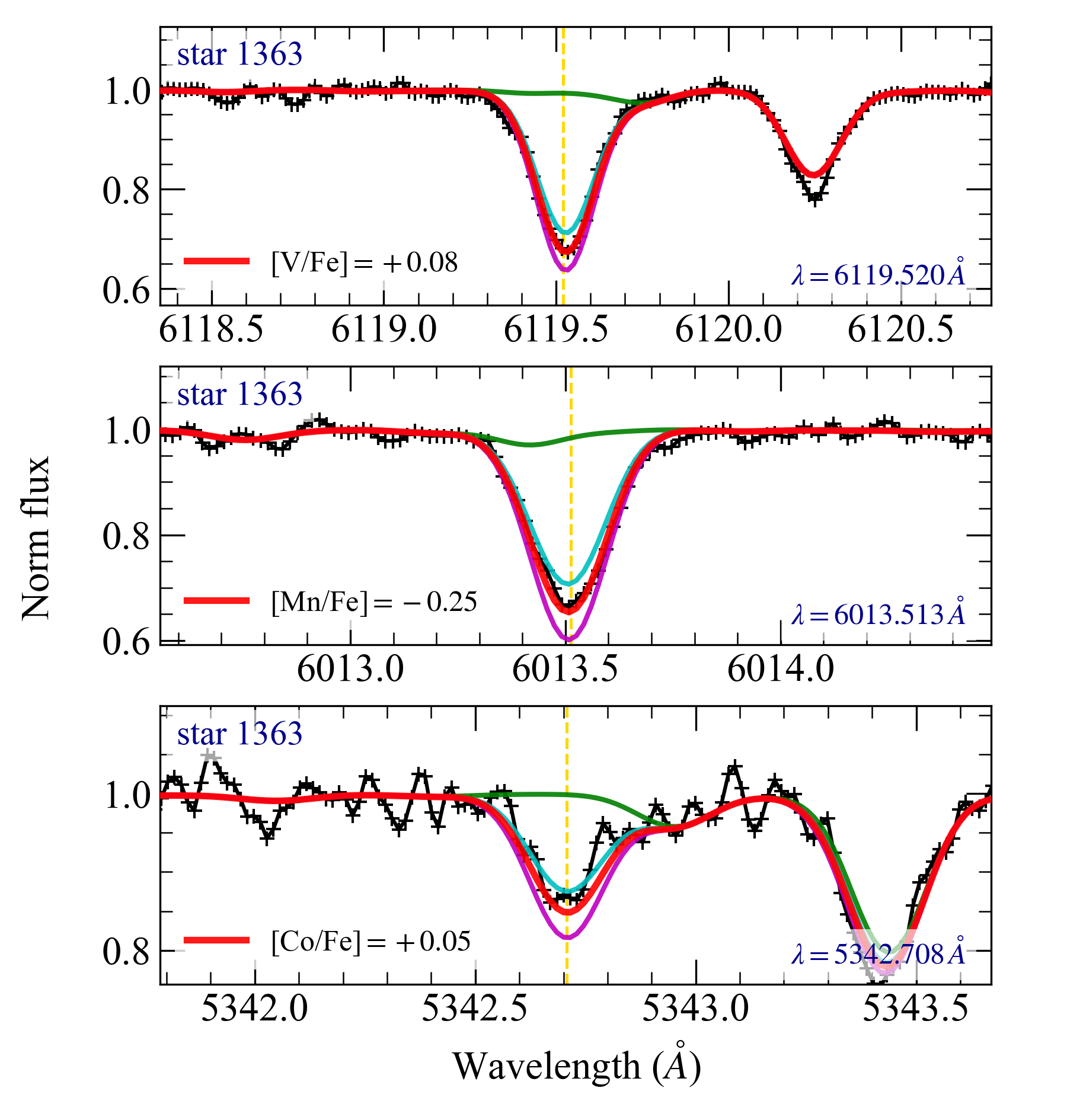}
    \caption{Same as figure \ref{fig:cno-fits} for V, Mn, and Co. The solid red line is the best fit, and the cyan and magenta lines show the best fit $\pm 0.15$, respectively.}
    \label{fig:ironpeak-fits1}
\end{figure}

\begin{figure}
    \centering
    \includegraphics[width=\columnwidth]{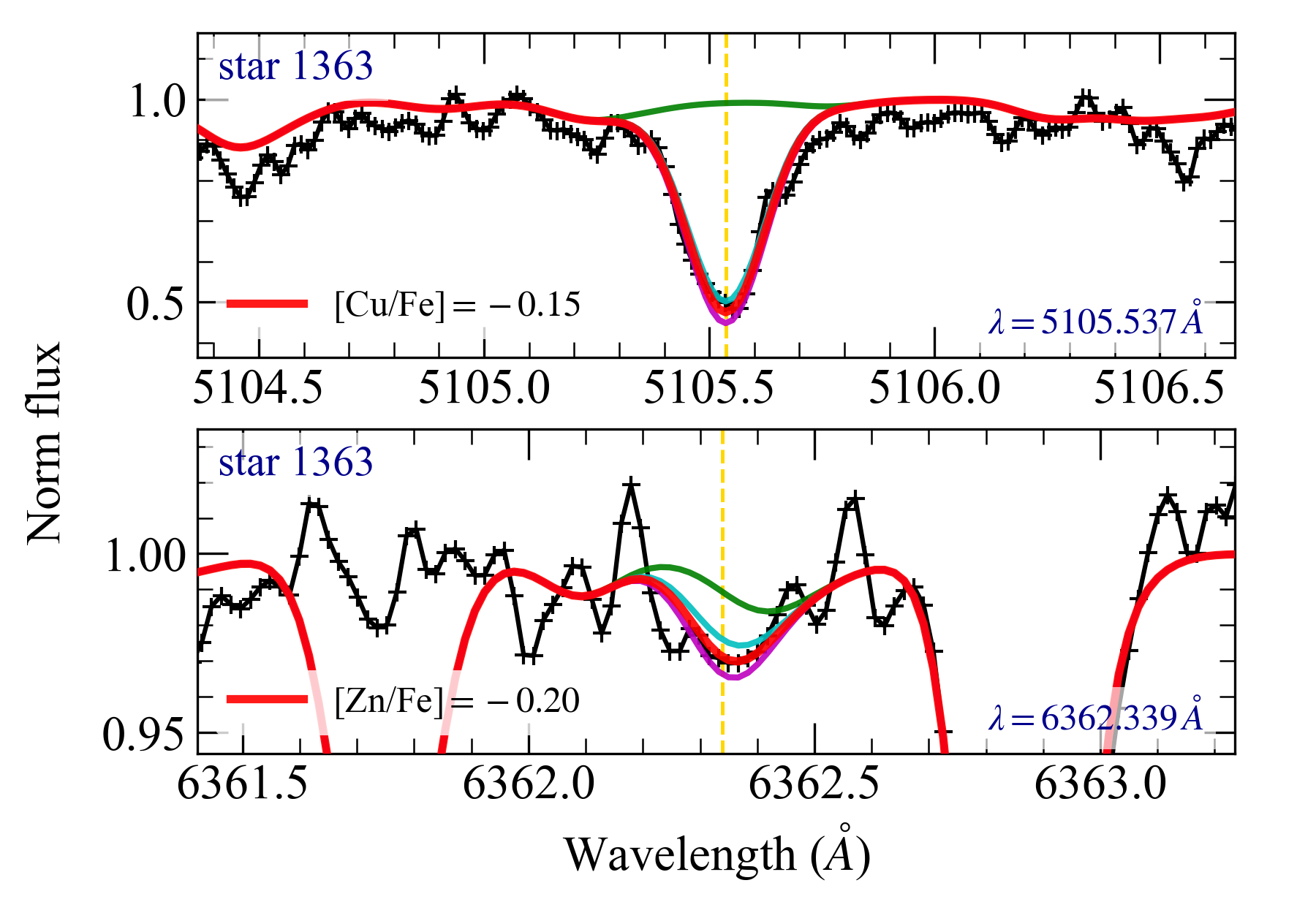}
    \caption{Same as figure \ref{fig:cno-fits} for Cu and Zn. The solid red line is the best fit, and the cyan and magenta lines show the best fit $\pm 0.15$, respectively.}
    \label{fig:ironpeak-fits2}
\end{figure}

We derived the abundances of the iron-peak elements V, Mn, Co, Cu, and Zn. While V and Mn are members of the lower iron-peak element group, Co, Cu, and Zn are considered to belong to the upper iron-peak group \citep{Woosley95}. The first group is mainly produced in type Ia supernovae (SNIa) with a contribution from core-collapse supernovae {\citep[][ and refereces therein]{nomoto13}}. In contrast, Co, Cu, and Zn are predominantly produced by core-collapse supernovae {\citep[][and references therein]{woosley02}}. The atomic lines were adopted from \cite{ernandes2018} and \cite{ernandes20}, together with their hyperfine structure. The spectral fitting results for V, Mn, and Co are shown in Figure \ref{fig:ironpeak-fits1} for star 1363, and Cu and Zn are  given in Figure \ref{fig:ironpeak-fits2} for star 1539.

\subsection{Heavy elements}

\begin{figure}
    \centering
    \includegraphics[width=\columnwidth]{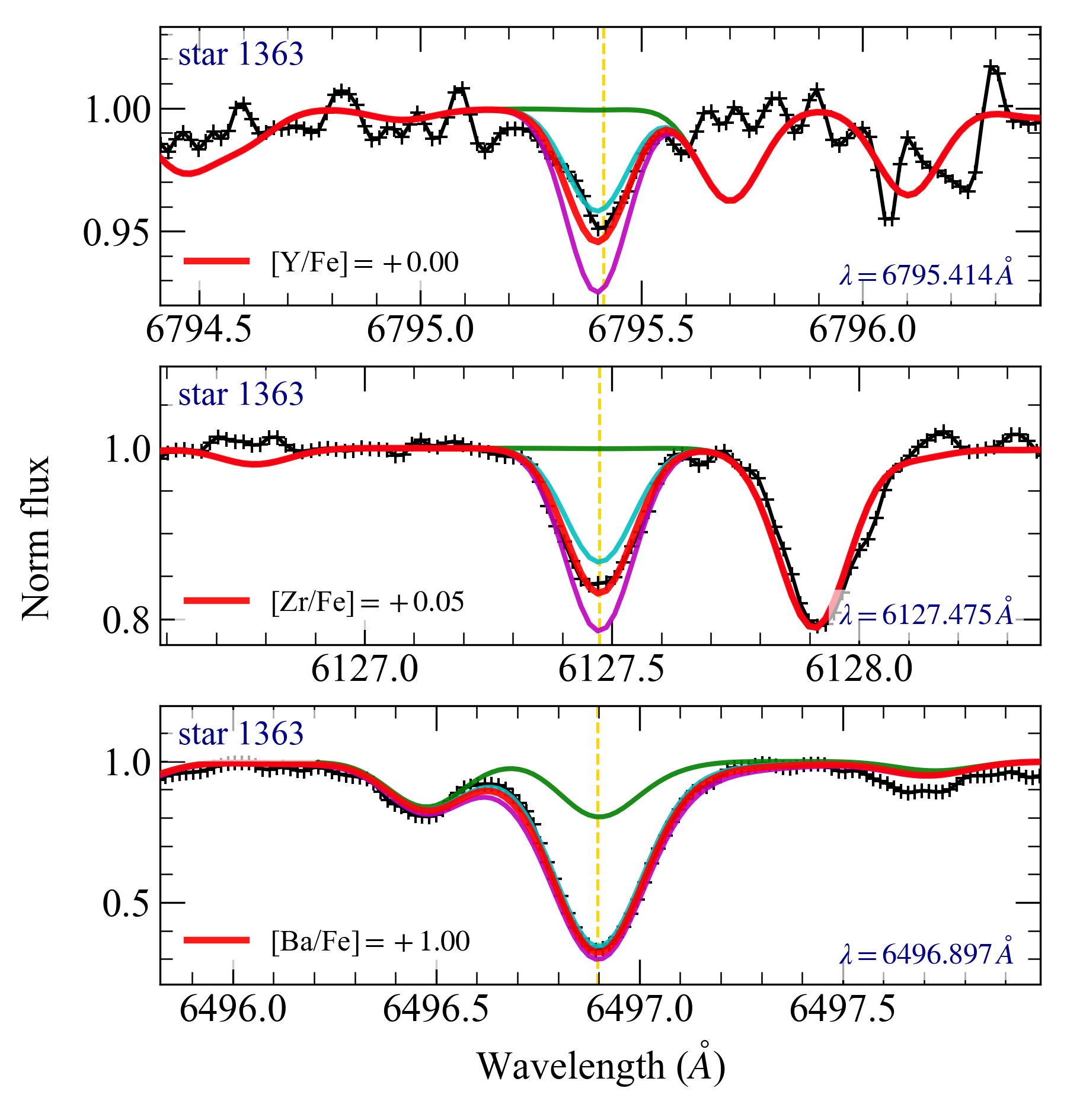}
    \caption{Same as figure \ref{fig:cno-fits} for Y, Zr, and Ba. The solid red line is the best fit, and the cyan and magenta lines show the best fit $\pm 0.15$, respectively.}
    \label{fig:heavy-fits1}
\end{figure}

\begin{figure}
    \centering
    \includegraphics[width=\columnwidth]{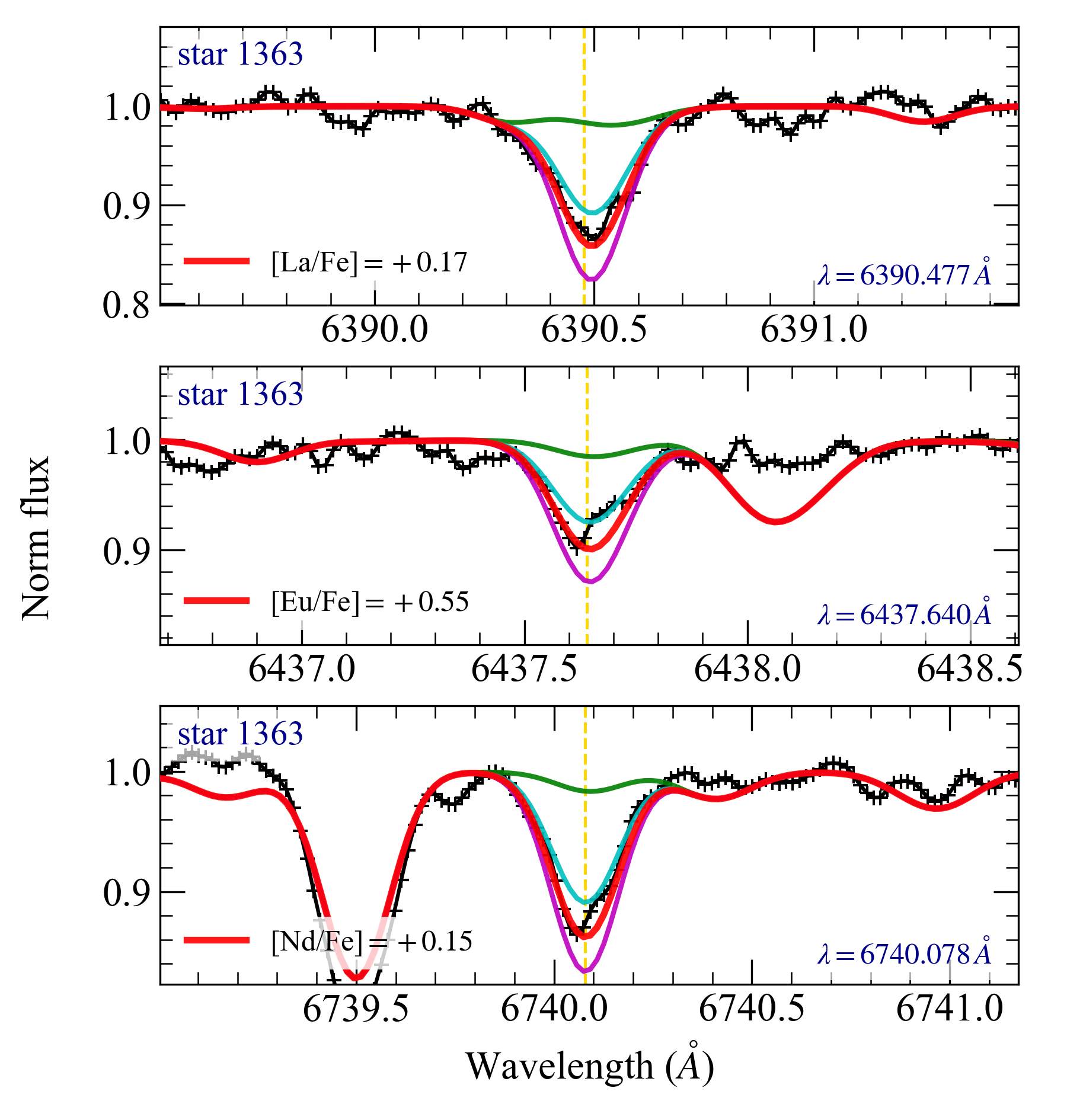}
    \caption{Same as figure \ref{fig:cno-fits} for La, Eu, and Nd. The solid red line is the best fit, and the cyan and magenta lines show the best fit $\pm 0.15$, respectively.}
    \label{fig:heavy-fits2}
\end{figure}

{ The abundances of the heavy neutron-capture s-elements Y, Zr, Ba, La, and Nd, and the r-element Eu {also} were derived. For Y, we measured the \ion{Y}{I} $6435.004$ \AA\, and the \ion{Y}{II} $6613.73$ \AA\, lines, and we assumed for the mean that the ionized species of Y contributes with $99\%$ to the abundance. For the barium abundance, we used the \ion{Ba}{II} lines $5853.675$ {\AA}, $6141.713$ {\AA}, and $6496.897$ {\AA}, with hyperfine structure from \cite{barbuy14}. The \ion{Zr}{I} $6127.47$ {\AA}, $6134.58$ {\AA}, $6140.535.58$ {\AA}, and $6143.25$ {\AA}, \ion{La}{II} $6262.287$ {\AA}, $6320.376$ {\AA}, and $6390.477$ {\AA}, \ion{Nd}{II} $6740.078$ {\AA}, $6790.372$ {\AA}, and $6549.525$ {\AA}, and \ion{Eu}{II} $6437.6$ {\AA} and $6645.1$ {\AA} were used for Zr, La, Nd, and Eu. The spectral fitting results for Y, Zr, Ba, La, Eu, and Nd are shown in Figures \ref{fig:heavy-fits1} and \ref{fig:heavy-fits2} for star 1363.}

\subsection{Errors}

 {The uncertainties in spectroscopic parameters are given in the last four columns of Table \ref{tab:mean_abunds} for star {133}. For each stellar parameter, we adopted the usual uncertainties for similar samples \citep{barbuy14,barbuy16,barbuy2018b}. 
 The sensitivities were computed by employing models with these modified parameters and recomputing lines of different elements considering changes of $\Delta T_{\rm eff}=+100$ K, $\Delta$log  g$ =+0.2$, $\Delta$v$_{\rm  t}$ $=  0.2$  km~s$^{-1}$. The given error is the difference between the new and the adopted abundance. The uncertainties due to non-LTE effects are negligible for these stellar parameters, as discussed in \cite{ernandes2018}. The same error analysis and estimations can be applied to other stars in our sample. {It is worth noting that star 133 has the lowest S/N of the four sample stars. The uncertainties given in Table \ref{tab:mean_abunds} can therefore be considered as upper limits. 
 The faint La lines appear to be more reliable than the strong Ba lines. Finally, it is important to note that the main uncertainties in stellar parameters are due to uncertainties in the $T_{\rm eff}$, as shown in Table \ref{tab:mean_abunds}. }

\begin{table*}
\centering
\caption{Abundances in the four UVES member stars. The mean values were computed considering all four stars ($<{\rm all}>$), considering only 1G stars ($<{\rm 1G}>$), and only 2G stars ($<{\rm 2G}>$). The last four columns show the abundance sensitivity due to variation in atmospheric parameters for star 15 (133) considering uncertainties of $\Delta$T$_{\rm eff}$ = 100 K, $\Delta$log g = 0.2, and $\Delta$v$_{\rm t}$ = 0.2 km s$^{-1}$, and the last column is the total error. These errors were taken into account when we composed the final reported abundances.}        
\label{tab:mean_abunds}      
\centering
\resizebox{\linewidth}{!}{%
\begin{tabular}{l|rrrrrrr|cccc}
\hline
\hline
\noalign{\vskip 0.1cm}
 [$X$/Fe]  &      star 1539      &       star 1363     &       star 1176     &      star 133      & $<{\rm all}>$  & $<{\rm 1G}>$   & $<{\rm 2G}>$      & $\Delta$T & $\Delta \log g$ & $\Delta v_{t}$ & $(\frac{1}{3}\sum$x$^{2})^{1/2}$ \\
           &      \multicolumn{2}{c}{1G}       &            \multicolumn{2}{c}{2G}               &                   &                   &                      &    K      &                 & kms$^{-1}$     & \\
\noalign{\vskip 0.1cm}
\hline
\hline
     C     &  $+0.10\pm0.05$   &  $+0.18\pm0.05$   &  $+0.00\pm0.05$   &  $-0.09\pm0.05$   &  $+0.05\pm0.11$   &  $+0.14\pm0.06$   &  $-0.04\pm0.07$      &  $+0.02$  &      $+0.03$    &      $+0.00$   & $+0.03$ \\
     N     &  $+0.21\pm0.05$   &  $+0.25\pm0.05$   &  $+0.87\pm0.05$   &  $+0.70\pm0.05$   &  $+0.51\pm0.29$   &  $+0.23\pm0.05$   &  $+0.78\pm0.10$      &  $+0.12$  &      $+0.08$    &      $+0.00$   & $+0.08$ \\
     O     &  $+0.43\pm0.05$   &  $+0.49\pm0.05$   &  $+0.37\pm0.05$   &  $+0.24\pm0.05$   &  $+0.38\pm0.11$   &  $+0.46\pm0.06$   &  $+0.30\pm0.08$      &  $+0.00$  &      $+0.03$    &      $+0.00$   & $+0.03$ \\
\hline
     Mg    &  $+0.33\pm0.05$   &  $+0.44\pm0.05$   &  $+0.38\pm0.05$   &  $+0.47\pm0.05$   &  $+0.41\pm0.07$   &  $+0.39\pm0.07$   &  $+0.42\pm0.07$      &  $+0.02$  &      $-0.02$    &      $-0.03$   & $+0.03$ \\
     Si    &  $+0.27\pm0.10$   &  $+0.25\pm0.12$   &  $+0.33\pm0.15$   &  $+0.28\pm0.27$   &  $+0.28\pm0.16$   &  $+0.26\pm0.11$   &  $+0.30\pm0.21$      &  $-0.02$  &      $-0.02$    &      $-0.07$   & $+0.04$ \\
     Ca    &  $+0.48\pm0.16$   &  $+0.47\pm0.10$   &  $+0.34\pm0.26$   &  $+0.57\pm0.12$   &  $+0.46\pm0.18$   &  $+0.47\pm0.13$   &  $+0.45\pm0.22$      &  $+0.26$  &      $+0.04$    &      $-0.08$   & $+0.16$ \\
     Ti    &  $+0.30\pm0.12$   &  $+0.34\pm0.12$   &  $+0.28\pm0.12$   &  $+0.38\pm0.10$   &  $+0.33\pm0.12$   &  $+0.32\pm0.12$   &  $+0.33\pm0.12$      &  $-0.03$  &      $+0.09$    &      $-0.06$   & $+0.06$ \\
\hline
     Na    &  $-0.29\pm0.08$   &  $-0.15\pm0.15$   &  $-0.22\pm0.15$   &  $+0.20\pm0.12$   &  $-0.11\pm0.23$   &  $-0.22\pm0.14$   &  $-0.01\pm0.25$      &  $+0.10$  &      $-0.00$    &      $-0.05$   & $+0.06$ \\
     Al    &  $-0.29\pm0.05$   &  $-0.15\pm0.15$   &  $<+0.30\pm0.05$   &  $<+0.30\pm0.05$   &  $<+0.04\pm0.28$   &  $-0.22\pm0.12$   &  $<+0.30\pm0.06$      &  $+0.08$  &      $-0.00$    &      $-0.02$   & $+0.05$ \\
\hline
      Y    &  $+0.20\pm0.07$   &  $-0.00\pm0.07$   &  $-0.00\pm0.07$   &       ---         &  $+0.06\pm0.12$   &  $+0.10\pm0.12$   &  $-0.00\pm0.07$      &  $+0.24$  &      $+0.09$    &      $-0.14$   & $+0.17$ \\
     Zr    &  $-0.06\pm0.08$   &  $+0.09\pm0.08$   &       ---         &  $-0.11\pm0.26$   &  $-0.02\pm0.16$   &  $+0.02\pm0.11$   &  $-0.11\pm0.26$      &  $+0.20$  &      $+0.02$    &      $-0.01$   & $+0.12$ \\
     Ba    &  $+0.84\pm0.17$   &  $+0.93\pm0.09$   &  $+0.92\pm0.19$   &  $+1.02\pm0.16$   &  $+0.93\pm0.17$   &  $+0.89\pm0.14$   &  $+0.97\pm0.19$      &  $+0.02$  &      $+0.03$    &      $-0.13$   & $+0.08$ \\
     La    &  $+0.08\pm0.12$   &  $+0.06\pm0.08$   &  $+0.10\pm0.07$   &  $+0.27\pm0.05$   &  $+0.13\pm0.12$   &  $+0.07\pm0.10$   &  $+0.19\pm0.11$      &  $+0.03$  &      $+0.09$    &      $-0.02$   & $+0.06$ \\
     Eu    &  $+0.53\pm0.05$   &  $+0.55\pm0.05$   &  $+0.57\pm0.08$   &  $+0.60\pm0.10$   &  $+0.56\pm0.07$   &  $+0.54\pm0.05$   &  $+0.59\pm0.09$      &  $-0.03$  &      $+0.07$    &      $-0.02$   & $+0.05$ \\
     Nd    &  $+0.47\pm0.06$   &  $+0.28\pm0.10$   &  $+0.06\pm0.08$   &  $-0.30\pm0.05$   &  $+0.13\pm0.30$   &  $+0.38\pm0.12$   &  $-0.12\pm0.19$      &  $+0.03$  &      $+0.09$    &      $-0.03$   & $+0.06$ \\
\hline      
      V    &  $+0.03\pm0.06$   &  $+0.19\pm0.10$   &  $-0.33\pm0.06$   &  $+0.00\pm0.08$   &  $-0.03\pm0.20$   &  $+0.11\pm0.11$   &  $-0.17\pm0.18$      &  $+0.20$  &      $+0.02$    &      $-0.07$   & $+0.12$ \\
     Mn    &  $-0.34\pm0.05$   &  $-0.42\pm0.10$   &  $-0.39\pm0.13$   &  $-0.43\pm0.08$   &  $-0.39\pm0.10$   &  $-0.38\pm0.09$   &  $-0.41\pm0.11$      &  $+0.11$  &      $-0.00$    &      $-0.02$   & $+0.06$ \\
     Co    &  $+0.03\pm0.05$   &  $+0.07\pm0.05$   &  $+0.07\pm0.09$   &  $+0.16\pm0.11$   &  $+0.08\pm0.09$   &  $+0.05\pm0.06$   &  $+0.11\pm0.11$      &  $+0.15$  &      $+0.04$    &      $-0.00$   & $+0.09$ \\
     Cu    &  $-0.35\pm0.05$   &  $-0.07\pm0.07$   &  $-0.12\pm0.17$   &  $-0.17\pm0.17$   &  $-0.18\pm0.16$   &  $-0.21\pm0.15$   &  $-0.15\pm0.18$      &  $+0.13$  &      $+0.04$    &      $-0.09$   & $+0.09$ \\
     Zn    &  $-0.30\pm0.05$   &  $-0.20\pm0.05$   &  $-0.30\pm0.05$   &  $-0.10\pm0.05$   &  $-0.23\pm0.10$   &  $-0.25\pm0.07$   &  $-0.20\pm0.11$      &  $+0.01$  &      $+0.03$    &      $-0.06$   & $+0.04$ \\
\hline     
\noalign{\vskip 0.2cm}
  [Fe/H]   &  $-1.34\pm0.15$   &  $-1.36\pm0.07$   &  $-1.48\pm0.17$   &  $-1.45\pm0.13$   &  $-1.39\pm0.08$   &  $-1.35\pm0.09$   &  $-1.46\pm0.13$      &  $+0.10$  &      $+0.10$    &      $+0.04$   & $+0.08$ \\  
\noalign{\vskip 0.2cm}

\noalign{\hrule}
 \hline
\end{tabular}}
\end{table*}  

\section{Dynamical properties}\label{sec:dynamics}

    In order to obtain the orbital parameters of NGC~6355, we employed an axisymmetric potential \cite{mcmillan17} adopting the Python package \texttt{galpy} \citep{bovy15}. We integrated a set of 1000 initial conditions forward  for 10 Gyr. The set was generated by using a MC algorithm adopting the observational uncertainties of the cluster data on proper motions $\mu_{\alpha}^{*}$ and $\mu_{\delta}$, heliocentric radial velocity, and the heliocentric distance. The \cite{mcmillan17} Galactic potential was adopted to compare our results with those of \cite{massari19} and to relate NGC~6355 with its plausible progenitor. A more realistic potential, including a contribution of the Galactic bar \citep{perez-villegas18,perezvillegas2020}, could provide a farther inward orbit for the GC members of the Galactic bulge. The orbital parameters are listed in Table \ref{tab:orbital-params}, including the values of the IOM. 
    
    Figure \ref{fig:orbits} shows the density probability map of the orbits of NGC~6355 in the $x-y$ and $R-z$ projections. The space region in which the orbits of NGC~6355 cross more frequently are shown in orange, and the black curves are the orbits considering the central values of the observational parameters. NGC~6355 is mostly confined within $\sim 2.6$ kpc and therefore has a high probability of belonging to the bulge component ($> 95 \%$) when we adopt the distance of $8.54\pm0.22$ kpc that we estimated in this work. Our new distance derivation indicates that the cluster NGC~6355 lies far inward based on its maximum height of $|z|<2.1$ kpc and the high eccentric orbit $>0.8$. {It may well be that this perigalactic distance is one the closest distances to the Galactic center.} 

\begin{table}
    \centering
    \caption{{Orbital parameters, velocities, and membership probabilities.}}
    \label{tab:orbital-params}
    \begin{tabular}{lrl}
    \hline
    \hline
    \noalign{\vskip 0.1cm}
     Parameter  &     Mean     &      Unit     \\
    \noalign{\vskip 0.1cm}
    \hline
      E                                &   $-2.31\pm0.03$             &    $\times10^{5}$km$^{2}$\,s$^{-2}$ \\
      L$_Z$                            &   $-31.28\pm24.42$           &    km\,s$^{-1}$kpc                  \\
      r$_{\rm peri}$                   &   $0.25\pm0.08$              &    kpc                              \\
      r$_{\rm apo}$                    &   $2.46\pm0.14$              &    kpc                              \\
      $|z|_{\rm max}$                  &   $1.91\pm0.08$              &    kpc                              \\
      ecc                              &   $0.82\pm0.05$              &    ---                              \\
      \hline
      $v_R$                            &   $-218.27\pm66.25$          &    km\,s$^{-1}$                     \\
      $v_{\phi}$                       &   $-192.39\pm34.54$          &    km\,s$^{-1}$                     \\
      \hline
      $\mathcal{P}_{\rm bulge}$        &   $95.00$                    &    \%                               \\
      $\mathcal{P}_{\rm disk}$         &   $5.00$                     &    \%                               \\
      $\mathcal{P}_{\rm inner\,halo}$  &   $0.00$                     &    \%                               \\
      $\mathcal{P}_{\rm outer\,halo}$  &   $0.00$                     &    \%                               \\
     \hline
     \hline
    \end{tabular}
\end{table}

\begin{figure}
    \centering
    \includegraphics[width=\columnwidth]{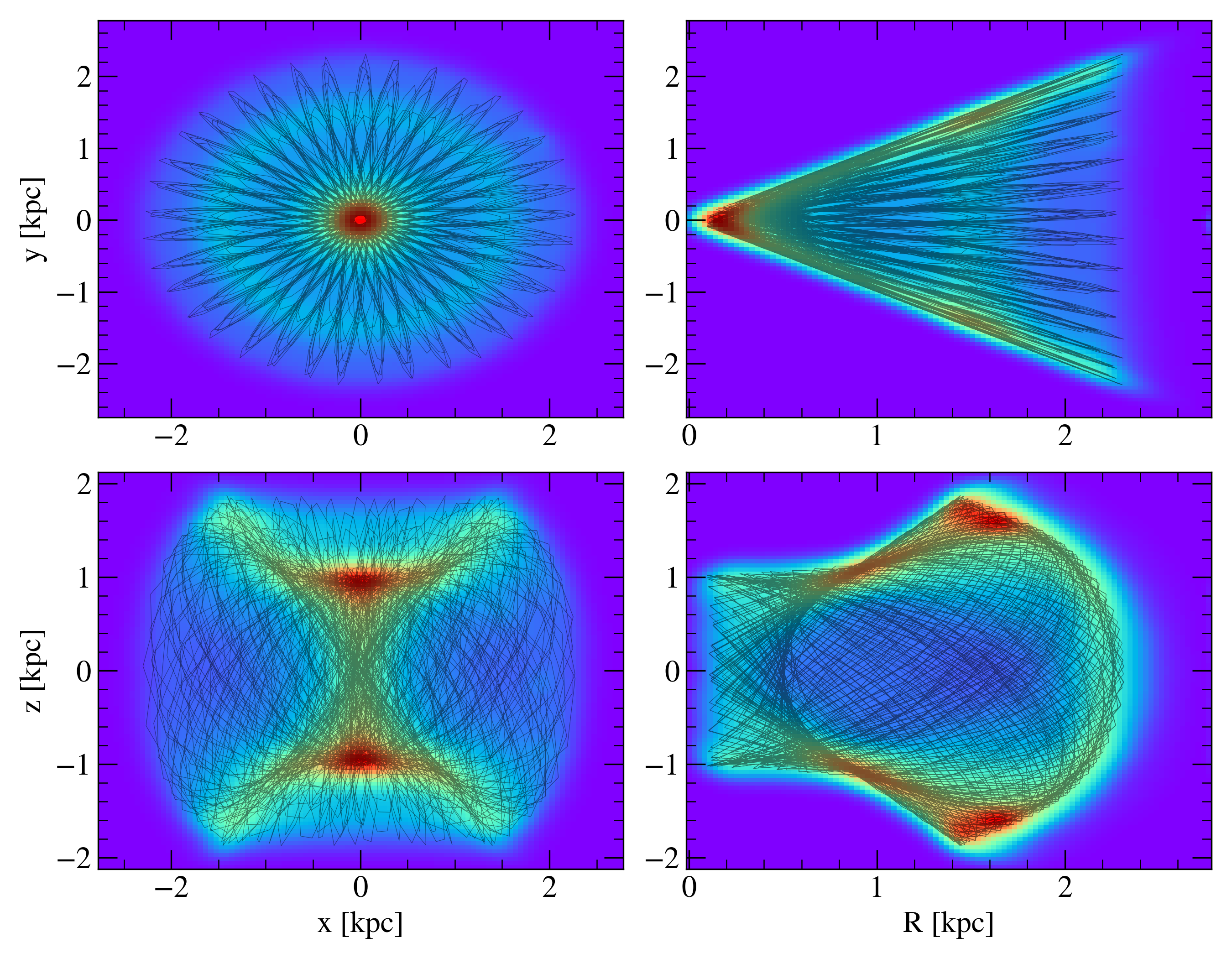}
    \caption{{Density probability map for the $x-y$ and $R-z$ projections of the set of orbits for NGC~6355. Orange corresponds to higher probabilities, and the black lines show the orbits using the main observational parameters.}}
    \label{fig:orbits}
\end{figure}

\section{Discussion}\label{sec:discussion}
\subsection{Kinematic classification}

{The orbital analysis shows that the orbit of NGC~6355 is compatible with a location at the Galactic bulge volume according to the classification of \citetalias{perezvillegas2020}, who presented the probability distribution of belonging to each Galactic component through the values of r$_{\rm apo}$ and $|z|_{\rm max}$  (Figure \ref{fig:orbits} and Table \ref{tab:orbital-params}). It is essential to mention that their classification is based on a Galactic potential that includes the contribution of the Galactic bar. Another robust Galactic potential, taking into account the friction dynamics in addition to the contribution of the Galactic bar, was applied by \cite{moreno22}. Their orbital parameters are essentially compatible with our results. The values of $E$ are precisely the same. The $L_Z$ and $r_{\rm peri}$ are compatible within $1\sigma$, while our value of $r_{\rm apo}$ is higher than that of \cite{moreno22}. This indicates that a more realistic Galactic potential confines NGC~6355 even more within the Galactic bulge volume. With the results using the \cite{mcmillan17} Galactic potential, NGC~6355 is a Galactic bulge GC with a probability of about $95\%$, and a $5\%$ probability of belonging to the Galactic disk.}

{After establishing that NGC~6355 currently is a member of the Galactic bulge, we investigated whether this cluster originated from the primordial material of the Galaxy or if it is a remnant of the first mergers of the MW. To do this, we studied the chemodynamical and photometric information derived in previous sections. }
\subsection{Comparison with bulge field stars}

{To study NGC~6355 in the context of the Galactic bulge, we compared the orbital parameters derived in this work with the field star population composed by the reduced proper motion (RPM) sample from \citetalias{queiroz21} and the bulge RR Lyrae from the OGLE Galaxy Variability Survey \citep{Soszynski2019}.}

{We matched the OGLE sample with APOGEE DR17, which already provides the abundances, radial velocities, and proper motions (previously obtained from Gaia EDR3). After this, the second sample was matched with Starhorse \citep{queiroz20} in order to obtain the distance values. Finally, the resulting sample consisted of $4132$ stars.}

{NGC~6355 has a relatively high $|Z|_{max}$ and $e$, placing it in cell {F} of Figure 20 in \citetalias{queiroz21}. This is reproduced here in the upper left panel of Figure \ref{fig:bulge} for the RPM sample and in the upper right panel for the RR Lyrae sample. The normalized population density as a function of [Fe/H] (MDF), $R_{mean}$ (mean between $r_{apo}$ and $r_{peri}$), and $v_{\phi}$ is shown in the three bottom panels of Figure \ref{fig:bulge}, respectively. Based on the MDF (lower left panel ), the RPM sample comprises the moderately metal-rich bulge MDF, while the RR Lyrae sample is the metal-poor tail one. As expected, NGC~6355, an old GC, is located together with the peak of the RR Lyrae MDF and $R_{mean}$ distribution. Nevertheless, the $v_{\phi}$ of the cluster is in the tail of both samples.}

\begin{figure*}[h]
    \centering
    \includegraphics[width=0.85\textwidth]{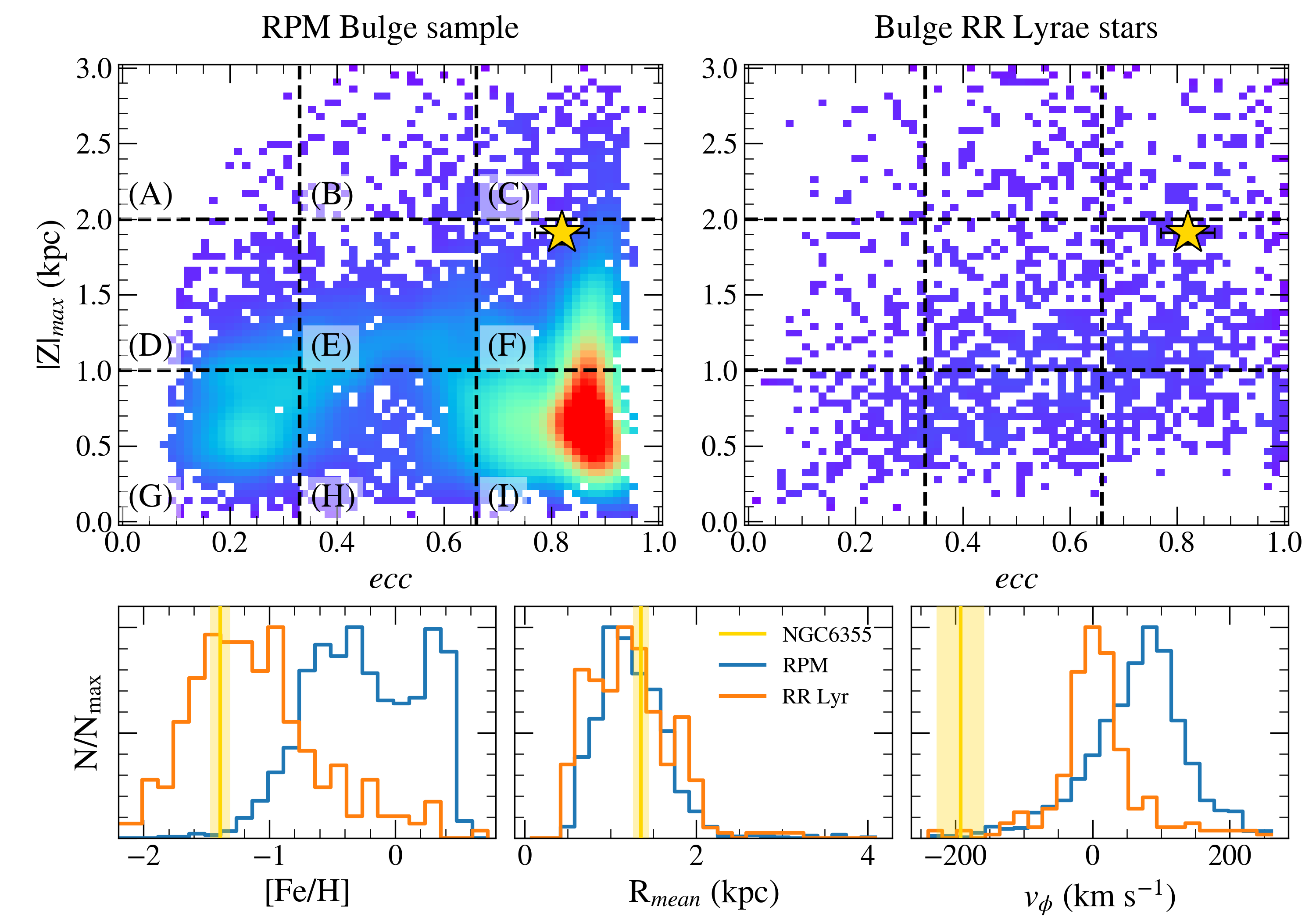}
    \caption{ NGC6355 compared with the RPM bulge sample of \citetalias{queiroz21} (left panel) and Galactic bulge RR Lyrae population (right panel). Upper panels: |Z|$_{max}$ as a function of the eccentricity plane divided into nine frames defined by the letter close to the horizontal lines. The golden star represents the locus of NGC~6355. The bottom panels show the population density of [Fe/H], R$_{mean}$, and $v_\phi$ for cell F. The gold lines represent the position of NGC~6355 in each panel, and the shaded gold region shows the $1\sigma$ distribution.}
    \label{fig:bulge}
\end{figure*}

{The comparison with the bulge field populations shows that NGC~6355 is likely an in-situ GC compatible with the old and metal-poor RR Lyrae component of the Galactic bulge.}
\subsection{Comparison with chemodynamical models}

{To investigate the chemical abundances in the context of nucleosynthesis, we compared our results with chemical evolution models. The models for O, Mg, Si, Ca, V, Mn, Co, Cu, and Zn were computed with the code described in \cite{friaca17} \citep[see also][ for V, Mn, Co, Cu, and Zn]{barbuy15,ernandes20,ernandes2022}.} The star formation rate (SFR) was found to be best suited with $\nu = 1$ Gyr$^{-1}$ in order to fit the abundances of a selected sample of bulge stars in \cite{razera22}. Therefore, we adopted this SFR for all elements. The SFR is the rate at which the available gas mass is turned into stars. Consequently, it measures the inverse of the system formation timescale: Our adopted $\nu=1.0$ Gyr$^{-1}$ represents a rather fast star formation of $1.0$ Gyr. The models assume a baryonic mass of $2\times10^9$ M$_\odot$, a dark halo mass $1.3\times10^{10}$ M$_\odot$, and the cosmological parameters from Planck Collaboration (2016). The bulge is considered a classical spheroidal component. Finally, the models project the chemical abundance distribution at different radius ranges:  r$< 0.5$ kpc {(dash-dotted line in Figure \ref{fig:abd-alphaels})}, $0.5<$ r $< 1$ kpc (dashed line), $1 <$ r $< 2$ kpc (dotted line), and $2<$ r $< 3$ kpc (solid line). For Na and Al, we used the \cite{kobayashi20} models for the Galactic bulge. These models also assume an SFR $\nu\sim 1.0$ Gyr$^{-1}$. 
 
{In addition to the chemodynamical models, in the following analysis, we also compare the abundance ratios obtained in this work with the bulge GCs Palomar 6 \citep{souza21},  HP~1 \citep{barbuy2018a}, NGC~6558 \citep{barbuy2018b}, and NGC~6522 \citep{barbuy2021}. Furthermore, we compare NGC~6355 with the RPM and RR Lyrae samples (for the case of $\alpha$ and odd-Z elements) inside cell F of Figure \ref{fig:bulge}.}

{Figure \ref{fig:abd-alphaels} shows the abundances of O, Mg, Si, and Ca as a function of [Fe/H]. To better illustrate the comparison, the mean locus of the RPM sample is shown as a solid black line. We derived [$\alpha$/Fe]$=+0.36\pm0.09$ considering all $\alpha$ elements O, Mg, Si, Ca, and Ti. Our mean [$\alpha$/Fe] is compatible within $1\sigma$ with the assumed value for the isochrone fitting. This reinforces the consistency of the analysis. In all cases, NGC~6355 is compatible with the RR Lyr locus. {Additionally, NGC~6355 is also compatible with the other GCs, execpt for Ca, in which the cluster is relatively richer than the others.}}

\begin{figure}
    \centering
    \includegraphics[width=\columnwidth]{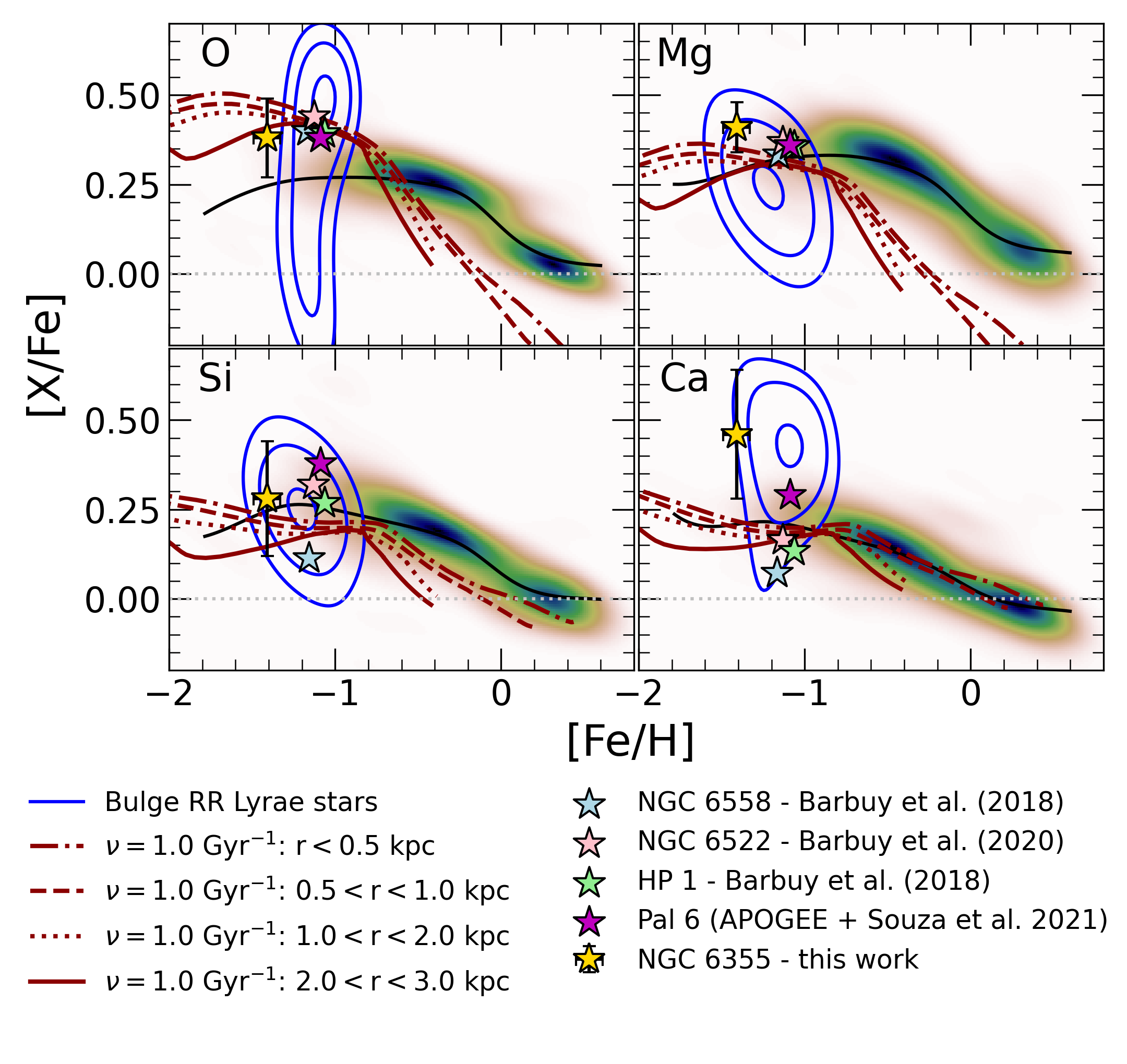}
    \caption{O, Mg, Si, and Ca abundance as a function of [Fe/H]. The KDE plot represents the RPM bulge selection from cell F, and the blue contours represent the RR Lyrae sample. The stars are abundances of bulge GCs: NGC 6558 (cyan), NGC 6522 (pink), HP1 (green), and Pal 6 (magenta). The golden star represents the mean abundance of NGC 6355. The chemodynamical evolution models are shown in different radii ranges:  r$< 0.5$ kpc (dash-dotted line), $0.5<$ r $< 1$ kpc (dashed line), $1 <$ r $< 2$ kpc (dotted line), and $2<$ r $< 3$ kpc (solid line).}
    \label{fig:abd-alphaels}
\end{figure}

{As a result of the presence of MPs, the spread in [Na/Fe] is higher than for the other elements. This effect can be observed in the top panel of Figure \ref{fig:abd-oddZ} with the discrepancy between the two models and the mean locus for the case of low metallicities. \cite{souza21} found that Pal~6 is not compatible with a bulge [Na/Fe]. They argued that the reason is due to the presence of a 2G star in their sample. The same effect is expected for [Al/Fe] because the Al abundance is a good indicator of the presence of 2G stars. The lower panel of the same figure shows the high error bars of [Al/Fe] for NGC~6355 that are due to the presence of two moderately Al-rich stars.}

\begin{figure}
    \centering
    \includegraphics[width=\columnwidth]{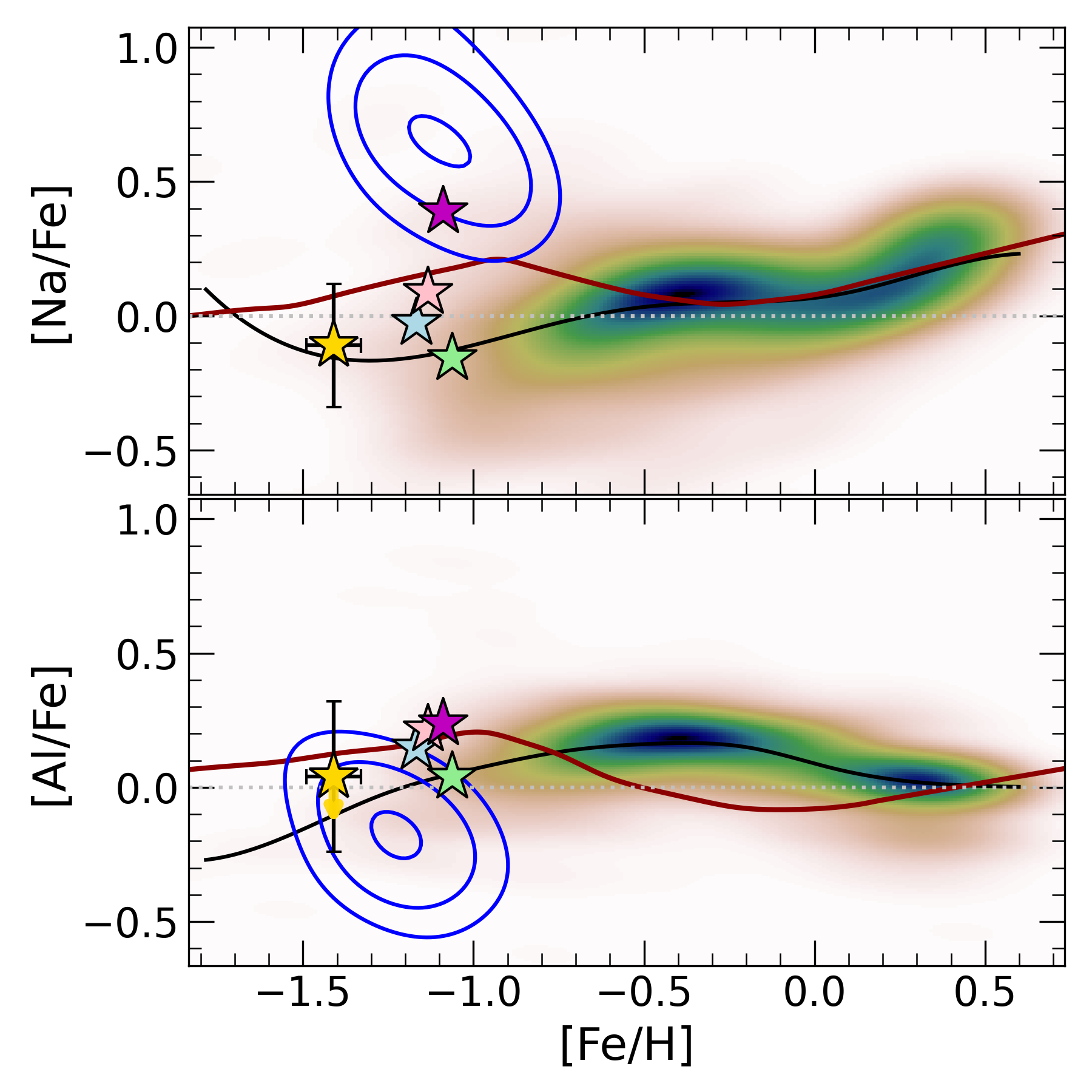}
    \caption{{Same as Figure \ref{fig:abd-alphaels} for odd-Z elements Na (upper) and (bottom). The solid red line is the chemical evolution model from \protect\cite{kobayashi20}.}}
    \label{fig:abd-oddZ}
\end{figure}

In Figure \ref{fig:iron-peak} we investigate the iron-peak elements V, Mn, Co, and Cu. To increase the bulge sample, we also compared our results with bulge GC stars from \cite{ernandes2018} and the bulge field stars from \cite{ernandes20}. The chemical evolution model fits NGC~6355 perfectly. {For Cu abundances, the selected bulge clusters have relatively lower values than NGC~6355, indicating a different possible scenario for its early evolution}. In the case of V (top left panel), the evolution model is shifted to lower abundances for all metallicities than the mean locus. The models suitably fit the abundances of NGC~6355, the selected clusters, and the bulge GC stars for Mn, Co, and Cu. 

\begin{figure}
    \centering
    \includegraphics[width=\columnwidth]{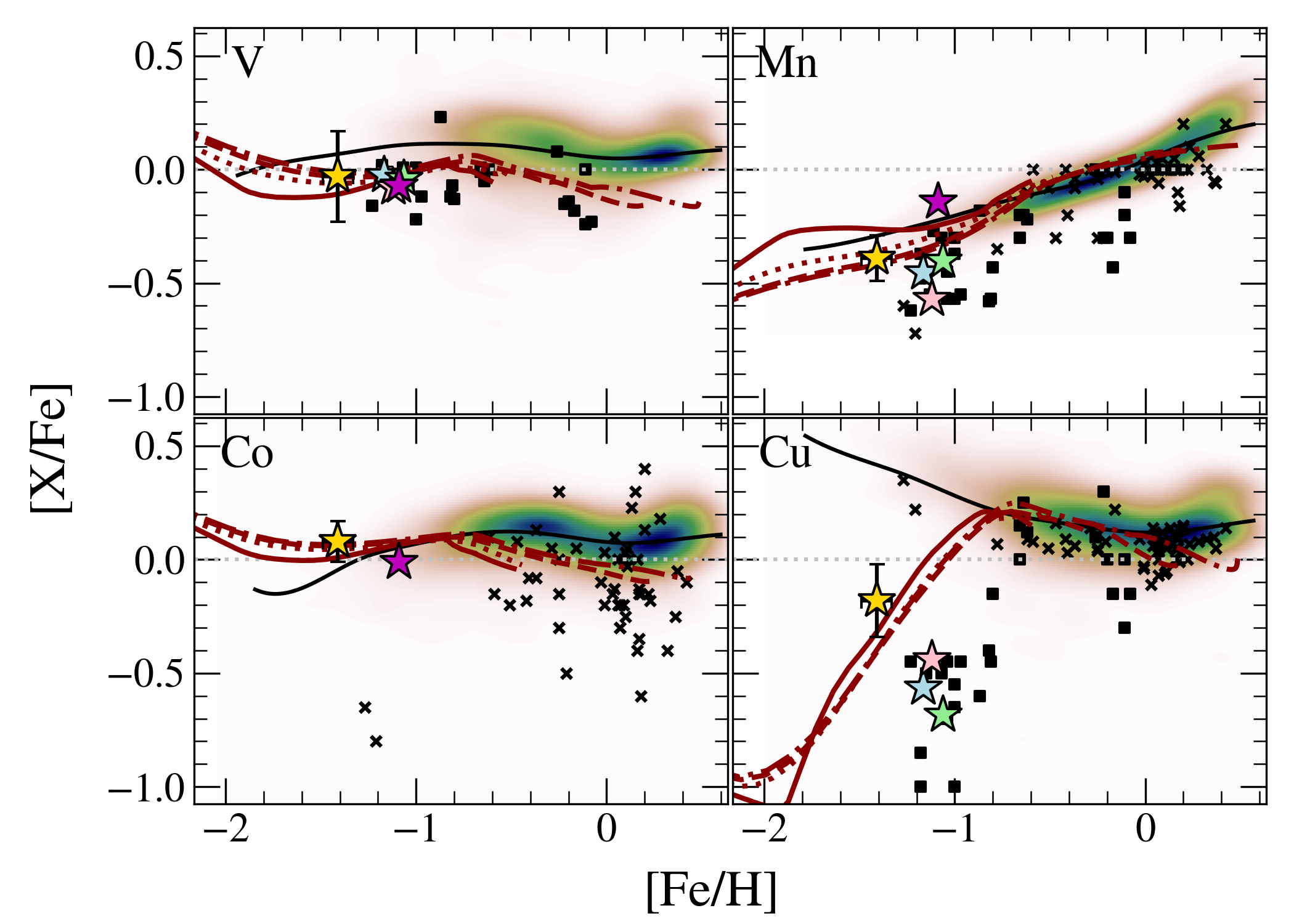}
    \caption{ Same as Figure \ref{fig:abd-alphaels} for V, Mn, Co, and Cu. The black squares are bulge GC stars from \protect\cite{ernandes2018}, and black crosses show bulge field stars from \protect\cite{ernandes20}. The chemodynamical evolution models are shown in different radius ranges:  r$< 0.5$ kpc (dash-dotted line), $0.5<$ r $< 1$ kpc (dashed line), $1 <$ r $< 2$ kpc (dotted line), and $2<$ r $< 3$ kpc (solid line).}
    \label{fig:iron-peak}
\end{figure}

{The Zn abundances derived in this work are based only on the line \ion{Zn}{I} $6362.339$ \AA. In Figure \ref{fig:zinc}, NGC~6355 is perfectly fitted by the models and is compatible with all reference clusters. Here it is worth noting that the models predict supersolar zinc abundances for metallicities above $-1.0$ and subsolar for values below $-1.0$. The low Zn as an indicator of an ex-situ origin was suggested only for the case of near-solar metal-rich stars \cite{minelli21}. }

\begin{figure}
    \centering
    \includegraphics[width=\columnwidth]{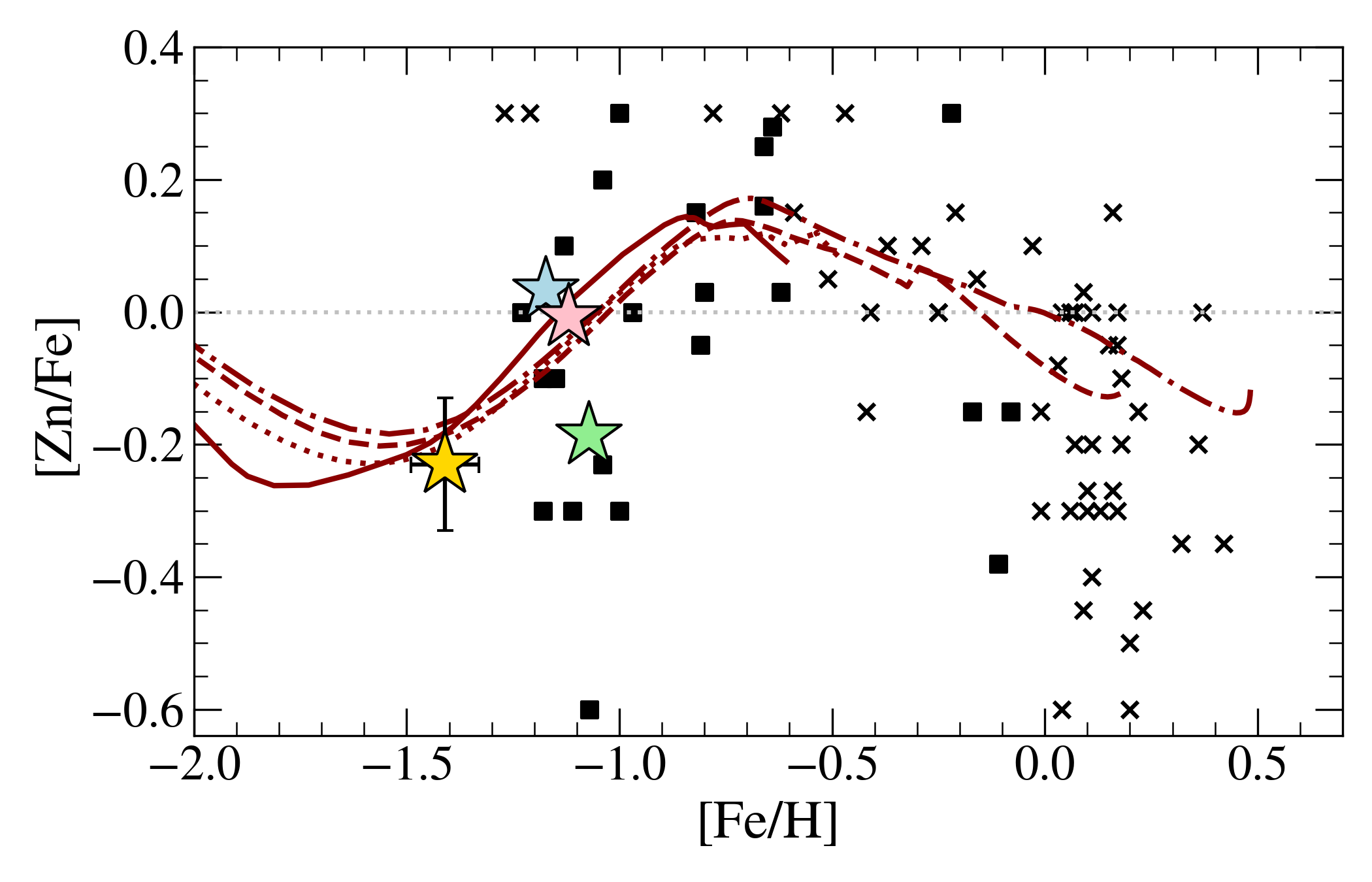}
    \caption{{Same as Figure \ref{fig:abd-alphaels} for Zn. The \protect\cite{friaca17} evolution models are shown in different radius ranges:  r$< 0.5$ kpc (dash-dotted line), $0.5<$ r $< 1$ kpc (dashed line), $1 <$ r $< 2$ kpc (dotted line), and $2<$ r $< 3$ kpc (solid line).}}
    \label{fig:zinc}
\end{figure}

{The comparison of heavy-element abundances of NGC~6355 with literature GCs is shown in Figure \ref{fig:heavy_pattern}. NGC~6355 abundances are compatible with HP~1 in almost all heavy elements except for Ba, for which NGC~6355 has higher values. {There is a rather large scatter in the abundances of n-capture elements, especially Y, Zr, and Ba. This pattern is better explained in \cite{chiappini11}, \cite{cescutti14}, and \cite{barbuy2018a}.}}

\begin{figure}
    \centering  
        \includegraphics[width=\columnwidth]{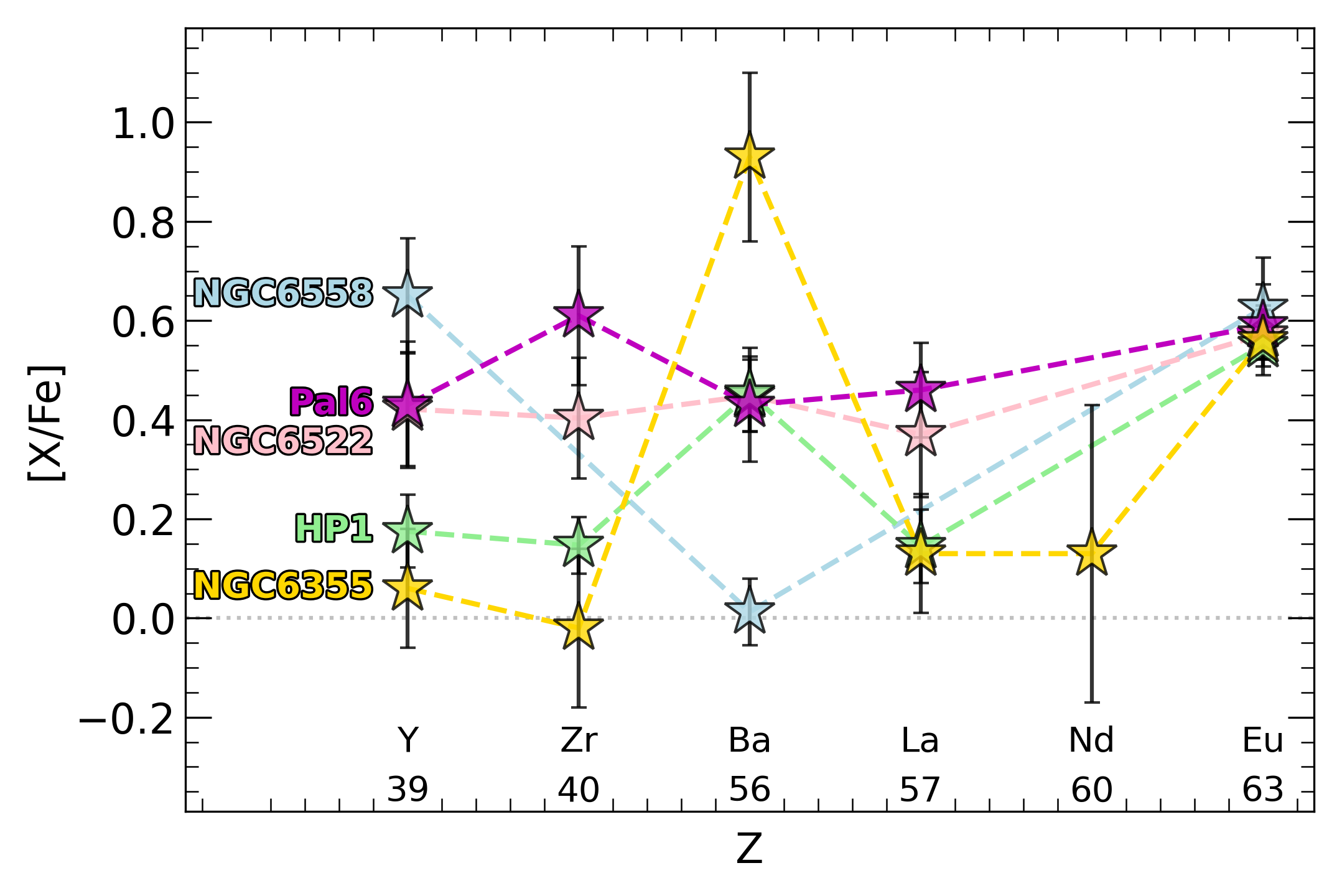}
    \caption{{Abundance pattern [X/Fe] vs. atomic number (Z) for heavy elements Y, Zr, Ba, La, Nd, and Eu. The colours are the same as in Figure \ref{fig:abd-alphaels}.}}
    \label{fig:heavy_pattern}
\end{figure}
\subsection{Analysis of abundance discriminators}

{The [Mg/Mn]-[Al/Fe] plane is often used in the context of the Galactic halo to split the original MW population {from} merger remnants \citep{hawkins15,limberg22} because the accreted population shows lower [Al/Fe] abundances and high $\alpha$ abundances due to the abrupt evolution interruptions of the merger progenitor. \cite{horta21} applied the same idea for a star sample located in the Galactic centre to find debris stars within the Galactic bulge. They called this inner Galaxy structure placed in the ex-situ portion of the [Mg/Mn]-[Al/Fe] plane Heracles and defined it as follows \citep{horta22}:}

\begin{align}
{\rm Heracles } = \left\{ \begin{array}{llll}  
                -2.60 \leq E/10^5 \leq -2.00\,{\rm km^2\,s^{-2} } \\ 
                e \geq 0.60 \\
                r^\dag \leq 4 kpc \longrightarrow \text{$^\dag$ Galactic centre distance} \\
                {\rm[Mg/Mn] > +0.25} \\
                {\rm[Mg/Mn] > 5 \times [Al/Fe] + 0.5 }. \\
                \end{array} \right. \label{eq:Heracles_def}
\end{align}

{In the context of the [Mg/Mn]-[Al/Fe] plane (Figure \ref{fig:mgmn_al}), the reference bulge GCs have no preferential positioning. However, Pal~6 and NGC~6355 show interesting behaviours. \cite{souza21} found that Pal~6 is an in-situ member. We confirm this through the [Mg/Mn]-[Al/Fe] plane, with APOGEE abundances for Pal~6 members, because this cluster is located perfectly in the high-$\alpha$ region. In contrast, the NGC~6355 is located at the border between Heracles and the \emph{in-situ high-$\alpha$} region. Two of our four stars present a maximum Al abundance of $+0.30$, which could be 2G stars \citep[2G][]{meszaros20,fernandez-trincado22}. Figure \ref{fig:anticorrs} shows the (anti-)correlations that indicate the presence of MPs. {We do not find a Mg-Al anticorrelation, although this is expected mainly for massive clusters because of the metallicity multimodality \citep{meszaros20}}. We can also observe a slight difference between the 1G and 2G [La/Fe] mean abundances. \cite{marino21} showed that it is not possible to separate the MPs using La abundances. However, the mean abundance value is higher for the anomalous than for the normal stars.}

{The mean on the [Mg/Mn]-[Al/Fe] plane changes in the diagonal direction going to the left or right circles of Figure \ref{fig:mgmn_al} when only the 1G or 2G stars are considered to compute the cluster mean abundance. Then, assuming the mean abundance of the 1G stars, NGC~6355 is placed inside the Heracles region on the [Mg/Mn]-[Al/Fe] plane. It is worth pointing out that \citetalias{queiroz21} showed that a sample of counter-rotating stars in the RPM sample presents no preferential location in the [Mg/Mn]-[Al/Fe] plane, suggesting that this region may not be entirely composed of accreted objects. Therefore, even though NGC~6355 is placed in the accreted region, this by itself does not signify an ex-situ origin.}

\begin{figure}
    \centering
    \includegraphics[width=1\columnwidth]{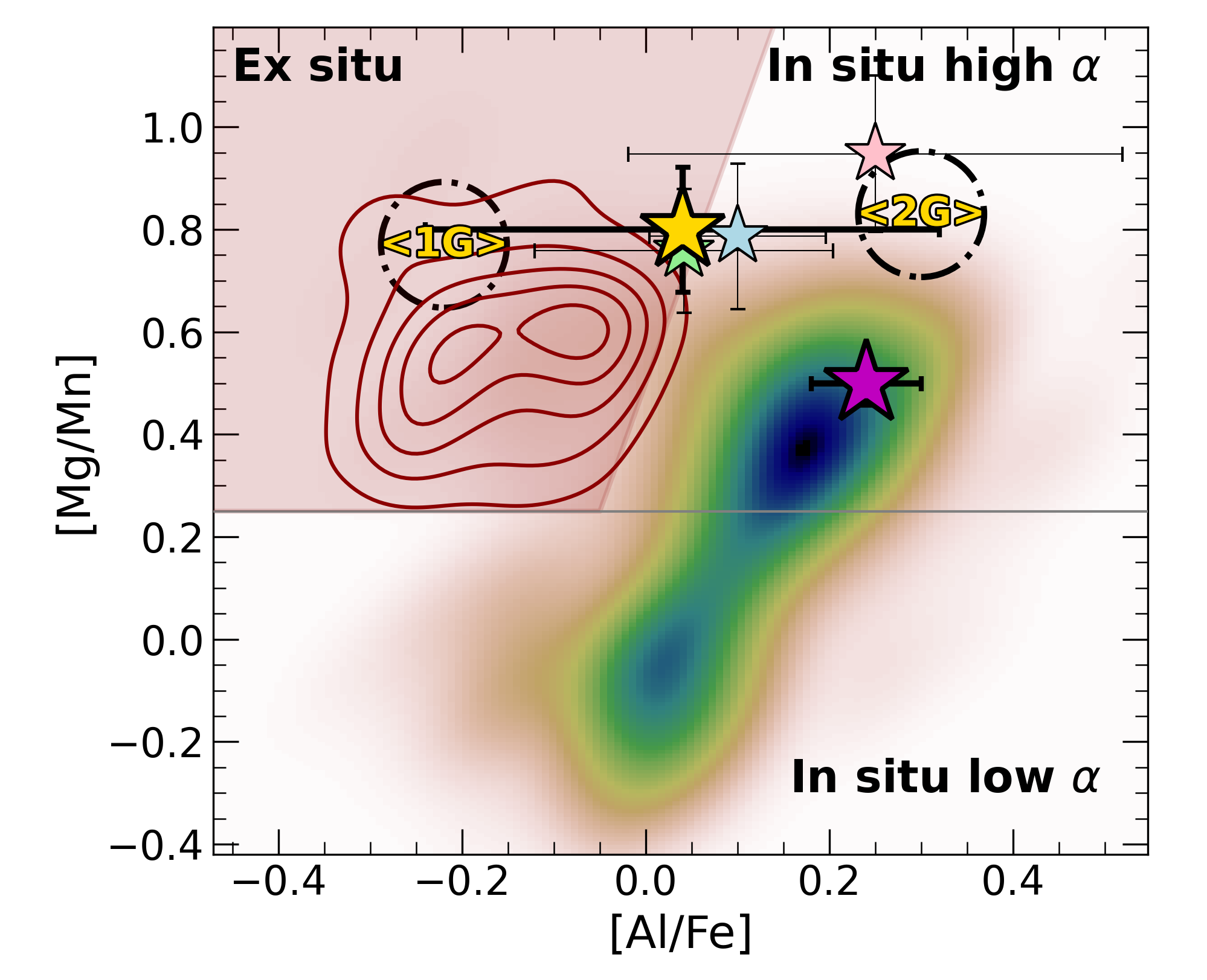}
    \caption{{ [Mg/Mn]-[Al/Fe] plane with the identification of Heracles in red contours. The colours and density map are the same as Figure \ref{fig:abd-oddZ}. The circles represent the mean considering only 1G (left) and 2G (right) stars in NGC~6355.}}
    \label{fig:mgmn_al}
\end{figure}
\subsection{Age-metallicity relation and integral-of-motion space}

{The AMR is an interesting tool for investigating the origin of the MW GCs \citep{massari19}. Figure \ref{fig:amr} shows the MW GCs AMR collected from \cite{kruijssen19}. We overplot the mean locus of the two branches (in-situ and ex-situ) as defined by \cite{forbes20}, which are defined as follows: }

\begin{equation}
Z = -p \ln\left(\frac{t}{t_f}\right){,}
\end{equation}
{where $Z$ is the mass metallicity, $p$ is the effective yield, and $t_f$ is the time to the initial formation of the system. The red line is the ex-situ population obtained using the parameters found by \cite{limberg22}. To represent the in-situ population, we only changed the parameters by hand to place the line above the older branch (blue line).}

{In Figure \ref{fig:amr}  we show NGC~6355 as the gold star together with the bulge GCs. It is clear that age is  hugely crucial for the progenitor of a GC classification. For the case of NGC~6355, the isochrone fitting considering the $T_{\rm eff}$ correction clearly indicates  an in-situ candidate. {Nevertheless, the uncertainty on age still gives NGC~6355 a low probability of having an ex-situ origin.}}

\begin{figure}
    \centering
    \includegraphics[width=\columnwidth]{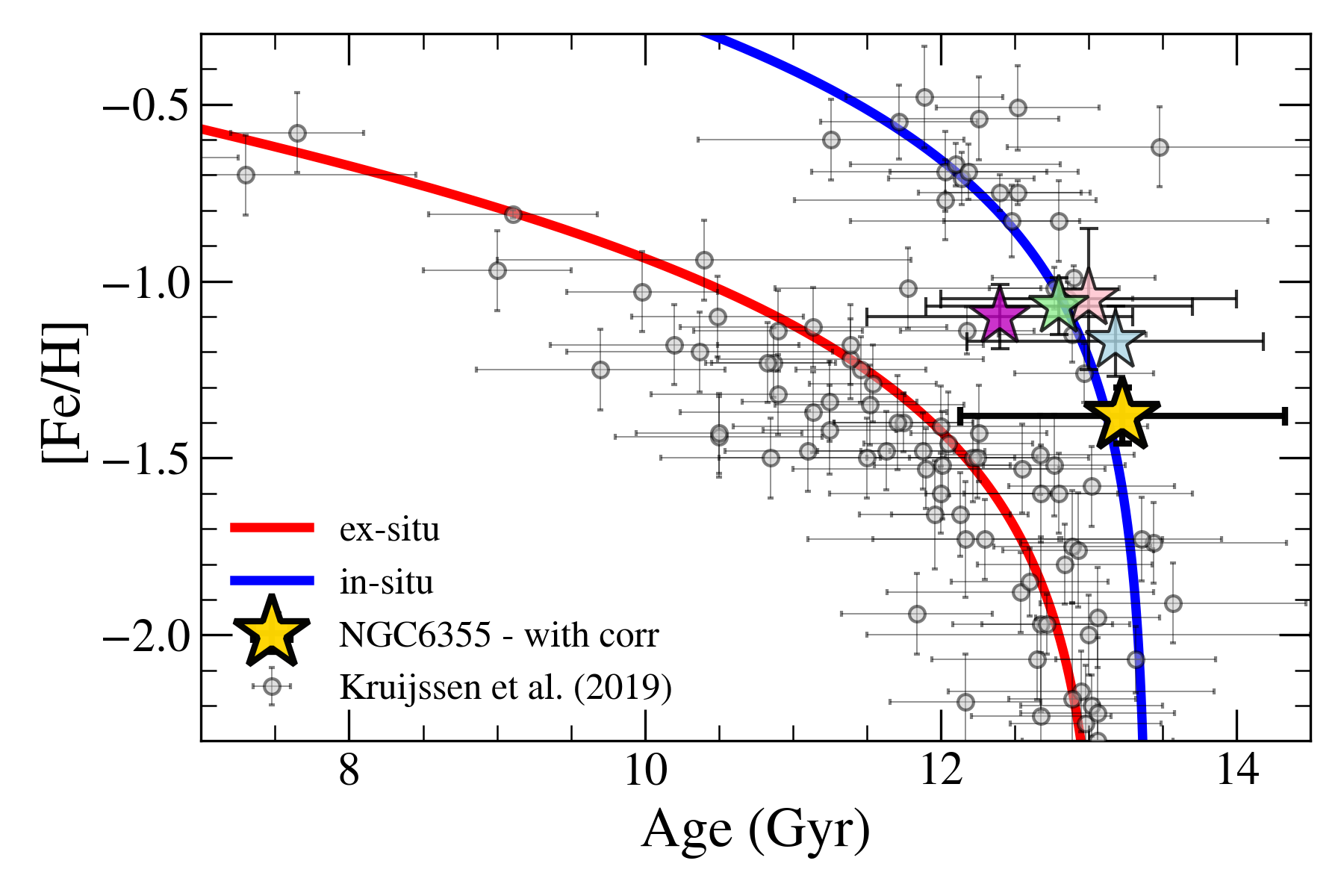}
    \caption{{AMR for the Galactic GCs system. The grey dots are the values reported in \protect\cite{kruijssen19}. For comparison criteria, the mean locus of in-situ and ex-situ populations are shown (see text for details). The symbols are the same as in Figure \ref{fig:abd-oddZ}. }}
    \label{fig:amr}
\end{figure}

{The dynamics based on the orbital parameters of Table \ref{tab:orbital-params} show that $E$ and $L_Z$ of NGC~6355 are $-2.31\pm0.03$ $\times10^{5} $km$^{2}$\,s$^{-2}$ and $-31.28\pm24.42$ km\,s$^{-1}$kpc, respectively. These values place the cluster in the low-energy and low absolute $L_Z$ region. \cite{horta21} also analysed the IOM space in the context of the inner Galaxy separating Heracles (as defined above) from the bulge selection. We reproduced their Figure 5 in our Figure \ref{fig:lz-e} and removed their high-E stars with $e<0.4$. The main-bulge progenitor is shown in the left panel, and the Heracles (supposedly ex-situ) progenitor is shown in the right panel. NGC~6355 and the selected GCs are placed almost in the same region {in the IOM space (Figure \ref{fig:lz-e})}. However, it is not possible to distinguish a specific region for each progenitor based on the contour curves alone.}
\begin{figure}
    \centering
    \includegraphics[width=\columnwidth]{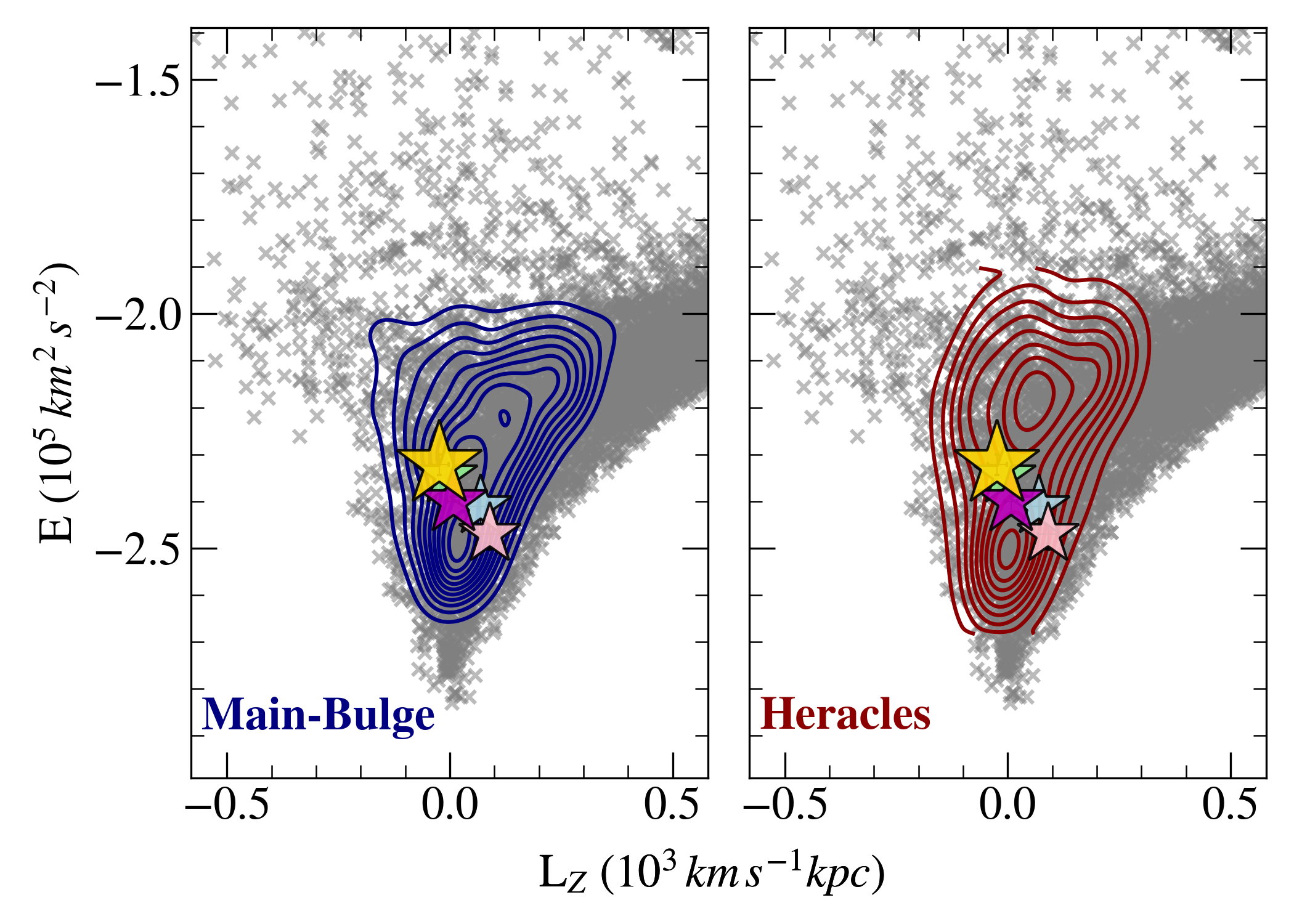}
    \caption{{IOM space for the bulge stars selected by \protect\cite{horta21}. The left panel shows the contours for the main-bulge progenitor stars. The right panel shows the contours of the Heracles progenitor. The stars are coulored as in Figure \ref{fig:abd-alphaels}. }}
    \label{fig:lz-e}
\end{figure}

{Although NGC~6355 has properties of ex-situ GCs such as the [Mg/Mn] and [Al/Fe] abundances, which are compatible with Heracles, we can confirm the in-situ origin of NGC~6355 because it is confined to the volume of the Galactic bulge. From the point of view of chemical  abundances, most of its element abundances follow the in-situ clusters and bulge RR Lyrae population, including its low Zn abundance, which appears to be compatible with the chemodynamical evolution models. The old age of NGC~6355 completes the in-situ scenario for the cluster because its age fits the predictions for the early evolution of the Galactic bulge perfectly.}

\section{Conclusions}\label{sec:conclusions}
{We analysed the globular cluster NGC~6355 in the context of the evolution of the Galactic bulge. This task required a deep and careful analysis, including photometry, chemical abundances, and dynamics. To do this, we gathered high-resolution spectroscopy from FLAMES-UVES, photometry from the HST F438W/F555W filters, and Galactic dynamics calculations.}

{The spectroscopic analysis resulted in a metallicity of [Fe/H]$=-1.39\pm0.08$ for NGC~6355. This means that the GC is one of the metal-poor clusters of the Galactic bulge. A mean [$\alpha$/Fe]$=+0.36\pm0.09$ was derived based on O, Mg, Si, and Ca abundances, indicating that NGC~6355 is characterized by enrichment from supernovae type II. We found good agreement among NGC~6355 abundances, bulge GCs, and the bulge RR Lyrae sample, except for the cases of Ca, Zn, and Ba, for which Ca and Ba appear to be enhanced, and Zn, which is deficient relative to the comparison clusters. At the same time, Ca is compatible with the RR Lyrae sample. We tested the hypothesis of a different origin for NGC~6355 with the [Mg/Mn]-[Al/Fe] plane and found that NGC~6355 is in the accreted region of the plane when we assume only the Al-depleted {(1G)} stars. We found that two of the four stars are Al-depleted and N-normal, while the others are N- and Al-rich. The detection of MPs  was confirmed through the (anti-)correlations Al-N/Na \citep{meszaros20}, and Na-O \citep{carretta09}. }

{From isochrone fitting, we derive an age of $13.2  \pm1.1$ Gyr, reinforced by the fact that the cluster has a BHB, similar to other old and moderately metal-poor bulge GCs such as NGC~6522, NGC~6558, HP~1, AL~3, Terzan~9, and UKS~1. This old age places NGC~6355 perfectly at the in-situ branch of the AMR. Because of the rather low metallicity of NGC~6355 relative to these other clusters, these results raise the question which would be the lowest metallicity for the family of metal-poor globular clusters of the Galactic bulge, as discussed in \cite{geisler22}.} 

{Finally, we investigated the IOM space. This is not conclusive because as observed by \cite{horta21}, the internal region of the Galaxy is dynamically mixed, and it is not possible to separate the ex-situ stars from the in-situ population. As evidence of this, \cite{kruijssen19} concluded that the low-energy progenitor called Kraken must have been formed at redshift $z=4$, or $\sim 12.3$ Gyr ago, indicating that the early merger of the Kraken galaxy with the Galaxy had enough time to mix dynamically.}

{In conclusion, according to our photochemodynamical analysis, NGC~6355 is currently a member of the Galactic bulge and seems to have originated from the main-bulge progenitor, with a low probability of an ex-situ origin.  More spectroscopic data of cluster stars could provide more consolidated abundance values, giving us more constraints on the NGC~6355 origin and its MPs formation scenario.}

\begin{acknowledgements}
    {We thank an anonymous referee for suggestions that have improved the quality of this paper}. SOS acknowledges a FAPESP PhD fellowship no. 2018/22044-3. HE acknowledges a CAPES PhD fellowship. SOS and MV acknowledge the support of the Deutsche Forschungsgemeinschaft (DFG, project number: 428473034).  A.P.-V. and S.O.S. acknowledge the DGAPA-PAPIIT grant IA103122. BB and ACS acknowledge grants from FAPESP, CNPq and CAPES - Financial code 001.
SO acknowledges partial support by the Università degli
Studi di Padova Progetto di Ateneo BIRD178590. S.C. acknowledges support
from Progetto Mainstream INAF (PI: S. Cassisi), from INFN (Iniziativa specifica TAsP), and from PLATO ASI-INAF agreement n.2015-019-R.1-2018.

\end{acknowledgements}

%
%

\clearpage
\onecolumn

\appendix

\section{Line lists}

\begin{longtable}{lcccccccc}
\caption{Equivalent widths for \ion{Fe}{I} and \ion{Fe}{II} lines.  }            
\label{tab:fe}\\      

\hline
\noalign{\smallskip}
\multirow{2}{*}{Ion} & $\lambda$ & $\chi_{ex}$  & \multirow{2}{*}{log$gf$} &     star 1539   &    star 1363   &     star 1176  &  star 133 \\ 
                     &  [\AA]  &  [eV]        &                          & \multicolumn{4}{c}{[m \AA]}\\ 
\hline\hline \endfirsthead
\multicolumn{8}{c}%
{{\bfseries \tablename\ \thetable{} -- continued}} \\ \hline
Ion & $\lambda$ & $\chi_{ex}$  & log$gf$ &     star 1539   &    star 1363   &     star 1176  &  star 133\\ 
\hline\hline \endhead

   \ion{Fe}{II} & $5991.38$ & $3.10$ & $-3.65$  &   ---    & $ 21.1$ & $ 33.4$ & $ 17.0$ \\ 
   \ion{Fe}{II} & $6084.11$ & $3.20$ & $-3.97$  &   ---    & $ 16.8$ & $ 13.2$ & $ 11.8$ \\ 
   \ion{Fe}{II} & $6149.25$ & $3.80$ & $-2.69$  &   ---    & $ 62.4$ & $ 28.1$ & $ 17.5$ \\ 
   \ion{Fe}{II} & $6247.56$ & $3.80$ & $-2.30$  &  $ 41.6$ &  ---    & $ 47.1$ & $ 29.9$ \\ 
   \ion{Fe}{II} & $6416.93$ & $3.80$ & $-2.64$  &  $ 29.7$ & $ 34.6$ & $ 25.5$ & $ 18.1$ \\ 
   \ion{Fe}{II} & $6432.68$ & $2.80$ & $-3.57$  &  $ 42.2$ & $ 42.4$ & $ 40.2$ & $ 33.1$ \\ 
   \ion{Fe}{II} & $6456.39$ & $3.90$ & $-2.31$  &   ---    & $ 48.2$ & $ 61.7$ & $ 40.8$ \\ 
   \ion{Fe}{II} & $6516.08$ & $2.80$ & $-3.31$  &  $ 53.8$ & $ 17.1$ & $ 47.0$ & $ 42.7$ \\ 
   \hline
    \ion{Fe}{I} & $5905.67$ & $4.60$ & $-0.73$  &   ---    & $ 30.3$ &  ---    & $ 27.2$ \\ 
    \ion{Fe}{I} & $5916.25$ & $2.40$ & $-2.97$  &  $ 87.1$ &  ---    & $ 55.6$ &  ---    \\ 
    \ion{Fe}{I} & $5927.79$ & $4.60$ & $-1.09$  &   ---    & $ 23.7$ & $ 17.1$ & $ 19.3$ \\ 
    \ion{Fe}{I} & $5929.67$ & $4.50$ & $-1.41$  &  $ 23.9$ & $ 17.7$ & $ 14.4$ & $ 16.0$ \\ 
    \ion{Fe}{I} & $5930.18$ & $4.60$ & $-0.23$  &  $ 66.3$ & $ 62.6$ & $ 46.7$ & $ 49.9$ \\ 
    \ion{Fe}{I} & $5934.65$ & $3.90$ & $-1.17$  &  $ 70.3$ &  ---    & $ 49.0$ & $ 54.2$ \\ 
    \ion{Fe}{I} & $5952.73$ & $3.90$ & $-1.44$  &  $ 50.9$ & $ 49.2$ & $ 38.9$ & $ 43.3$ \\ 
    \ion{Fe}{I} & $5956.69$ & $0.80$ & $-4.60$  &  $119.8$ &  ---    & $ 78.4$ &  ---    \\ 
    \ion{Fe}{I} & $5975.35$ & $4.80$ & $-0.69$  &  $ 33.6$ & $ 35.7$ & $ 23.6$ &  ---    \\ 
    \ion{Fe}{I} & $5983.69$ & $4.50$ & $-1.47$  &   ---    & $ 44.4$ & $ 34.0$ &  ---    \\ 
    \ion{Fe}{I} & $5987.06$ & $4.80$ & $-0.43$  &  $ 39.7$ &  ---    & $ 25.3$ &  ---    \\ 
    \ion{Fe}{I} & $6003.01$ & $3.80$ & $-1.12$  &  $ 74.9$ & $ 74.5$ & $ 54.2$ & $ 65.3$ \\ 
    \ion{Fe}{I} & $6005.54$ & $2.50$ & $-3.61$  &  $ 37.8$ & $ 41.6$ & $ 19.2$ & $ 24.9$ \\ 
    \ion{Fe}{I} & $6008.56$ & $3.80$ & $-0.99$  &  $ 84.3$ &  ---    & $ 60.1$ & $ 67.0$ \\ 
    \ion{Fe}{I} & $6020.17$ & $4.60$ & $-0.27$  &  $ 73.3$ &  ---    & $ 50.0$ & $ 54.5$ \\ 
    \ion{Fe}{I} & $6024.05$ & $4.50$ & $-0.12$  &  $ 82.2$ & $ 85.0$ & $ 67.4$ & $ 72.0$ \\ 
    \ion{Fe}{I} & $6027.06$ & $4.00$ & $-1.09$  &  $ 55.7$ & $ 57.8$ & $ 34.0$ & $ 38.6$ \\ 
    \ion{Fe}{I} & $6054.08$ & $4.30$ & $-2.31$  &   ---    &  ---    &  ---    &  ---    \\ 
    \ion{Fe}{I} & $6065.48$ & $2.60$ & $-1.53$  &  $148.7$ &  ---    & $124.5$ &  ---    \\ 
    \ion{Fe}{I} & $6078.49$ & $4.80$ & $-0.32$  &  $ 43.5$ & $ 45.5$ & $ 26.9$ & $ 37.8$ \\ 
    \ion{Fe}{I} & $6079.01$ & $4.60$ & $-1.12$  &  $ 21.2$ & $ 23.2$ & $ 13.6$ & $ 17.0$ \\ 
    \ion{Fe}{I} & $6082.71$ & $2.20$ & $-3.57$  &  $ 67.4$ & $ 69.1$ & $ 33.1$ & $ 50.0$ \\ 
    \ion{Fe}{I} & $6093.64$ & $4.60$ & $-1.50$  &  $ 20.3$ & $ 14.6$ & $ 20.1$ & $ 12.8$ \\ 
    \ion{Fe}{I} & $6137.70$ & $2.50$ & $-1.40$  &   ---    &  ---    & $137.6$ &  ---    \\ 
    \ion{Fe}{I} & $6151.62$ & $2.10$ & $-3.30$  &  $ 82.6$ &  ---    & $ 53.7$ & $ 72.7$ \\ 
    \ion{Fe}{I} & $6165.36$ & $4.10$ & $-1.47$  &  $ 32.4$ & $ 30.1$ & $ 16.6$ & $ 17.6$ \\ 
    \ion{Fe}{I} & $6180.21$ & $2.70$ & $-2.59$  &  $ 82.4$ & $ 84.8$ & $ 49.6$ & $ 64.2$ \\ 
    \ion{Fe}{I} & $6187.99$ & $3.90$ & $-1.72$  &  $ 40.0$ & $ 37.3$ & $ 22.4$ & $ 31.8$ \\ 
    \ion{Fe}{I} & $6213.44$ & $2.20$ & $-2.48$  &  $122.4$ &  ---    & $ 96.4$ &  ---    \\ 
    \ion{Fe}{I} & $6219.29$ & $2.20$ & $-2.43$  &  $131.7$ &  ---    & $109.2$ &  ---    \\ 
    \ion{Fe}{I} & $6220.78$ & $3.80$ & $-2.46$  &  $ 10.3$ &  ---    &  ---    &  ---    \\ 
    \ion{Fe}{I} & $6226.73$ & $3.80$ & $-2.22$  &  $ 17.5$ & $ 23.1$ & $ 13.8$ & $ 16.1$ \\ 
    \ion{Fe}{I} & $6252.57$ & $2.40$ & $-1.69$  &   ---    &  ---    & $134.3$ &  ---    \\ 
    \ion{Fe}{I} & $6254.25$ & $2.20$ & $-2.44$  &  $133.7$ &  ---    & $114.8$ &  ---    \\ 
    \ion{Fe}{I} & $6265.14$ & $2.10$ & $-2.55$  &  $128.2$ &  ---    & $110.6$ &  ---    \\ 
    \ion{Fe}{I} & $6270.23$ & $2.80$ & $-2.46$  &  $ 71.7$ & $ 76.1$ & $ 42.6$ & $ 58.0$ \\ 
    \ion{Fe}{I} & $6271.28$ & $3.30$ & $-2.70$  &  $ 27.4$ & $ 27.3$ & $ 15.2$ & $ 16.9$ \\ 
    \ion{Fe}{I} & $6301.51$ & $3.60$ & $-0.72$  &  $104.0$ &  ---    & $ 86.7$ &  ---    \\ 
    \ion{Fe}{I} & $6311.50$ & $2.80$ & $-3.14$  &  $ 43.6$ & $ 42.7$ & $ 23.3$ & $ 28.9$ \\ 
    \ion{Fe}{I} & $6315.31$ & $4.10$ & $-1.23$  &  $ 45.3$ & $ 41.9$ & $ 28.3$ &  ---    \\ 
    \ion{Fe}{I} & $6315.81$ & $4.00$ & $-1.71$  &  $ 29.2$ & $ 28.7$ & $ 10.3$ & $ 18.2$ \\ 
    \ion{Fe}{I} & $6335.34$ & $2.20$ & $-2.18$  &  $142.7$ &  ---    & $122.2$ &  ---    \\ 
    \ion{Fe}{I} & $6344.16$ & $2.40$ & $-2.92$  &  $ 91.2$ &  ---    & $ 62.8$ & $ 72.9$ \\ 
    \ion{Fe}{I} & $6380.75$ & $4.10$ & $-1.38$  &  $ 42.2$ & $ 35.9$ & $ 26.3$ & $ 33.5$ \\ 
    \ion{Fe}{I} & $6392.54$ & $2.20$ & $-4.03$  &  $ 36.9$ & $ 37.5$ &  ---    &  ---    \\ 
    \ion{Fe}{I} & $6393.61$ & $2.40$ & $-1.43$  &   ---    &  ---    & $140.8$ &  ---    \\ 
    \ion{Fe}{I} & $6411.66$ & $3.60$ & $-0.60$  &  $115.8$ &  ---    & $ 98.5$ & $101.7$ \\ 
    \ion{Fe}{I} & $6419.94$ & $4.70$ & $-0.24$  &  $ 57.5$ & $ 55.9$ & $ 40.6$ & $ 45.1$ \\ 
    \ion{Fe}{I} & $6421.35$ & $2.20$ & $-2.03$  &   ---    &  ---    & $132.5$ &  ---    \\ 
    \ion{Fe}{I} & $6430.86$ & $2.10$ & $-2.01$  &   ---    &  ---    & $132.8$ &  ---    \\ 
    \ion{Fe}{I} & $6469.21$ & $4.80$ & $-0.77$  &   ---    &  ---    &  ---    &  ---    \\ 
    \ion{Fe}{I} & $6475.63$ & $2.50$ & $-2.94$  &  $ 82.5$ &  ---    & $ 52.8$ & $ 68.5$ \\ 
    \ion{Fe}{I} & $6546.25$ & $2.70$ & $-1.54$  &  $136.6$ &  ---    & $116.3$ &  ---    \\ 
    \ion{Fe}{I} & $6569.22$ & $4.70$ & $-0.42$  &  $ 54.3$ & $ 55.8$ & $ 42.5$ & $ 46.4$ \\ 
    \ion{Fe}{I} & $6581.21$ & $1.40$ & $-4.68$  &  $ 52.6$ & $ 63.0$ & $ 16.7$ & $ 39.1$ \\ 
    \ion{Fe}{I} & $6593.87$ & $2.40$ & $-2.42$  &  $125.9$ &  ---    & $ 98.8$ &  ---    \\ 
    \ion{Fe}{I} & $6597.56$ & $4.80$ & $-1.07$  &  $ 21.0$ & $ 21.1$ &  ---    & $ 14.9$ \\ 
    \ion{Fe}{I} & $6608.04$ & $2.20$ & $-4.03$  &  $ 37.5$ & $ 37.9$ &  ---    &  ---    \\ 
    \ion{Fe}{I} & $6609.12$ & $2.50$ & $-2.69$  &  $ 97.6$ &  ---    & $ 64.0$ & $ 81.1$ \\ 
    \ion{Fe}{I} & $6627.54$ & $4.50$ & $-1.68$  &   ---    & $ 14.5$ &  ---    & $ 11.2$ \\ 
    \ion{Fe}{I} & $6678.00$ & $2.60$ & $-1.42$  &   ---    &  ---    & $134.5$ &  ---    \\ 
 \\ \hline \hline \\            
\end{longtable}

\begin{longtable}{lccccrrrr}
\caption{Line-by-line abundances ratios in the six UVES sample stars for the odd-Z (Na and Al), alpha- (Mg, Si, Ca) $+$ Ti, iron-peak (V, Mn, Co, Cu, and Zn), and heavy elements (Y, Zr, Ba, La, Nd, and Eu).}            
\label{tab:lines_ratios}\\   
\hline
\noalign{\smallskip}
\multirow{2}{*}{Species} & $\lambda$    & \phantom{-}\phantom{-}{ $\chi_{ex}$ }&  \multirow{2}{*}{log$gf$} &       star 1539   &    star 1363   &     star 1176  &  star 133 \\
                     &  [\AA]  &  [eV]        &                          & \multicolumn{4}{c}{[X/Fe]}\\ 
\hline\hline \endfirsthead
\multicolumn{8}{c}%
{{\bfseries \tablename\ \thetable{} -- continued}} \\ \hline
Species & $\lambda$    & \phantom{-}\phantom{-}{ $\chi_{ex}$ }&  log$gf$ &     star 1539   &    star 1363   &     star 1176  &  star 133  \\ 
\hline\hline \endhead

\ion{Mg}{I}   &  $6318.720$  &  $5.11$  & $-2.36$  &  $+0.35$  &  $+0.47$  &  $+0.38$  &  $+0.47$ \\ 
\ion{Mg}{I}   &  $6319.242$  &  $5.11$  & $-2.80$  &  $+0.32$  &  $+0.41$  &    ---    &    ---   \\ 
\ion{Si}{I}   &  $5665.555$  &  $4.92$  & $-2.04$  &  $+0.20$  &  $+0.20$  &  $+0.25$  &  $+0.40$ \\ 
\ion{Si}{I}   &  $5666.690$  &  $5.62$  & $-1.74$  &    ---    &    ---    &  $+0.56$  &  $+0.10$ \\ 
\ion{Si}{I}   &  $5690.425$  &  $4.93$  & $-1.87$  &  $+0.35$  &  $+0.33$  &  $+0.40$  &  $+0.30$ \\ 
\ion{Si}{I}   &  $5948.545$  &  $5.08$  & $-1.30$  &  $+0.30$  &  $+0.30$  &  $+0.30$  &  $+0.35$ \\ 
\ion{Si}{I}   &  $6142.494$  &  $5.62$  & $-1.50$  &  $+0.45$  &  $+0.30$  &    ---    &  $-0.40$ \\ 
\ion{Si}{I}   &  $6145.020$  &  $5.61$  & $-1.45$  &  $+0.30$  &  $+0.50$  &  $+0.15$  &  $+0.50$ \\ 
\ion{Si}{I}   &  $6155.142$  &  $5.62$  & $-0.85$  &  $+0.25$  &  $+0.19$  &  $+0.35$  &  $+0.30$ \\ 
\ion{Si}{I}   &  $6237.328$  &  $5.61$  & $-1.01$  &  $+0.06$  &  $+0.14$  &  $+0.25$  &  $+0.35$ \\ 
\ion{Si}{I}   &  $6243.823$  &  $5.61$  & $-1.30$  &  $+0.26$  &  $+0.21$  &  $+0.35$  &  $+0.35$ \\ 
\ion{Si}{I}   &  $6414.987$  &  $5.87$  & $-1.13$  &  $+0.23$  &  $+0.06$  &  $+0.56$  &  $+0.10$ \\ 
\ion{Si}{I}   &  $6721.844$  &  $5.86$  & $-1.17$  &    ---    &    ---    &  $+0.10$  &  $+0.69$ \\ 
\ion{Ca}{I}   &  $5601.277$  &  $2.53$  & $-0.52$  &  $+0.43$  &  $+0.47$  &  $+0.09$  &  $+0.68$ \\ 
\ion{Ca}{I}   &  $5867.562$  &  $2.93$  & $-1.55$  &  $+0.26$  &  $+0.31$  &  $+0.35$  &  $+0.45$ \\ 
\ion{Ca}{I}   &  $6156.030$  &  $2.52$  & $-2.39$  &    ---    &    ---    &    ---    &    ---   \\ 
\ion{Ca}{I}   &  $6102.723$  &  $1.88$  & $-0.79$  &  $+0.60$  &  $+0.50$  &  $+0.70$  &  $+0.70$ \\ 
\ion{Ca}{I}   &  $6122.217$  &  $1.89$  & $-0.20$  &  $+0.50$  &  $+0.50$  &  $+0.70$  &  $+0.70$ \\ 
\ion{Ca}{I}   &  $6161.295$  &  $2.51$  & $-1.02$  &  $+0.36$  &  $+0.44$  &  $+0.20$  &  $+0.44$ \\ 
\ion{Ca}{I}   &  $6162.167$  &  $1.90$  & $-1.09$  &  $+0.30$  &  $+0.30$  &  $+0.70$  &  $+0.50$ \\ 
\ion{Ca}{I}   &  $6166.440$  &  $2.52$  & $-0.90$  &  $+0.15$  &  $+0.30$  &  $+0.09$  &  $+0.30$ \\ 
\ion{Ca}{I}   &  $6169.044$  &  $2.52$  & $-0.54$  &  $+0.61$  &  $+0.50$  &  $+0.05$  &  $+0.55$ \\ 
\ion{Ca}{I}   &  $6169.564$  &  $2.52$  & $-0.27$  &  $+0.65$  &  $+0.58$  &  $+0.15$  &  $+0.71$ \\ 
\ion{Ca}{I}   &  $6439.080$  &  $2.52$  & $+0.30$  &  $+0.75$  &  $+0.55$  &  $+0.70$  &  $+0.75$ \\ 
\ion{Ca}{I}   &  $6455.605$  &  $2.52$  & $-1.35$  &  $+0.30$  &  $+0.44$  &  $+0.20$  &  $+0.53$ \\ 
\ion{Ca}{I}   &  $6464.679$  &  $2.52$  & $-2.10$  &  $+0.60$  &  $+0.70$  &  $+0.55$  &  $+0.70$ \\ 
\ion{Ca}{I}   &  $6493.788$  &  $2.52$  & $-2.44$  &  $+0.50$  &  $+0.50$  &  $+0.55$  &  $+0.50$ \\ 
\ion{Ca}{I}   &  $6499.654$  &  $2.52$  & $-0.85$  &  $+0.50$  &  $+0.50$  &  $+0.10$  &  $+0.50$ \\ 
\ion{Ca}{I}   &  $6572.779$  &  $0.00$  & $-4.32$  &  $+0.60$  &  $+0.49$  &  $-0.01$  &  $+0.55$ \\ 
\ion{Ca}{I}   &  $6717.687$  &  $2.71$  & $-0.61$  &  $+0.50$  &  $+0.50$  &  $+0.25$  &  $+0.50$ \\ 
\ion{Ti}{I}   &  $5922.108$  &  $1.05$  & $-1.46$  &  $+0.51$  &  $+0.55$  &  $+0.18$  &  $+0.48$ \\ 
\ion{Ti}{I}   &  $5941.750$  &  $1.05$  & $-1.50$  &  $+0.36$  &  $+0.40$  &  $+0.28$  &  $+0.33$ \\ 
\ion{Ti}{I}   &  $5965.825$  &  $1.88$  & $-0.42$  &  $+0.30$  &  $+0.45$  &  $+0.08$  &  $+0.34$ \\ 
\ion{Ti}{I}   &  $5978.539$  &  $1.87$  & $-0.53$  &  $+0.40$  &  $+0.30$  &  $+0.13$  &  $+0.37$ \\ 
\ion{Ti}{I}   &  $6064.623$  &  $1.05$  & $-1.94$  &  $+0.31$  &  $+0.30$  &  $+0.05$  &  $+0.28$ \\ 
\ion{Ti}{I}   &  $6091.169$  &  $2.27$  & $-0.42$  &  $+0.14$  &  $+0.26$  &    ---    &  $+0.23$ \\ 
\ion{Ti}{I}   &  $6126.214$  &  $1.07$  & $-1.43$  &  $+0.30$  &  $+0.40$  &  $+0.10$  &  $+0.30$ \\ 
\ion{Ti}{I}   &  $6258.110$  &  $1.44$  & $-0.36$  &  $+0.10$  &  $+0.15$  &  $+0.19$  &  $+0.10$ \\ 
\ion{Ti}{I}   &  $6261.106$  &  $1.43$  & $-0.48$  &  $+0.30$  &  $+0.50$  &  $+0.10$  &  $+0.30$ \\ 
\ion{Ti}{I}   &  $6312.240$  &  $1.46$  & $-1.60$  &  $+0.24$  &  $+0.40$  &  $+0.09$  &  $+0.35$ \\ 
\ion{Ti}{I}   &  $6336.113$  &  $1.44$  & $-1.74$  &  $+0.20$  &  $+0.34$  &  $+0.26$  &  $+0.18$ \\ 
\ion{Ti}{I}   &  $6554.238$  &  $1.44$  & $-1.22$  &  $+0.10$  &  $+0.15$  &  $+0.06$  &  $+0.26$ \\ 
\ion{Ti}{I}   &  $6556.077$  &  $1.46$  & $-1.07$  &  $+0.30$  &  $+0.40$  &  $+0.22$  &  $+0.32$ \\ 
\ion{Ti}{I}   &  $6599.113$  &  $0.90$  & $-2.09$  &  $+0.28$  &  $+0.30$  &  $+0.10$  &  $+0.30$ \\ 
\ion{Ti}{I}   &  $6743.127$  &  $0.90$  & $-1.73$  &  $+0.20$  &  $+0.26$  &  $-0.19$  &  $+0.20$ \\ 
\ion{Ti}{II}  &  $5418.751$  &  $1.58$  & $-2.13$  &  $+0.30$  &  $+0.38$  &  $+0.30$  &  $+0.43$ \\ 
\ion{Ti}{II}  &  $6491.580$  &  $2.06$  & $-2.10$  &  $+0.30$  &  $+0.38$  &  $+0.30$  &  $+0.43$ \\ 
\ion{Ti}{II}  &  $6559.576$  &  $2.05$  & $-2.35$  &  $+0.26$  &  $+0.30$  &  $+0.23$  &  $+0.36$ \\ 
\ion{Ti}{II}  &  $6606.970$  &  $2.06$  & $-2.85$  &  $+0.35$  &  $+0.30$  &  $+0.31$  &  $+0.30$ \\ 
\hline
\ion{Na}{I}   &  $5682.633$  &  $2.10$  & $-0.71$  &  $-0.32$  &  $-0.30$  &  $-0.00$  &  $+0.31$ \\ 
\ion{Na}{I}   &  $5688.194$  &  $2.10$  & $-1.40$  &  $-0.35$  &  $-0.30$  &  $-0.30$  &  $+0.20$ \\ 
\ion{Na}{I}   &  $6154.230$  &  $2.10$  & $-1.56$  &  $-0.15$  &  $-0.00$  &    ---    &  $+0.30$ \\ 
\ion{Na}{I}   &  $6160.753$  &  $2.10$  & $-1.26$  &  $-0.35$  &  $-0.00$  &  $-0.35$  &  $+0.00$ \\ 
\ion{Al}{I}   &  $6696.185$  &  $4.02$  & $-1.58$  &  $-0.30$  &  $-0.30$  &  $<+0.30$  &  $<+0.30$ \\ 
\ion{Al}{I}   &  $6698.673$  &  $3.14$  & $-1.65$  &  $-0.29$  &  $+0.00$  &  $<+0.30$  &  $<+0.30$ \\ 
\hline
\ion{V}{I}    &  $5703.560$  &  $1.05$  & $-0.21$  &  $+0.05$  &  $+0.17$  &    ---    &  $+0.14$ \\ 
\ion{V}{I}    &  $6081.440$  &  $1.05$  & $-0.58$  &  $-0.02$  &  $+0.17$  &    ---    &  $+0.02$ \\ 
\ion{V}{I}    &  $6090.220$  &  $1.08$  & $-0.16$  &  $+0.05$  &  $+0.14$  &  $-0.32$  &  $+0.08$ \\ 
\ion{V}{I}    &  $6119.520$  &  $1.06$  & $-0.47$  &  $-0.05$  &  $+0.08$  &  $-0.26$  &  $-0.05$ \\ 
\ion{V}{I}    &  $6199.190$  &  $0.29$  & $-1.48$  &  $+0.05$  &  $+0.17$  &    ---    &  $-0.11$ \\ 
\ion{V}{I}    &  $6243.100$  &  $0.30$  & $-0.88$  &  $+0.11$  &  $+0.38$  &    ---    &  $+0.05$ \\ 
\ion{V}{I}    &  $6251.820$  &  $0.29$  & $-1.44$  &  $+0.11$  &  $+0.32$  &  $-0.41$  &  $-0.05$ \\ 
\ion{V}{I}    &  $6274.650$  &  $0.27$  & $-1.72$  &  $-0.05$  &  $+0.11$  &    ---    &  $-0.08$ \\ 
\ion{Mn}{I}   &  $5394.669$  &  $0.00$  & $-3.55$  &  $-0.40$  &  $-0.50$  &  $-0.40$  &  $-0.50$ \\ 
\ion{Mn}{I}   &  $6013.513$  &  $3.07$  & $-0.40$  &  $-0.30$  &  $-0.25$  &  $-0.30$  &  $-0.30$ \\ 
\ion{Mn}{I}   &  $6016.640$  &  $3.07$  & $-0.22$  &  $-0.35$  &  $-0.50$  &  $-0.60$  &  $-0.40$ \\ 
\ion{Mn}{I}   &  $6021.800$  &  $3.08$  & $-0.10$  &  $-0.30$  &  $-0.45$  &  $-0.25$  &  $-0.50$ \\ 
\ion{Co}{I}   &  $5212.691$  &  $3.51$  & $-0.11$  &    ---    &  $+0.15$  &    ---    &  $+0.15$ \\ 
\ion{Co}{I}   &  $5301.047$  &  $1.71$  & $-2.00$  &  $+0.00$  &  $+0.10$  &    ---    &  $+0.10$ \\ 
\ion{Co}{I}   &  $5342.708$  &  $4.02$  & $+0.69$  &  $+0.05$  &  $+0.05$  &    ---    &  $+0.00$ \\ 
\ion{Co}{I}   &  $5454.572$  &  $4.07$  & $+0.24$  &    ---    &  $+0.10$  &  $+0.20$  &  $+0.10$ \\ 
\ion{Co}{I}   &  $5647.234$  &  $2.28$  & $-1.56$  &  $+0.05$  &  $+0.05$  &  $+0.00$  &  $+0.30$ \\ 
\ion{Co}{I}   &  $6188.996$  &  $1.71$  & $-2.45$  &  $+0.00$  &  $+0.00$  &  $+0.00$  &  $+0.30$ \\ 
\ion{Cu}{I}   &  $5105.537$  &  $1.39$  & $-1.52$  &  $-0.40$  &  $-0.15$  &  $-0.30$  &  $-0.35$ \\ 
\ion{Cu}{I}   &  $5218.197$  &  $3.82$  & $+0.00$  &  $-0.30$  &  $+0.00$  &  $+0.05$  &  $+0.00$ \\ 
\ion{Zn}{I}   &  $6362.339$  &  $5.79$  & $-0.30$  &  $-0.30$  &  $-0.20$  &  $+0.30$  &  $-0.10$ \\ 
\hline
\ion{Y}{I}    &  $6435.004$  &  $0.07$  & $-0.82$  &  $-0.30$  &  $-0.32$  &  $-0.30$  &  $+0.00$ \\ 
\ion{Y}{II}   &  $6795.414$  &  $1.74$  & $-1.19$  &  $+0.20$  &  $+0.00$  &  $+0.00$  &    ---   \\ 
\ion{Zr}{I}   &  $6127.475$  &  $0.15$  & $-1.06$  &  $-0.08$  &  $+0.05$  &    ---    &  $-0.08$ \\ 
\ion{Zr}{I}   &  $6134.585$  &  $0.00$  & $-1.42$  &  $+0.05$  &  $+0.20$  &    ---    &  $+0.23$ \\ 
\ion{Zr}{I}   &  $6140.535$  &  $0.52$  & $-1.60$  &    ---    &    ---    &    ---    &  $-0.50$ \\ 
\ion{Zr}{I}   &  $6143.252$  &  $0.07$  & $-1.10$  &  $-0.14$  &  $+0.02$  &    ---    &  $-0.08$ \\ 
\ion{Ba}{II}  &  $5853.675$  &  $0.60$  & $-1.10$  &  $+0.92$  &  $+1.00$  &  $+1.00$  &  $+1.05$ \\ 
\ion{Ba}{II}  &  $6141.713$  &  $0.70$  & $-0.08$  &  $+0.60$  &  $+0.80$  &  $+0.65$  &  $+0.80$ \\ 
\ion{Ba}{II}  &  $6496.897$  &  $0.60$  & $-0.32$  &  $+1.00$  &  $+1.00$  &  $+1.10$  &  $+1.20$ \\ 
\ion{La}{II}  &  $6262.287$  &  $0.40$  & $-1.60$  &  $+0.00$  &  $+0.00$  &  $+0.17$  &  $+0.26$ \\ 
\ion{La}{II}  &  $6320.376$  &  $0.17$  & $-1.56$  &  $+0.00$  &  $+0.00$  &  $+0.14$  &  $+0.30$ \\ 
\ion{La}{II}  &  $6390.477$  &  $0.32$  & $-1.41$  &  $+0.25$  &  $+0.17$  &  $+0.00$  &  $+0.26$ \\ 
\ion{Nd}{II}  &  $6740.078$  &  $0.06$  & $-1.53$  &  $+0.45$  &  $+0.15$  &  $+0.17$  &  $-0.30$ \\ 
\ion{Nd}{II}  &  $6790.372$  &  $0.18$  & $-1.77$  &  $+0.55$  &  $+0.40$  &  $+0.00$  &  $-0.30$ \\ 
\ion{Nd}{II}  &  $6549.525$  &  $0.06$  & $-2.01$  &  $+0.40$  &  $+0.30$  &  $+0.00$  &    ---   \\ 
\ion{Eu}{II}  &  $6437.640$  &  $1.32$  & $-0.32$  &  $+0.55$  &  $+0.55$  &  $+0.65$  &  $+0.50$ \\ 
\ion{Eu}{II}  &  $6645.064$  &  $1.38$  & $+0.12$  &  $+0.50$  &  $+0.55$  &  $+0.50$  &  $+0.70$ \\ 
\\ \hline \hline \\            
\end{longtable} 

\end{document}